\documentclass[prd,preprint,tightenlines,floatfix,
showpacs,preprintnumbers,nofootinbib,eqsecnum,superscriptaddress]{revtex4}

\usepackage[T1]{fontenc}		

\usepackage{amsmath,amsfonts,amssymb,amstext,mathrsfs}
\usepackage{mathpazo}

\usepackage[dvips]{graphicx}
\usepackage{epsf,float}
\usepackage{revsymb}

\usepackage{dcolumn}
\usepackage{braket}
\usepackage{color,xcolor}
\usepackage{graphicx}
\usepackage{subfigure}
\usepackage{multirow}
\usepackage{tabularx}
\usepackage{pstricks}
\usepackage[section]{placeins}
\usepackage{booktabs}
\usepackage{array}

\renewcommand\slash[1]{\not \! #1}

\newcommand{\ba}{\mbox{\boldmath $a$}}
\newcommand{\bb}{\mbox{\boldmath $b$}}

\newcommand{\bba}{\mbox{\boldmath $b_{1}$}}
\newcommand{\bbb}{\mbox{\boldmath $b_{2}$}}
\newcommand{\bbi}{\mbox{\boldmath $b_{i}$}}

\newcommand{\bp}{\mbox{\boldmath $p$}}

\newcommand{\bpp}{\mbox{\boldmath $p_{3}$}}
\newcommand{\bppbar}{\mbox{\boldmath $p_{4}$}}

\newcommand{\bka}{\mbox{\boldmath $k_{1}$}}
\newcommand{\bkb}{\mbox{\boldmath $k_{2}$}}
\newcommand{\bkj}{\mbox{\boldmath $k_{j}$}}

\newcommand{\bhp}{\mbox{\boldmath $\hat{p}$}}

\newcommand{\bhpp}{\mbox{\boldmath $\hat{p}_{3}$}}
\newcommand{\bhppbar}{\mbox{\boldmath $\hat{p}_{4}$}}

\newcommand{\bsigma}{\mbox{\boldmath $\sigma$}}

\newcommand{\bea}{\mbox{\boldmath $e_{x}$}}
\newcommand{\beb}{\mbox{\boldmath $e_{y}$}}
\newcommand{\bec}{\mbox{\boldmath $e_{z}$}}

\newcommand{\bepsilona}{\mbox{\boldmath $\epsilon_{1}$}}
\newcommand{\bepsilonb}{\mbox{\boldmath $\epsilon_{2}$}}
\newcommand{\bepsilonj}{\mbox{\boldmath $\epsilon_{j}$}}

\usepackage{hyperref}

\begin{document}

\title{From the \boldmath{$\gamma \gamma \to p \bar{p}$} reaction 
to the production of \boldmath{$p \bar{p}$} pairs \\ 
in ultraperipheral ultrarelativistic heavy-ion collisions at the LHC}

\author{Mariola K{\l}usek-Gawenda}
 \email{Mariola.Klusek@ifj.edu.pl}
\affiliation{Institute of Nuclear Physics Polish Academy of Sciences, Radzikowskiego 152, PL-31-342 Krak\'ow, Poland}

\author{Piotr Lebiedowicz}
 \email{Piotr.Lebiedowicz@ifj.edu.pl}
\affiliation{Institute of Nuclear Physics Polish Academy of Sciences, Radzikowskiego 152, PL-31-342 Krak\'ow, Poland}

\author{Otto Nachtmann}
 \email{O.Nachtmann@thphys.uni-heidelberg.de}
\affiliation{Institut f\"ur Theoretische Physik, Universit\"at Heidelberg,
Philosophenweg 16, D-69120 Heidelberg, Germany}

\author{Antoni Szczurek
\footnote{Also at \textit{Faculty of Mathematics and Natural Sciences, University of Rzesz\'ow, Pigonia 1, PL-35-310 Rzesz\'ow, Poland}.}}
\email{Antoni.Szczurek@ifj.edu.pl}
\affiliation{Institute of Nuclear Physics Polish Academy of Sciences, Radzikowskiego 152, PL-31-342 Krak\'ow, Poland}

\begin{abstract}
In this paper we consider the production of proton-antiproton pairs
in two-photon interactions in electron-positron and heavy-ion collisions.
We try to understand the dependence of the total cross section 
on the photon-photon c.m. energy as well as corresponding angular
distributions measured by the Belle Collaboration 
for the $\gamma \gamma \to p \bar{p}$ process.
To understand the Belle data we include the proton-exchange, 
the $f_2(1270)$ and $f_2(1950)$ $s$-channel exchanges, 
as well as the hand-bag mechanism. 
The helicity amplitudes for the $\gamma \gamma \to f_{2} \to p \bar{p}$ 
process are written explicitly based on a Lagrangian approach.
The parameters of vertex form factors are adjusted to the Belle data. 
Having described the angular distributions 
for the $\gamma \gamma \to p \bar{p}$ process
we present first predictions for the ultraperipheral, ultrarelativistic, heavy-ion reaction
$^{208}\!Pb \, ^{208}\!Pb \to$ $^{208}\!Pb \, ^{208}\!Pb \,p \bar{p}$.
Both, the total cross section and several differential distributions 
for experimental cuts corresponding to the ALICE, ATLAS, CMS, and LHCb
experiments are presented.
We find the total cross section 100~$\mu$b for the ALICE cuts,
160~$\mu$b for the ATLAS cuts, 500~$\mu$b for the CMS cuts,
and 104~$\mu$b taking into account the LHCb cuts.
This opens a possibility to study the $\gamma \gamma \to p \bar{p}$ process at the LHC.
\end{abstract}


\maketitle

\section{Introduction}
\label{sec:intro}

The baryon pair production via $\gamma \gamma$ fusion was measured at
electron-positron colliders by various experimental groups: 
CLEO \cite{Artuso:1993xk} at CESR,
VENUS \cite{Hamasaki:1997cy} at TRISTAN, 
OPAL \cite{Abbiendi:2002bxa} and L3 \cite{Achard:2003jc} at LEP, 
and Belle \cite{Kuo:2005nr} at KEKB.
In the latter experiment the $\gamma \gamma \to p \bar{p}$ cross sections
were extracted from the $e^+ e^- \to e^+ e^- p \bar p$ reaction
for the $\gamma \gamma$ center-of-mass (c.m.) 
energy range of $2.025 < W_{\gamma\gamma} < 4$~GeV
and in the c.m. angular range of $|\cos\theta|<0.6$.

QCD predictions for $\gamma \gamma \to p \bar{p}$
were first calculated in \cite{Farrar:1985gv,Farrar:1988vz} 
using the leading twist nucleon wave functions determined 
from QCD sum rules, see e.g. \cite{Chernyak:1984bm}. 
The calculated cross sections from the leading-twist QCD terms
turned out to be about one order of magnitude smaller than the experimental data.
To explain these experimental observations,
various phenomenological approaches were suggested.
For example, in the diquark model, 
which is a variant of the leading-twist approach,
see e.g. \cite{Berger:2002vc} and references therein,
the proton was considered to be a quark-diquark system
and a diquark form factor was introduced.
In the hand-bag approach, see e.g. \cite{Diehl:2002yh},
the $\gamma \gamma \to p \bar{p}$ amplitude
was factorized into a hard $\gamma \gamma \to q \bar{q}$
subprocess and form factors describing a soft $q \bar{q} \to p \bar{p}$ transition. 
The transition form factors could not be calculated from first
principles in QCD and were, therefore, determined phenomenologically.
The pQCD-inspired phenomenological models have more chances to describe 
the absolute size of the cross section for $W_{\gamma\gamma} > 2.5$~GeV,
however, they contain a number of free parameters that are fitted to data.
Moreover, most data were taken at energies which are rather low for
the kinematic requirements of large $s$, $|t|$, $|u|$
in the hand-bag approach.

The low center-of-mass energy region of $\gamma \gamma \to p \bar{p}$
may be dominated by $s$-channel resonance contributions.
One of the effective approaches used for this region 
is the Veneziano model \cite{Odagiri:2004mn}.
While a reasonable $\sigma (W_{\gamma\gamma})$ dependence was obtained
without adjustable parameters,
the agreement of the model with 
the angular distributions was only qualitative.

In a recent calculation \cite{Ahmadov:2016sdg} only the proton exchange
contribution was considered.
But we think that this calculation has some problems
as we shall discuss below in Sec.~\ref{sec:nucleon_exchange}.

In our approach we wish to include all important theory ingredients
in order to achieve a quantitative description of the Belle data.
Then we present our predictions for the production of $p \bar{p}$ pairs 
in the ultraperipheral, ultrarelativistic, heavy-ion collisions at the LHC.
To describe the dynamics of the $\gamma \gamma \to p \bar{p}$ process
we take into account not only the nonresonant proton exchange 
contribution but also the $s$-channel tensor meson exchange contributions
and the hand-bag mechanism.
A measurement of the $^{208}\!Pb \, ^{208}\!Pb \to$ $^{208}\!Pb \, ^{208}\!Pb \,p \bar{p}$ reaction
will provide further information on the two-photon interactions involved
and, thus, will allow further tests of existing theoretical approaches.

\section{The $\gamma \gamma \to p \bar{p}$ reaction}
\label{sec:gamgam_ppbar}

\begin{figure}[!ht]
(a)\includegraphics[width=0.32\textwidth]{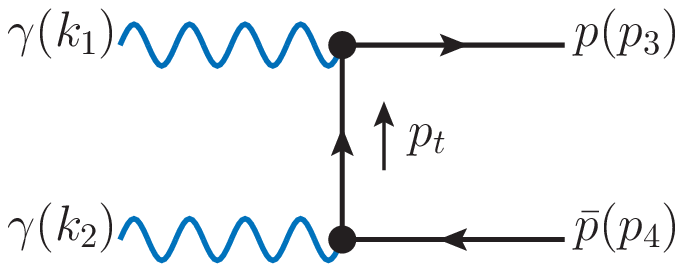}
(b)\includegraphics[width=0.32\textwidth]{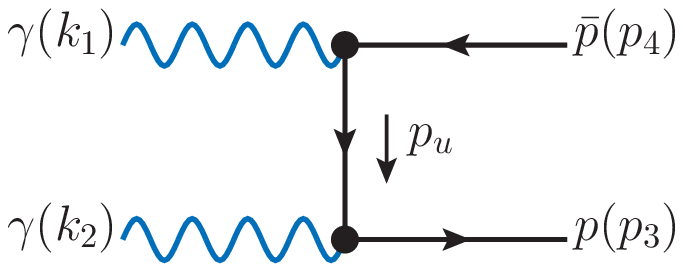}
(c)\includegraphics[width=0.34\textwidth]{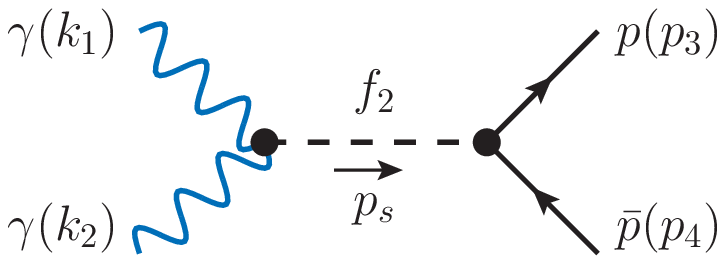}
(d)\includegraphics[width=0.4\textwidth]{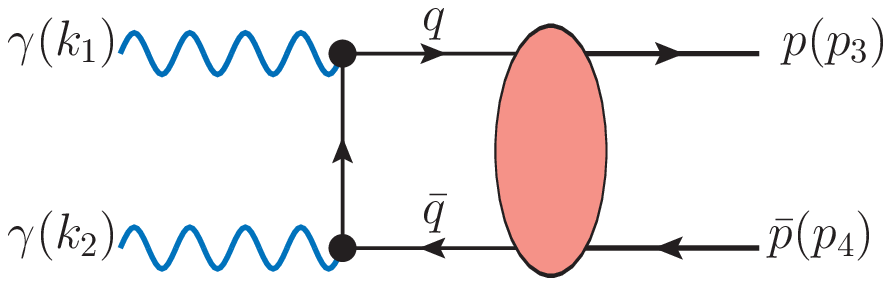}
  \caption{\label{fig:diagram_2to2}
  \small
Diagrams for the production of $p \bar{p}$ in $\gamma \gamma$ collisions.
We consider the $t$- and $u$-channel proton exchange
(diagrams (a) and (b), respectively),
the exchange of $f_{2}$ meson in the $s$-channel (diagram (c)),
and the hand-bag mechanism (diagram (d)
plus the one with the photon vertices interchanged).
Here $f_{2}$ stands generically for a $J^{PC} = 2^{++}$ meson.
}
\end{figure}

We consider the reaction (see Fig.~\ref{fig:diagram_2to2})
\begin{eqnarray}
&&\gamma(k_{1},\epsilon_{1}) + \gamma(k_{2},\epsilon_{2}) \to p(p_{3},s_{3}) + \bar{p}(p_{4},s_{4}) \,,
\nonumber\\
&&s_{3}, s_{4} \in \lbrace 1/2, -1/2 \rbrace \,,
\label{2to2_reaction}
\end{eqnarray}
where the momenta, the polarization vectors of the photons, and
the helicity indices for proton and antiproton
are indicated in brackets.
In the following we shall calculate the $\cal T$-matrix element
for the reaction (\ref{2to2_reaction}),
\begin{eqnarray}
\langle p(p_{3},s_{3}), \bar{p}(p_{4},s_{4}) |{\cal T}| \gamma(k_{1},\epsilon_{1}), \gamma(k_{2},\epsilon_{2}) \rangle 
&=&{\cal M}^{\mu\nu}(p_{3},p_{4};k_{1},k_{2}) \,\epsilon_{1 \mu} \,\epsilon_{2 \nu} \nonumber \\
&\equiv & {\cal M}_{\gamma \gamma \to p \bar{p}}
\,,
\label{2to2_amp1}
\end{eqnarray}
for nonresonant proton exchange, exchange of spin 2 mesons in the $s$-channel,
and for the hand-bag mechanism.
We note that gauge invariance requires
\begin{eqnarray}
&&{\cal M}^{\mu\nu}(p_{3},p_{4};k_{1},k_{2}) \,k_{1 \mu} = 0\,, \nonumber \\
&&{\cal M}^{\mu\nu}(p_{3},p_{4};k_{1},k_{2}) \,k_{2 \nu} = 0\,.
\label{2to2_amp2}
\end{eqnarray}
Since the photons are bosons we must have
\begin{eqnarray}
{\cal M}^{\mu\nu}(p_{3},p_{4};k_{1},k_{2}) = 
{\cal M}^{\nu\mu}(p_{3},p_{4};k_{2},k_{1}) \,.
\label{2to2_amp2_aux}
\end{eqnarray}
%

The kinematical variables used in the present paper are
(see Fig.~\ref{fig:diagram_2to2})
\begin{eqnarray}
&&s = (k_{1} + k_{2})^{2} = (p_{3} + p_{4})^{2} = W_{\gamma \gamma}^{2}\,, \nonumber \\
&&t = (k_{1} - p_{3})^{2} = (k_{2} - p_{4})^{2}\,,\nonumber \\
&&u = (k_{1} - p_{4})^{2} = (k_{2} - p_{3})^{2}\,,\nonumber \\
&&s+t+u = 2m_{p}^{2} \,;\label{2to2_kinematic_app} \\
&&p_{s} = k_{1} + k_{2} = p_{3} + p_{4}\,,\nonumber \\
&&p_{t} = k_{2} - p_{4} = p_{3} - k_{1}\,,\nonumber \\
&&p_{u} = k_{1} - p_{4} = p_{3} - k_{2}\,;\\
&&p_{s}^{2} = s\,,\; p_{t}^{2} = t\,,\; p_{u}^{2} = u\,.
\label{2to2_kinematic}
\end{eqnarray}
We shall work in the c.m. frame of the reaction (\ref{2to2_reaction});
see Fig.~\ref{fig:cm_system} in Appendix~\ref{section:Appendix1}.
For the incoming photons we use the polarization vectors (\ref{B_31})
and the helicity spinors for the proton are as in (\ref{B_02}) -- (\ref{B_04})
with $\theta \to \theta$, $\phi \to 0$.
The helicity spinors for the antiproton
are obtained from (\ref{B_20}) and (\ref{B_10}), (\ref{B_11}),
with $\theta \to \pi-\theta$, $\phi \to \pi$.

There are 16 helicity amplitudes
\begin{eqnarray}
&&\langle 
p(\bpp,s_{3}), \bar{p}(\bppbar,s_{4})
|{\cal T}| 
\gamma(\bka,m_{1}), \gamma(\bkb,m_{2})\rangle \equiv
\langle 2s_{3},2s_{4}|{\cal T}| m_{1},m_{2} \rangle \,.
\label{hel_amp}
\end{eqnarray}
Here $s_{3}$, $s_{4} \in \{1/2,-1/2 \}$ 
and $m_{1}$, $m_{2} \in \{1,-1 \}\,$
are the helicity labels of proton, antiproton and the photons,
respectively. We have also introduced
a convenient shorthand notation for the amplitudes.
Using rotational, parity and charge-conjugation invariance
one finds that only 6 of the 16 helicity amplitudes
are independent which we denote by $\psi_{1}(s,t)$, ..., $\psi_{6}(s,t)$;
see (\ref{B_44}) and Table~\ref{tab:helicity_amp} of Appendix~\ref{section:Appendix1}.

The unpolarized differential cross section 
for the reaction (\ref{2to2_reaction}) is given by
\begin{eqnarray}
\frac{d\sigma}{d\cos\theta} =
\frac{1}{32 \pi s} \;
\frac{|\bpp|}{|\bka|} \;
\frac{1}{4} \sum_{ {\rm spins}}
|{\cal M}_{\gamma \gamma \to p \bar{p}}|^{2} \,,
\label{xsecttion_diff_2to2}
\end{eqnarray}
where $s$ is the invariant mass squared of the $\gamma \gamma$ system,
$\theta$ denotes the angle of the outgoing nucleon relative to the beam
direction in the c.m. frame, see Fig.~\ref{fig:cm_system}
in Appendix~\ref{section:Appendix1},
and $\bka$ and $\bpp$ are the c.m. 3-momenta
of the initial photon and final nucleon, respectively; see (\ref{B_29}).

\subsection{Nonresonant proton exchange contribution}
\label{sec:nucleon_exchange}

The amplitude for the proton exchange mechanism
[see the diagrams (a) and (b) in Fig.~\ref{fig:diagram_2to2}]
is written as
\begin{eqnarray}
{\cal M}^{(p\;\rm{exchange})}_{\rm{bare}}
&=&
(-i)\,
\epsilon_{1 \mu}
\epsilon_{2 \nu}
\,\bar{u}(p_{3}) 
\Big(
i\Gamma^{(\gamma pp)\,\mu}(p_{3},p_{t})
\frac{i(\slash{p}_{t} + m_{p})}{t - m_{p}^2+i\epsilon}
i\Gamma^{(\gamma pp)\,\nu}(p_{t},-p_{4}) \nonumber\\
&&
+
i\Gamma^{(\gamma pp)\,\nu}(p_{3},p_{u})
\frac{i(\slash{p}_{u} + m_{p})}{u - m_{p}^2+i\epsilon}
i\Gamma^{(\gamma pp)\,\mu}(p_{u},-p_{4}) \Big)
v(p_{4}) 
\,.
\label{amplitude_2to2}
\end{eqnarray}
Here we use the free proton propagator for the internal proton lines
and the photon-proton vertex function 
as for on-shell protons respectively antiprotons.
This photon-proton vertex function is,
with $q = p' - p$, given by
\begin{equation}\label{F1_normalisation}
\begin{split}
i\Gamma^{(\gamma pp)}_{\mu}(p',p) &= 
-ie\left[ \gamma_{\mu} F_{1}(q^{2})+ \frac{i}{2 m_{p}}
\sigma_{\mu \nu} q^{\nu} F_{2}(q^{2}) \right]
\,;
\end{split}
\end{equation}
see e.g. (3.26) of \cite{Ewerz:2013kda}.
In (\ref{F1_normalisation}) $\sigma_{\mu \nu} =\dfrac{i}{2}[\gamma_{\mu},\gamma_{\nu}]$,
$F_{1}$ and $F_{2}$ are the Dirac and Pauli form factors of the proton,
respectively. For real photons we have
$F_{1}(0) = 1$ and $F_{2}(0) = \kappa_{p} = 1.7928$,
where $\kappa_{p}$ is the anomalous magnetic moment of the proton.
The amplitude (\ref{amplitude_2to2}) satisfies the gauge-invariance relations (\ref{2to2_amp2}) and the Bose-symmetry relation (\ref{2to2_amp2_aux}). 

Of course, the virtual protons in the diagrams of Fig.~\ref{fig:diagram_2to2}~(a) and (b)
are off shell. Their propagators will, in general, not be the ones of free protons
and the photon-proton vertex functions also will have an off-shell dependence.
We take these off-shell dependences into account
via multiplication of the amplitude (\ref{amplitude_2to2}) 
by an extra form factor.
We adopt here the scheme used in previous works 
\cite{Poppe:1986dq, Szczurek:2002bn, Klusek-Gawenda:2013rtu, Lebiedowicz:2014bea}
and set
\begin{eqnarray}
F(t,u,s) = \frac{\left[F(t)\right]^{2} + [F(u)]^{2}}{1+[\tilde{F}(s)]^{2}}\,,
\label{ff_tus}
\end{eqnarray}
with the exponential parametrizations
\begin{eqnarray}\label{ff_Poppe}
F(t) &=& \exp \left( \frac{t-m_{p}^{2}}{\Lambda_{p}^{2}} \right)\,, \nonumber \\
F(u) &=& \exp \left( \frac{u-m_{p}^{2}}{\Lambda_{p}^{2}} \right)\,,  \\
\tilde{F}(s) &=& \exp \left( \frac{-(s-4m_{p}^{2})}{\Lambda_{p}^{2}} \right)\,. \nonumber
\end{eqnarray}
The parameter $\Lambda_{p}$ should be fitted to the experimental data.
Note that the form factor $F(t)$ is normalized to unity for $t = m_{p}^{2}$.

Our complete result for the nonresonant proton exchange contribution reads, therefore, 
\begin{eqnarray}\label{amplitude_2to2_complete}
{\cal M}^{(p\;\rm{exchange})} = {\cal M}^{(p\;\rm{exchange})}_{\rm{bare}}\,
F(t,u,s) \,.
\end{eqnarray}
The multiplication of the ``bare'' amplitude with a common form factor
guarantees that the gauge-invariance relations (\ref{2to2_amp2})
are satisfied for ${\cal M}^{(p\;\rm{exchange})}$.
Also the Bose-symmetry relation (\ref{2to2_amp2_aux})
is satisfied
\footnote{The amplitude for $\gamma \gamma \to p \bar{p}$
considered in Eqs. (8) - (10) of \cite{Ahmadov:2016sdg}
does not satisfy the Bose-symmetry relation (\ref{2to2_amp2_aux}).
Therefore, this amplitude and the corresponding cross section,
(16) and (20) of \cite{Ahmadov:2016sdg}, cannot correspond to reality.}
by (\ref{amplitude_2to2_complete}) since
${\cal M}^{(p\;\rm{exchange})}_{\rm{bare}}$ satisfies (\ref{2to2_amp2_aux})
and the form factor $F(t,u,s)$ is symmetric under the exchange
$t \leftrightarrow u$; see (\ref{amplitude_2to2}) and (\ref{ff_tus}).
\subsection{$f_{2}$ meson contributions}
\label{sec:resonances}

In this section we discuss the contributions from the $s$-channel exchange
of $J^{PC} = 2^{++}$ mesons, generically denoted by $f_{2}$ in diagram (c)
of Fig.~\ref{fig:diagram_2to2}. 
In the following we shall take into account
the $f_{2}(1270)$ and $f_{2}(1950)$ resonances.
That is, in the formulas $f_{2}$ stands for any of these resonances.
In the final calculations their contributions are summed.

The amplitude for the $p\bar{p}$ production through 
the $s$-channel exchange of a tensor meson $f_{2}$
[the corresponding diagram is shown in Fig.~\ref{fig:diagram_2to2}~(c)]
is written as
\begin{eqnarray}
{\cal M}^{(f_{2}\;\rm{exchange})} =
(-i)\,
\bar{u}(p_{3}) 
i\Gamma^{(f_{2} p\bar{p})\,\alpha \beta}(p_{3},p_{4}) \,
v(p_{4}) \,
i\Delta^{(f_{2})}_{\alpha \beta,\kappa \lambda}(p_{s}) \,
i\Gamma^{(f_{2} \gamma \gamma)\, \mu \nu \kappa \lambda}(k_{1},k_{2}) \,  
\epsilon_{1 \mu}
\epsilon_{2 \nu}
\,. \nonumber \\
\label{f2_resonance}
\end{eqnarray}

The $f_{2} \gamma \gamma$ vertex is given as 
\begin{equation}\label{3.29new}
\begin{split}
i \Gamma_{\mu\nu\kappa\lambda}^{(f_2 \gamma \gamma)} (k_1,k_2) 
=i 
\left[
2 a_{f_2 \gamma\gamma}\, F_{a}^{(f_2 \gamma \gamma)}(p_{s}^2) \,\Gamma_{\mu\nu\kappa\lambda}^{(0)}(k_1,k_2) 
- b_{f_2 \gamma\gamma}\, F_{b}^{(f_2 \gamma \gamma)}(p_{s}^2) \,\Gamma_{\mu\nu\kappa\lambda}^{(2)}(k_1,k_2) 
\right] \,,
\end{split}
\end{equation}
with two rank-four tensor functions,
\begin{eqnarray}
\label{3.15}
&&\Gamma_{\mu\nu\kappa\lambda}^{(0)} (k_1,k_2) =
\Big[(k_1 \cdot k_2) g_{\mu\nu} - k_{2\mu} k_{1\nu}\Big] 
\Big[k_{1\kappa}k_{2\lambda} + k_{2\kappa}k_{1\lambda} - 
\frac{1}{2} (k_1 \cdot k_2) g_{\kappa\lambda}\Big] \,,\\
\label{3.16}
&&\Gamma_{\mu\nu\kappa\lambda}^{(2)} (k_1,k_2) = \,
 (k_1\cdot k_2) (g_{\mu\kappa} g_{\nu\lambda} + g_{\mu\lambda} g_{\nu\kappa} )
+ g_{\mu\nu} (k_{1\kappa} k_{2\lambda} + k_{2\kappa} k_{1\lambda} ) \nonumber \\
&& \qquad \qquad \qquad \quad - k_{1\nu} k_{2 \lambda} g_{\mu\kappa} - k_{1\nu} k_{2 \kappa} g_{\mu\lambda} 
- k_{2\mu} k_{1 \lambda} g_{\nu\kappa} - k_{2\mu} k_{1 \kappa} g_{\nu\lambda} 
\nonumber \\
&& \qquad \qquad \qquad \quad - [(k_1 \cdot k_2) g_{\mu\nu} - k_{2\mu} k_{1\nu} ] \,g_{\kappa\lambda} \,;
\end{eqnarray}
see (3.39) and (3.18) -- (3.22) of \cite{Ewerz:2013kda}.
In our case we have $k_1^2 =k_2^2 =0$.

For the $f_{2}(1270)$ meson, the coupling constants
$a_{f_2 \gamma\gamma}$ and
$b_{f_2 \gamma\gamma}$ are estimated in Secs. 5.3 and 7.2 of \cite{Ewerz:2013kda}.
In the case of the $f_{2}(1950)$ meson the numerical values of the $a$ and $b$ parameters
will be obtained here from a fit to the Belle data \cite{Kuo:2005nr}.
In (\ref{3.29new}) we have introduced form factors
$F_{a,b}^{(f_2 \gamma \gamma)}(p_{s}^2)$
describing the $s$ dependence of the $f_{2} \gamma \gamma$ coupling.
These form factors will be particularly important for the diagram
Fig.~\ref{fig:diagram_2to2}~(c) with $f_{2}(1270)$ exchange
since in $p\bar{p}$ production this meson significantly contributes 
but only far off shell.

Let us now discuss in detail the $f_{2} p\bar{p}$ vertex.
From the $l$-$S$ analysis, presented in Appendix~\ref{section:Appendix2},
we know that there are two independent couplings
corresponding to $(l,S) = (1,1)$ and $(3,1)$.
In accord with this we choose two coupling Lagrangians, (\ref{A01}) and (\ref{A02}) below, 
which correspond to two linearly independent combinations of the two $(l,S)$ possibilities;
see Appendix~\ref{section:Appendix2}. We set
\begin{eqnarray}
&&{\cal L}'^{(1)}_{f_{2}pp}(x) =
- \frac{g_{f_{2}pp}^{(1)}}{M_{0}} {f_{2\,\kappa\lambda}}(x)
\frac{i}{2} \bar{\psi}_p(x)
\left[ 
  \gamma^\kappa  \stackrel{\leftrightarrow}{\partial^\lambda}
+ \gamma^\lambda \stackrel{\leftrightarrow}{\partial^\kappa}
- \frac{1}{2} g^{\kappa\lambda} \gamma^\rho
                 \stackrel{\leftrightarrow}{\partial_\rho} 
\right] \psi_p(x)\,,
\label{A01}\\
&&{\cal L}'^{(2)}_{f_{2}pp}(x) =
\frac{g_{f_{2}pp}^{(2)}}{M_{0}^{2}} {f_{2\,\kappa\lambda}}(x)
\bar{\psi}_p(x) 
\left[ \stackrel{\leftrightarrow}{\partial^\kappa}
       \stackrel{\leftrightarrow}{\partial^\lambda}
- \frac{1}{4} g^{\kappa\lambda} \stackrel{\leftrightarrow}{\partial^\rho} 
                                \stackrel{\leftrightarrow}{\partial_\rho} 
\right] \psi_p(x)\,,
\label{A02}
\end{eqnarray}
where $\psi_p(x)$ and $f_{2}(x)$ are the proton and $f_{2}$ meson field operators, respectively.
The corresponding vertices, including form factors, are
\begin{eqnarray}
&&i\Gamma_{\kappa\lambda}^{(f_{2} p\bar{p})(1)}(p_{3},p_{4})=
-i \frac{g_{f_{2} pp}^{(1)}}{M_{0}} 
\left[ \frac{1}{2} \gamma_{\kappa}(p_{3}-p_{4})_{\lambda} 
     + \frac{1}{2} \gamma_{\lambda}(p_{3}-p_{4})_{\kappa} 
     - \frac{1}{4} g_{\kappa\lambda} ( \slash{p}_{3} - \slash{p}_{4} )
\right] \nonumber\\ 
&&\qquad \qquad \qquad \qquad \;\, \times F^{(f_{2}p\bar{p})(1)}[(p_{3}+p_{4})^{2}]\,,
\label{A03}\\
&&i\Gamma_{\kappa\lambda}^{(f_{2} p\bar{p})(2)}(p_{3},p_{4})=
-i \frac{g_{f_{2} pp}^{(2)}}{M_{0}^{2}} 
\left[ (p_{3}-p_{4})_{\kappa} (p_{3}-p_{4})_{\lambda} 
     - \frac{1}{4} g_{\kappa\lambda} (p_{3} - p_{4})^{2}
\right] \nonumber\\ 
&&\qquad \qquad \qquad \qquad \;\, \times F^{(f_{2}p\bar{p})(2)}[(p_{3}+p_{4})^{2}]\,.
\label{A04}
\end{eqnarray}
Here $g_{f_{2}pp}^{(j)}$ ($j=1,2$) are dimensionless coupling constants
and $M_{0} \equiv 1$~GeV.
The complete $f_{2} p \bar{p}$ vertex function is given by
\begin{eqnarray}
i\Gamma_{\kappa\lambda}^{(f_{2} p\bar{p})}(p_{3},p_{4})=
\sum_{j=1,2} i\Gamma_{\kappa\lambda}^{(f_{2} p\bar{p})(j)}(p_{3},p_{4})\,.
\label{A05}
\end{eqnarray}

For the $f_{2}$ propagator we use the simple formula
\begin{eqnarray}
i\Delta_{\alpha \beta,\kappa \lambda}^{(f_{2})}(p_{s})&=&
iP^{(2)}_{\alpha \beta,\kappa \lambda}(p_{s}) \, \Delta^{(2)}(p_{s}^{2}) \nonumber\\
&=&
i\Big[ 
\frac{1}{2} 
\left( \hat{g}_{\alpha \kappa} \hat{g}_{\beta \lambda}   + \hat{g}_{\alpha \lambda} \hat{g}_{\beta \kappa}  \right)
-\frac{1}{3} 
\hat{g}_{\alpha \beta} \hat{g}_{\kappa \lambda} 
\Big] 
\frac{1}{p_{s}^{2}-m_{f_{2}}^2+i m_{f_{2}} \Gamma_{f_{2}}}\,, 
\label{prop_f2}
\end{eqnarray}
where $\hat{g}_{\mu \nu} = -g_{\mu \nu} + p_{s \mu} p_{s \nu} / p_{s}^2$.
$\Gamma_{f_{2}}$ 
is the total decay width of the $f_{2}$ resonance
and $m_{f_{2}}$ its mass.
For a more detailed analysis we should use a model for the $f_{2}$ propagator 
along the lines considered in \cite{Ewerz:2013kda}; 
see (3.6) -- (3.8) and Appendix~A of \cite{Ewerz:2013kda}.

With the expressions from Appendix~\ref{section:Appendix1}
we get the helicity amplitudes
for the reaction $\gamma \gamma \to f_{2} \to p \bar{p}$,
using the notation of (\ref{B_41})
and $\varepsilon = (\varepsilon_{rs})$
as defined in (\ref{B_16}), as follows
\begin{eqnarray}
&&\langle 2s_{3}, 2s_{4} |{\cal T}| + , + \rangle =
  \langle 2s_{3}, 2s_{4} |{\cal T}| - , - \rangle \nonumber\\
&&= -\frac{1}{2} s^{2} \sqrt{s - 4 m_{p}^{2}} \,\Delta^{(2)}(s) \,a_{f_{2} \gamma \gamma}\,
F_{a}^{(f_{2} \gamma \gamma)}(s)\nonumber\\
&& \quad \times 
\Big\lbrace 
\frac{g_{f_{2}pp}^{(1)}}{M_{0}} F^{(f_{2}p \bar{p})(1)}(s) 
\Big[ -2 m_{p} \Big( \cos^{2}\theta - \frac{1}{3} \Big) \,\delta_{s_{3}s_{4}} 
      -\sqrt{s}\sin\theta \cos\theta \,\varepsilon_{s_{3}s_{4}}
\Big]\nonumber\\
&& \qquad +
\frac{g_{f_{2}pp}^{(2)}}{M_{0}^{2}} F^{(f_{2}p \bar{p})(2)}(s) \,
(s-4m_{p}^{2}) \Big( \cos^{2}\theta - \frac{1}{3} \Big) \,\delta_{s_{3}s_{4}}
\Big\rbrace
\,,
\label{hel0_ppbar}\\
&&\langle 2s_{3}, 2s_{4} |{\cal T}| \pm , \mp \rangle \nonumber\\
&&= -\frac{1}{2} s \sqrt{s - 4 m_{p}^{2}} \,\Delta^{(2)}(s) \,b_{f_{2} \gamma \gamma}\, 
F_{b}^{(f_{2} \gamma \gamma)}(s)\nonumber\\
&& \quad \times 
\Big\lbrace 
\frac{g_{f_{2}pp}^{(1)}}{M_{0}} F^{(f_{2}p \bar{p})(1)}(s) 
\Big[ -2 m_{p} \sin^{2}\theta \,\delta_{s_{3}s_{4}} 
      +\sqrt{s}\sin\theta \cos\theta \,\varepsilon_{s_{3}s_{4}}
      \pm\sqrt{s}\sin\theta \,\delta_{s_{3},-s_{4}} 
\Big]\nonumber\\
&& \qquad +
\frac{g_{f_{2}pp}^{(2)}}{M_{0}^{2}} F^{(f_{2}p \bar{p})(2)}(s) \,
(s-4m_{p}^{2}) \sin^{2}\theta\, \delta_{s_{3}s_{4}}
\Big\rbrace
\,.
\label{hel2_ppbar}
\end{eqnarray}
Note the different $s$ dependences in (\ref{hel0_ppbar}) and (\ref{hel2_ppbar})
that are due to the different dimensions of 
$a_{f_{2} \gamma \gamma}$ and $b_{f_{2} \gamma \gamma}$.
Using different functional forms for the form factors $F_{a}$ and $F_{b}$
these $s$ dependences could be adjusted to experimental data.

In the calculation we assume the same form for $F_{a}$ and $F_{b}$
\begin{eqnarray}
F_{a}^{(f_{2} \gamma \gamma)}(s) = F_{b}^{(f_{2} \gamma \gamma)}(s) = F^{(f_{2} \gamma \gamma)}(s)\,.
\label{formfactor_f2gamgam}
\end{eqnarray}
A convenient ansatz for such a form factor is the exponential one
(see (4.22) of \cite{Lebiedowicz:2016ioh})
\begin{eqnarray}
F^{(f_{2} \gamma \gamma)}(s) = 
\exp\Big(-\frac{(s-m_{f_{2}}^{2})^{2}}{\Lambda_{f_{2},exp}^{4}}\Big)
\label{formfactor_f2}
\end{eqnarray}
with $\Lambda_{f_{2}}$ a parameter of the order 1 -- 2~GeV.
Alternatively, we can use
\begin{eqnarray}
F^{(f_{2} \gamma \gamma)}(s) = 
\frac{\Lambda_{f_{2},pow}^{4}}{\Lambda_{f_{2},pow}^{4}+(s-m_{f_{2}}^{2})^{2}}\,.
\label{formfactor_f2_secondform}
\end{eqnarray}
The form factors (\ref{formfactor_f2}) and (\ref{formfactor_f2_secondform})
are normalized to $F^{(f_{2} \gamma \gamma)}(m_{f_{2}}^{2}) = 1$.
For the $f_{2} p \bar{p}$ form factors we assume
\begin{eqnarray}
F^{(f_{2} p \bar{p})(1)}(s) = F^{(f_{2} p \bar{p})(2)}(s) = F^{(f_{2} \gamma \gamma)}(s)\,.
\label{formfactor_f2pp}
\end{eqnarray}
The numerical values of the form factor parameters will be adjusted to the Belle
experimental data.
\subsection{Hand-bag approach}
\label{sec:hand_bag}

The hand-bag contribution to $\gamma \gamma \to B \bar{B}$ processes
was described in detail in \cite{Diehl:2002yh}.
The hand-bag amplitude can be written in terms of the hard scattering
kernel for $\gamma \gamma \to q \bar{q}$
and a soft matrix element describing the $q \bar{q} \to p \bar{p}$ transition.
Their c.m. helicity amplitudes, which we denote by $\widetilde{\cal M}$, 
are written in terms of the light-cone helicity amplitudes ${\cal A}$
(see Eq.~(30) in \cite{Diehl:2002yh}) as
\begin{equation}
\begin{split}
\widetilde{\cal M}_{s_{3} s_{4}, m_{1} m_{2}} 
= {\cal A}_{s_{3} s_{4}, m_{1} m_{2}} +
\frac{m_{p}}{\sqrt{s}}
\Big[ 
 2s_{3}{\cal A}_{-s_{3}s_{4},m_{1}m_{2}}
+2s_{4}{\cal A}_{s_{3}-s_{4},m_{1}m_{2}}
\Big] 
+ {\cal O}(m_{p}^{2}/s)
\,.
\end{split}
\label{lc_amplitudes_trans}
\end{equation}
The light-cone helicity amplitudes, 
including terms suppressed only by $m_{p}/\sqrt{s}$, 
read \cite{Diehl:2002yh}
\begin{equation}
\begin{split}
&{\cal A}_{s_{3}s_{4},+-} = -(-1)^{s_{3}-s_{4}}{\cal A}_{-s_{3}-s_{4},-+}=
\\
&\qquad 
4 \pi \alpha_{em} \frac{s}{\sqrt{tu}}\,
\Big\lbrace 
\delta_{s_{3},-s_{4}} \frac{t-u}{s} R_{V}(s)
+ 2s_{3}\delta_{s_{3},-s_{4}} \Big[ R_{A}(s)+R_{P}(s) \Big]
- \frac{\sqrt{s}}{2m_{p}} \delta_{s_{3}s_{4}} R_{P}(s)
\Big\rbrace 
\,.
\end{split}
\label{lc_amplitudes}
\end{equation}
The authors of \cite{Diehl:2002yh} argue that
the amplitudes with identical photon helicities
will be nonzero only at next-to-leading order in $\alpha_{s}$, 
in analogy to the photon helicity flip transitions 
in large-angle Compton scattering \cite{Huang:2001ej}.
Note that for zero mass the light-cone helicity amplitudes (\ref{lc_amplitudes})
are identical with the helicity amplitudes (\ref{lc_amplitudes_trans}),
but not if the mass is finite.
The $q \bar q \to p \bar p$ transition form factors 
$R_V(s)$, $R_A(s)$ and $R_P(s)$
were determined phenomenologically in \cite{Diehl:2002yh}.
In our calculation we neglect the term with $R_V(s)$
and assume
$\frac{\sqrt{s}}{2 m_{p}} \left|\frac{R_{P}(s)}{R_{A}(s)}\right| = 0.37$
(see formula (45) from \cite{Diehl:2002yh}).
In addition we take $R_{A}(s)$ and $R_{P}(s)$ as real and positive.
We parametrize $R_{A}(s) = C_{A}/s$ (in parameter set A) 
with $C_{A}$ a parameter of dimension GeV$^{2}$ or 
$R_{A}(s) = \tilde{C}_{A}/s^{2}$ (in parameter set B) 
with $\tilde{C}_{A}$ a parameter of dimension GeV$^{4}$
which we shall determine from a fit to the Belle data in Sec.~\ref{subsec:res_data};
see Table~\ref{table:parameters}.
Note that the $s$-dependence of $R_{A}$ with $C_{A}$ is different (less steep) 
than in \cite{Diehl:2002yh}, where only the hand-bag contribution
was fitted to rather old experimental data.
In \cite{Diehl:2002yh} different phase conventions compared to ours are used.
Taking this into account we find
\begin{equation}
\begin{split}
&\langle 2s_{3}, 2s_{4} |{\cal T}| + , - \rangle_{hb} =
2 s_{4} \widetilde{\cal M}_{s_{3} s_{4}, +-} \,,\\
&\langle 2s_{3}, 2s_{4} |{\cal T}| - , + \rangle_{hb} =
2 s_{4} \widetilde{\cal M}_{s_{3} s_{4}, -+} \,;
\end{split}
\label{hb_amplitudes}
\end{equation}
see Appendix~\ref{section:Appendix3}.

The hand-bag helicity amplitudes (\ref{hb_amplitudes})
must be added coherently within our approach (see previous subsections).
At small momentum transfer $|t|$ or $|u|$ 
the hand-bag and proton-exchange mechanisms compete 
and it would be a double counting to include both of them simultaneously. 
We emphasize, however, that in regions of small $|t|$ or $|u|$
the hand-bag approach has to be taken with a grain of salt.
To avoid in addition double counting 
(we include explicitly the proton-exchange mechanism) we suggest
to multiply the hand-bag amplitudes by a purely phenomenological factor:
\begin{equation}
F_{corr}(t,u) = \Big( 1-\exp \Big(\frac{t}{\Lambda_{hb}^2}\Big) \Big) 
                \Big( 1-\exp \Big(\frac{u}{\Lambda_{hb}^2}\Big) \Big)
\label{corr_hb}
\end{equation}
with an extra free parameter $\Lambda_{hb}$.
Its role is to cut off the region of small $|t|$ and $|u|$
where the hand-bag approach does not apply.
As a consequence it also reduces the hand-bag contribution 
to the cross section at low $\sqrt{s}$ in the whole angular range.

\section{Nuclear reaction}
\label{sec:nuclear_theory}

Now we will present theoretical formulas for the nuclear reaction
\begin{eqnarray}
^{208}\!Pb + ^{208}\!\!Pb \to ^{208}\!\!Pb + ^{208}\!\!Pb + p + \bar{p}\,.
\label{nuclear_reaction}
\end{eqnarray}
\begin{figure}[!ht]
\includegraphics[width=0.4\textwidth]{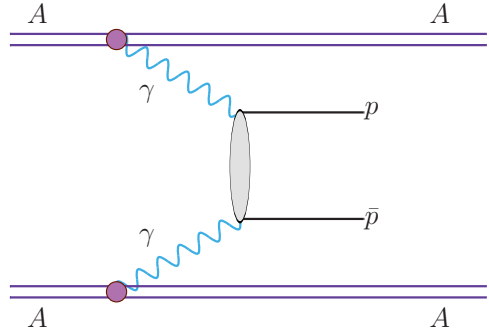}
\caption{\label{fig:AA_AAppbar}
\small
Diagram representing proton-antiproton production in ultrarelativistic 
ultraperipheral collisions (UPC) of heavy ions.}
\end{figure}

We focus on the processes for ultraperipheral collisions (UPC) of heavy ions,
see the diagram shown in Fig.~\ref{fig:AA_AAppbar}.
The nuclear cross section is calculated in the equivalent photon approximation (EPA) 
in the impact parameter space.
This approach allows to take into account the transverse distance between the colliding nuclei.
The total (phase space integrated) cross section 
is expressed through the five-fold integral
\begin{eqnarray}
\sigma_{AA \to AA p\bar{p}} \left( \sqrt{s_{AA}} \right)&=& 
\int \sigma_{\gamma\gamma \to p\bar{p}}(W_{\gamma\gamma})  
N(\omega_1, \bba) N(\omega_2, \bbb) 
S^2_{abs}(\bb) \nonumber \\
&& \times  \frac{W_{\gamma\gamma}}{2}
dW_{\gamma \gamma} \, d{\rm Y}_{p\bar{p}} \, d\overline{b}_x \, d\overline{b}_y 
\, 2 \pi \,b\,db \,.
\label{eq:sig_nucl_tot}
\end{eqnarray}  
Above, $b = |\bb|$ is the impact parameter, i.e., the distance between colliding nuclei
in the plane perpendicular to their direction of motion.
$W_{\gamma\gamma}=\sqrt{4\omega_1\omega_2}$ is the invariant
mass of the $\gamma\gamma$ system and $\omega_i$, $i=1,2$,
is the energy of the photon which is emitted from the first or second nucleus, respectively.
${\rm Y}_{p\bar{p}}=\frac{1}{2}({\rm y}_{p} + {\rm y}_{\bar{p}})$
is the rapidity of the $p\bar{p}$ system. 
The quantities 
$\overline{b}_x = (b_{1x}+b_{2x})/2$, $\overline{b}_y=(b_{1y}+b_{2y})/2$
are given in terms of $b_{ix}$, $b_{iy}$ which are the components of 
the $\bba$ and $\bbb$ vectors 
which mark a point (distance from first and second nucleus) 
where photons collide and particles are produced. 
The diagram illustrating these quantities in the impact parameter
space can be found in \cite{KlusekGawenda:2010kx}. 

In Ref.~\cite{KlusekGawenda:2010kx} the dependence of the photon flux    
$N\left( \omega_i, \bbi \right)$ on the charge form factors of the colliding nuclei
was shown explicitly.
In our calculations we use the so-called realistic form factor 
which is the Fourier transform of the charge distribution in the nucleus.
A more detailed discussion of this issue is given in~\cite{KlusekGawenda:2010kx}.

The presence of the absorption factor $S^{2}_{abs}(\bb)$ in (\ref{eq:sig_nucl_tot})
assures that we consider only peripheral collisions, 
when the nuclei do not undergo nuclear breakup.
In the first approximation this geometrical factor
can be expressed as
\begin{equation}
S^{2}_{abs}(\bb) = \theta(|\bb|-(R_{A}+R_{B})) = \theta(|\bba-\bbb|-(R_{A}+R_{B}))\,,
\label{survival_factor}
\end{equation}
where the sum of the radii of the two nuclei occurs.

In our present study we calculate also distributions in kinematical variables
of each of the produced particles (for details 
how it is handled see \cite{Klusek-Gawenda:2016euz}).
Then one can impose easily experimental cuts on (pseudo)rapidities and transverse momenta.

\section{Results for the $\gamma \gamma \to p \bar{p}$ reaction}
\label{sec:results_ppbar}

First we will show some features of the
proton-exchange mechanism and the $s$-channel tensor meson exchanges.
We will show the dependence of the cross section on the photon-photon energy 
and the angular distributions of individual helicity components.
Then we will confront the model results with the experimental data
and adjust the model parameters.

\subsection{Proton exchange mechanism}

\begin{figure}[!ht]
\includegraphics[width=0.45\textwidth]{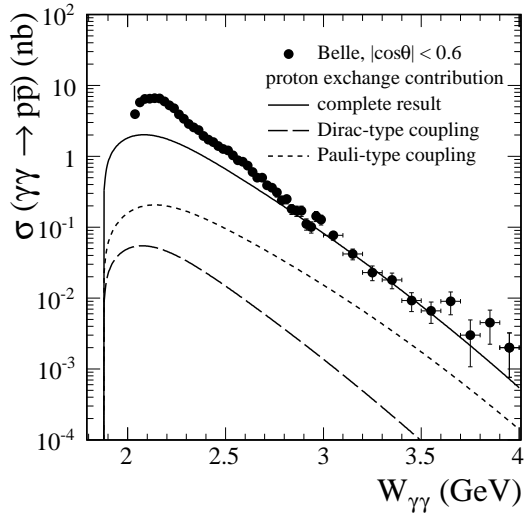}
\caption{\label{fig:p_exchange}
\small
The $\gamma \gamma \to p \bar{p}$ cross section as 
a function of photon-photon energy $W_{\gamma \gamma} \equiv \sqrt{s}$.
We present the results for the nonresonant contribution (see Sec.~\ref{sec:nucleon_exchange})
for $\Lambda_{p} = 1.1$~GeV in (\ref{ff_Poppe}).
The solid line represents the complete result with both Dirac- and Pauli-type couplings
included in the amplitude.
Other combinations of electromagnetic couplings in the $\gamma NN$ vertices
are also shown: only Dirac couplings, and only Pauli couplings
at the two vertices in Figs.~\ref{fig:diagram_2to2}~(a) and (b).
The Belle experimental data from \cite{Kuo:2005nr} are shown for comparison.
}
\end{figure}

In Fig.~\ref{fig:p_exchange} we show that the proton exchange mechanism alone 
cannot describe the energy-dependence of the cross sections measured by Belle \cite{Kuo:2005nr}.
We show results for the Dirac- or Pauli-type couplings separately
and when both couplings in the $\gamma NN$ vertices are taken into account.
We can see that the complete result indicates a large interference effect 
of Dirac and Pauli terms in the amplitudes.
Clearly, the proton exchange contribution is not sufficient to describe the Belle data.

\begin{figure}[!ht]
\includegraphics[width=0.45\textwidth]{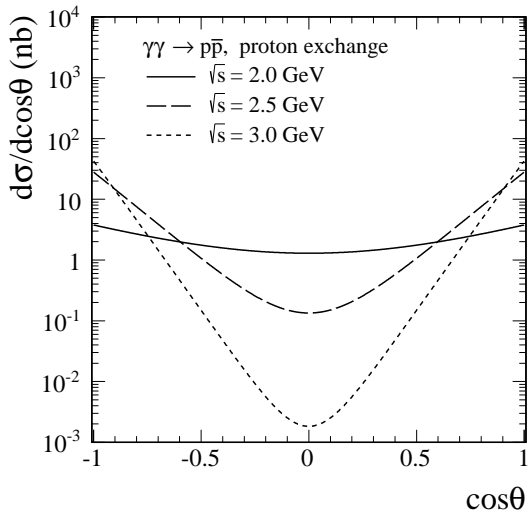}
  \caption{\label{fig:dsig_dz_ppbar}
  \small
The angular distributions for $\sqrt{s}$ = 2.0, 2.5 and 3.0~GeV
for the nonresonant proton-exchange mechanism.
}
\end{figure}

In Fig.~\ref{fig:dsig_dz_ppbar} we show
the unpolarized differential cross section
$d\sigma/d\cos\theta$ for three different $\gamma \gamma$ c.m. energies.
As one gets closer to $\sqrt{s} = 2 m_{p}$, the threshold energy,
the angular distributions become flatter and flatter.

\begin{figure}[!ht]
\includegraphics[width=0.45\textwidth]{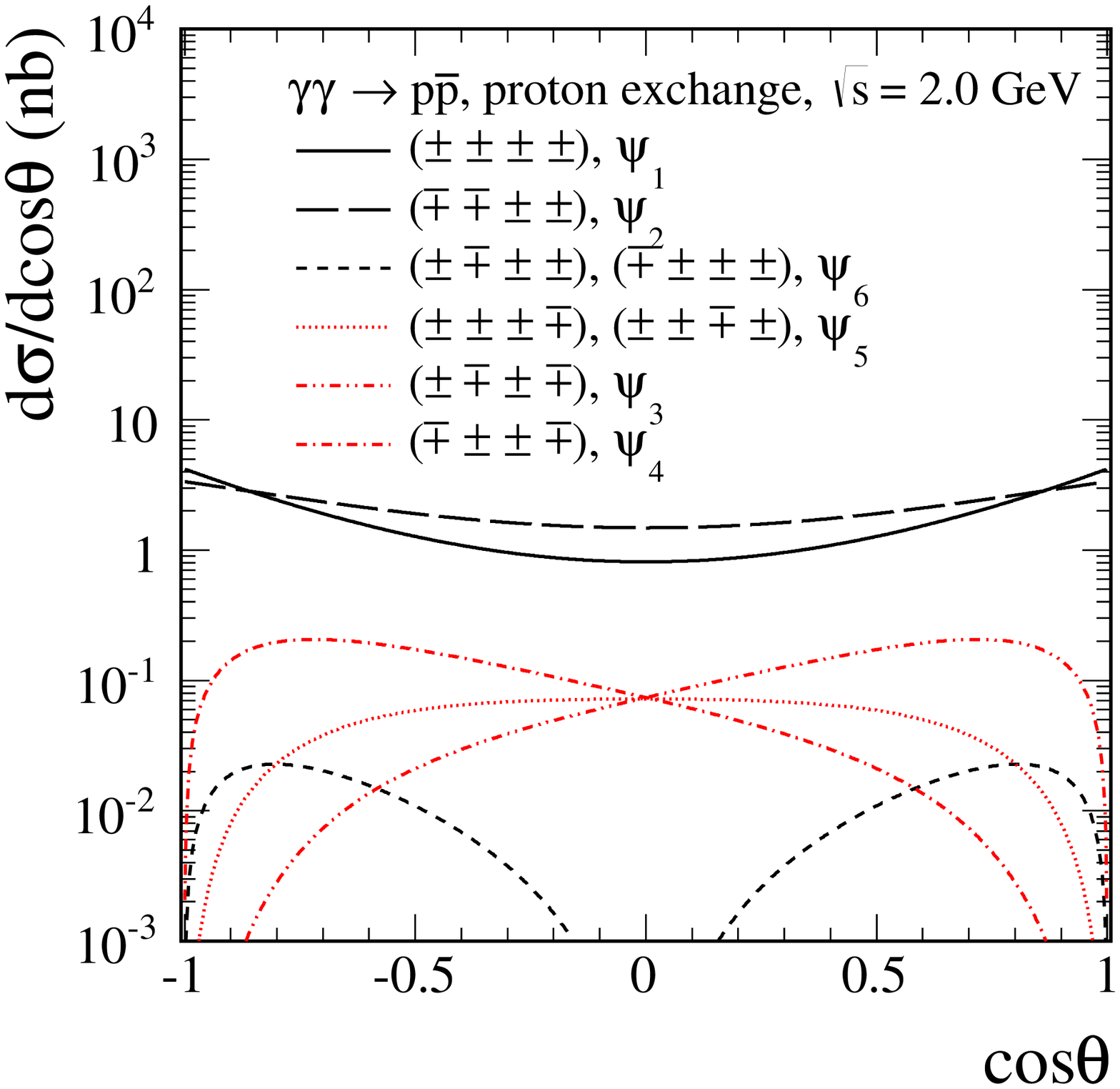}
\includegraphics[width=0.45\textwidth]{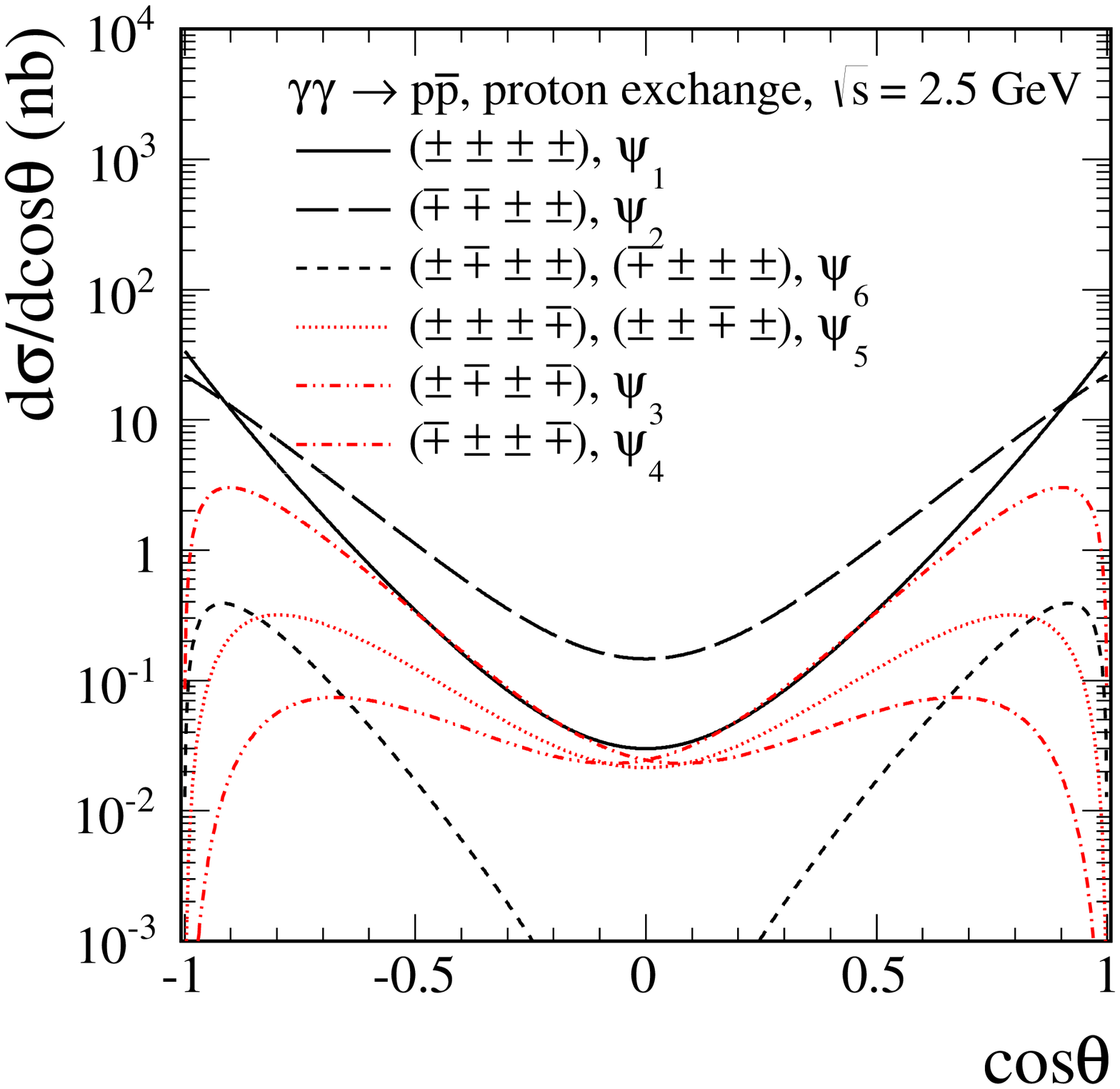}
  \caption{\label{fig:dsig_dz_ppbar_deco}
  \small
The helicity components of $d\sigma/d\cos\theta$ as a function of $\cos\theta$
for the proton exchange mechanism
for $\sqrt{s} = 2.0$ (the left panel) and 2.5~GeV (the right panel).
Contributions of different helicities $(2 s_{3} \,2s_{4} \,m_{1} \,m_{2})$
of the photons and baryons are shown.
}
\end{figure}

In Fig.~\ref{fig:dsig_dz_ppbar_deco} we present 
the helicity dependence of the differential cross section.
We label the results for different helicity terms as
$(2 s_{3} \,2s_{4} \,m_{1} \,m_{2})$ for 
$\langle 2s_{3},2s_{4}|{\cal T}| m_{1},m_{2} \rangle$
as defined in (\ref{B_41}).
One can see the dominance of the 
$(\pm \pm \pm \pm)$ and $(\mp \mp \pm \pm)$ contributions
over the $(2 s_{3} \,2s_{4} \,\pm \mp)$ ones (see the red lines).
In terms of the $\psi_{j}$ ($j=1,...,6$) from (\ref{B_44}) 
and Table~\ref{tab:helicity_amp} of Appendix~\ref{section:Appendix1}
we find dominance of the amplitudes $\psi_{1}$ and $\psi_{2}$.
Furthermore we see that the contributions of the amplitudes
$\psi_{3}$, $\psi_{4}$, $\psi_{5}$ and $\psi_{6}$
are suppressed in the forward and backward directions, $\cos\theta = \pm1$.
This is clear from angular momentum conservation.
For $\psi_{3}$, $\psi_{4}$, and $\psi_{5}$, 
the state of the two photons has $J_{z} = \pm 2$.
This cannot be reached by proton-antiproton produced in the 
forward or backward direction where we only get $J_{z} = 0$ or $\pm 1$.
For $\psi_{6}$ the two-photon state has $J_{z} = 0$
and the two-baryon state in forward and backward direction has $J_{z} = +1$ and -1,
respectively. We have again a mismatch.
The contributions of four helicity states
$(+ - + +)$, $(- + - -)$, $(- + + +)$, $(+ - - -)$
vanish when only the Dirac-type coupling in the $\gamma NN$ vertices is included.
That is, the amplitude $\psi_{6}$ vanishes in this case.

\subsection{$f_{2}$ meson contributions}

The Belle experimental angular distributions \cite{Kuo:2005nr}, 
at least at low energies,
cannot be described solely with the proton-exchange mechanism 
discussed in Sec.~\ref{sec:nucleon_exchange}.
It seems that a mechanism is missing.
A resonant $s$-channel contribution is a reasonable option
for a second mechanism (see also \cite{Klusek-Gawenda:2013rtu}
for the $\gamma \gamma \to \pi \pi$ reactions).

In Table~\ref{table:table} we have listed resonances that 
decay into $\gamma \gamma$ and $p \bar{p}$ and which,
therefore, may contribute to the reaction (\ref{2to2_reaction}).
In principle, also subthreshold resonances, 
such as $f_2(1270)$, may play some, even an important, role.
It is worth to mention that our knowledge 
about the $f_2(1950)$ resonance comes from the BES \cite{Ablikim:2006pt} 
and the CLEO \cite{Alexander:2010vd} analyses 
for $\psi(2S) \to \gamma p \bar{p}$ radiative decays.
In \cite{Alexander:2010vd} the authors include also the $f_{2}(2150) \to p\bar{p}$ contribution
in order to describe 
the $M_{p \bar{p}}$ and $M_{p \gamma}$ invariant mass distributions.
For $\psi(2S) \to \gamma p \bar{p}$ a stringent upper limit
for the threshold resonance
${\cal B}(\psi(2S) \to \gamma R_{thr})\times {\cal B}(R_{thr} \to p \bar{p}) <1.6\times10^{-5}$
at 90\% confidence level was found \cite{Alexander:2010vd}.
\begin{table}[!h]
\caption{A list of resonances that may contribute to the $\gamma\gamma \to p\bar{p}$ reaction.
Here we listed also the subthreshold $f_2(1270)$ resonance.
The meson masses, their total widths $\Gamma$ and branching fractions
are taken from PDG~\cite{Olive:2016xmw}.
}
\begin{tabular}{|l|l|l|l|l|}
\hline
Meson			& $m$ (MeV) 	& $\Gamma$ (MeV)		& $\Gamma_{p\bar{p}}/\Gamma$	& $\Gamma_{\gamma\gamma}/\Gamma$ \\ \hline\hline
$f_2(1270)$	&	$1275.5\pm0.8$ 			& $186.7^{+2.2}_{-2.5}$	&  & $(1.42\pm0.24) \times 10^{-5}$ \\ \hline
$f_2(1950)$	&	$1 944\pm12$ 			& $472 \pm18$ 				& seen & seen \\ \hline
$\eta_c(1S)$		& $2 983 \pm 0.5$ 		& $31.8\pm 0.8$ 		& $(1.50\pm0.16)\times 10^{-3}$	& $(1.59\pm0.13) \times 10^{-4}$ \\ 
$\chi_{c0}(1P)$	& $3 414.75\pm0.31$		& $10.5\pm0.6$ 		& $(2.25\pm0.09) \times 10^{-4}$	& $(2.23\pm0.13) \times 10^{-4}$ \\
$\chi_{c2}(1P)$	& $3 556.20\pm0.09$ 		& $1.93\pm0.11$ 		& $(7.5\pm0.4) \times 10^{-5}$		& $(2.74\pm0.14) \times 10^{-4}$ \\
$\eta_c(2S)$		& $3 639.2\pm1.2$ 		& $11.3^{+3.2}_{-2.9}$ & $<2 \times 10^{-3}$			& $(1.9\pm1.3) \times 10^{-4}$ \\
\hline
\end{tabular}
\label{table:table}
\end{table}

In our paper we consider only the $f_{2}$ meson exchanges in the $s$-channel. 
In general also the $c\bar{c}$ mesons (e.g. $\eta_c(1S)$, $\chi_{c0}(1P)$)
may contribute to the reaction (\ref{2to2_reaction}).
The charmonium states have rather small total widths (see Table~\ref{table:table})
thus they will appear in the invariant mass distribution as rather narrow peaks;
see \cite{Lebiedowicz:2017cuq} for the $\gamma \gamma \to \gamma \gamma$ reaction.
Even interference effects with other mechanisms may be important in this context.
This goes, however, beyond the scope of the present paper and will be studied elsewhere.

Now we will discuss the helicity structure of $\gamma \gamma \to p \bar{p}$ 
from the contribution of the $s$-channel
(below-threshold or above-threshold) $f_2$ resonances in our
Lagrangian approach; see Sec.~\ref{sec:resonances}.

\begin{figure}[!ht]
\includegraphics[width=0.45\textwidth]{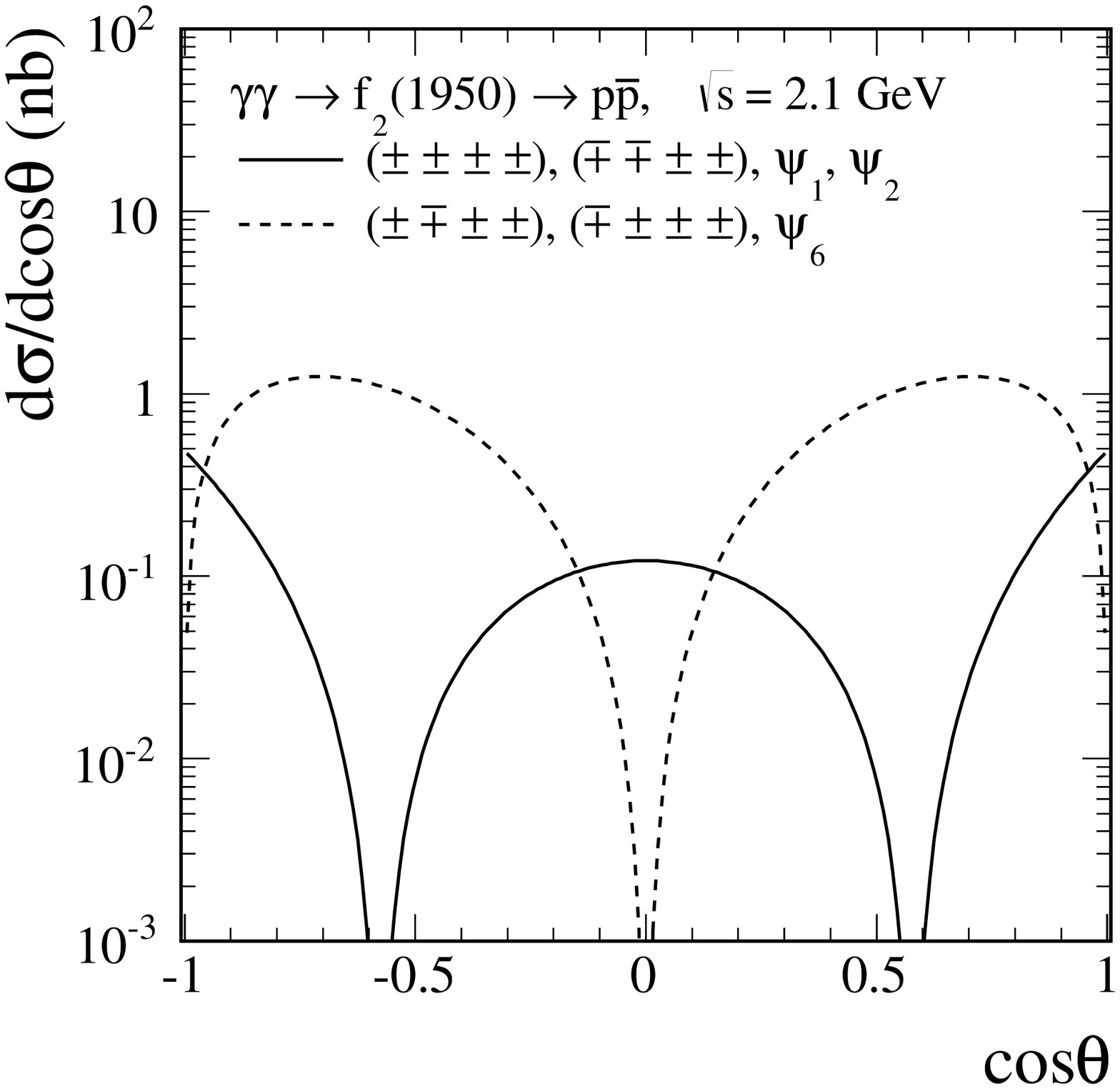}
\includegraphics[width=0.45\textwidth]{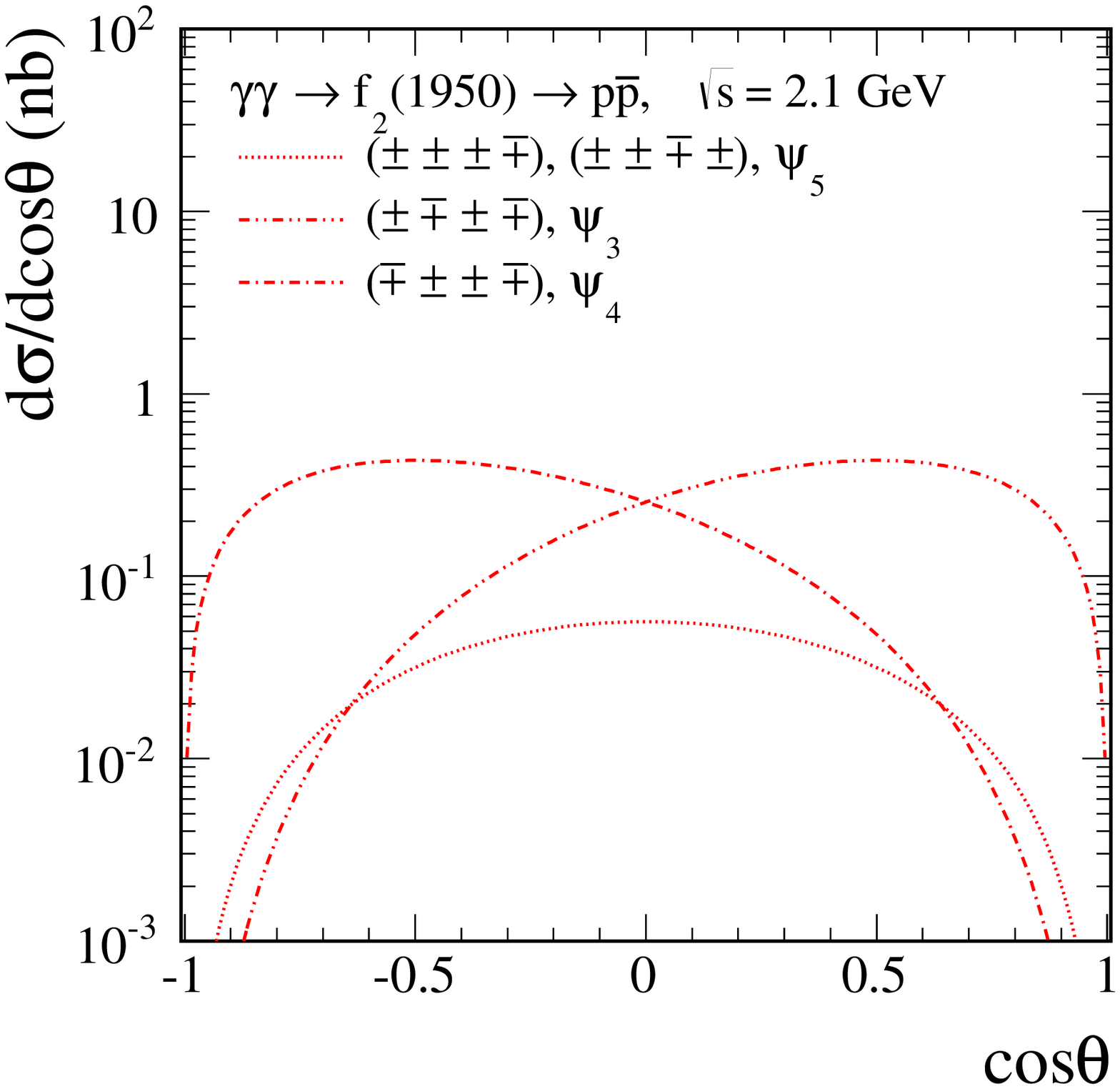}
  \caption{\label{fig:dsig_dz_helicity_structure}
  \small
The helicity components of the differential cross sections $d\sigma/d\cos\theta$
for the $\gamma \gamma \to f_{2}(1950) \to p \bar{p}$ reaction
for $\sqrt{s} = 2.1$~GeV.
Here, the coupling constants are fixed arbitrarily, for $j=1,2$:
$a_{f_{2} \gamma\gamma} g^{(j)}_{f_{2} pp} = 
\frac{e^{2}}{4 \pi} \,1$~GeV$^{-3}$ (the left panel) and
$b_{f_{2} \gamma\gamma} g^{(j)}_{f_{2} pp} = 
\frac{e^{2}}{4 \pi} \,1$~GeV$^{-1}$ (the right panel).
The calculations have been done for $\Lambda_{f_{2},pow} = 1.15$~GeV
in (\ref{formfactor_f2_secondform}).
}
\end{figure}

In Fig.~\ref{fig:dsig_dz_helicity_structure} we show the contributions of
different helicities for the two $\gamma \gamma \to f_2$ couplings in (\ref{3.29new}),
$a_{f_{2} \gamma \gamma}$ (left panel) 
and $b_{f_{2} \gamma \gamma}$ (right panel).
There are five independent helicity contributions
since here the contributions of the amplitudes $\psi_{1}$ and $\psi_{2}$
turn out to be the same; see (\ref{hel0_ppbar}),
(\ref{B_44}) and Table~\ref{tab:helicity_amp} of Appendix~\ref{section:Appendix1}.
Only the distributions that are proportional to $(\cos^{2}\theta - 1/3)^{2}$,
see (\ref{hel0_ppbar}) and (\ref{hel2_ppbar}),
(this corresponds to the solid line in the left panel) 
are favored by the Belle experimental data;
see Figs.~\ref{fig:dsig_dz_3mech_nohb} and \ref{fig:dsig_dz_3mech} below.
Here the cutoff parameter of form factors ($\Lambda_{f_{2},pow}$) and 
the products of coupling constants ($a_{f_{2} \gamma \gamma} g^{(j)}_{f_{2} pp}$ and
$b_{f_{2} \gamma \gamma} g^{(j)}_{f_{2} pp}$)
are fixed arbitrarily.

\subsection{Comparison with the Belle data}
\label{subsec:res_data}

Here we wish to demonstrate that it is possible 
to describe the Belle data taking into account
the $t$- and $u$-channel proton exchanges,
the $s$-channel tensor meson exchanges,
and the hand-bag mechanism
discussed in Sec.~\ref{sec:gamgam_ppbar}.
In the following we shall take in our calculation 
a coherent sum of all the above amplitudes.

\begin{figure}[!ht]
(a)\includegraphics[width=0.47\textwidth]{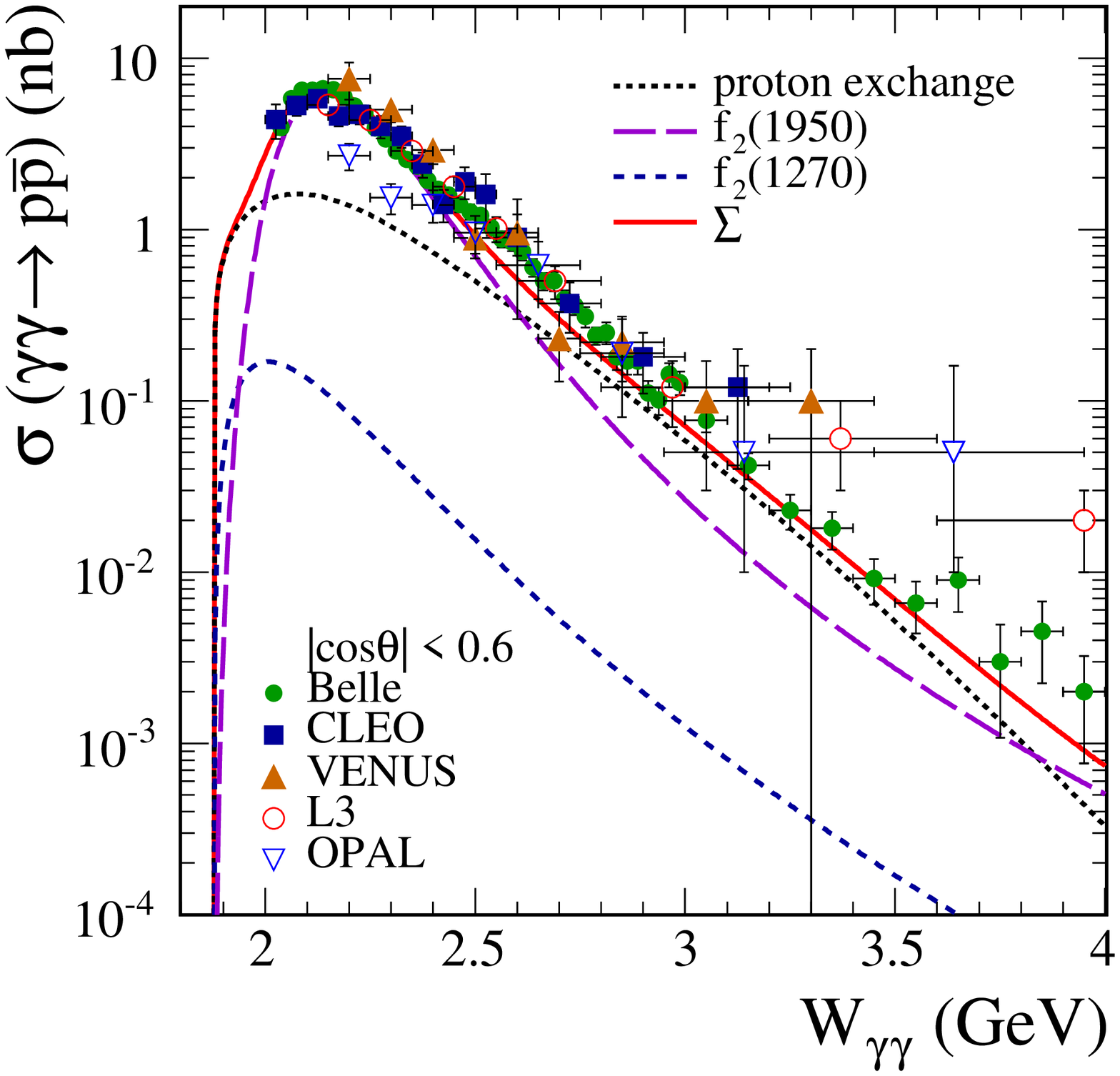}\\
(b)\includegraphics[width=0.47\textwidth]{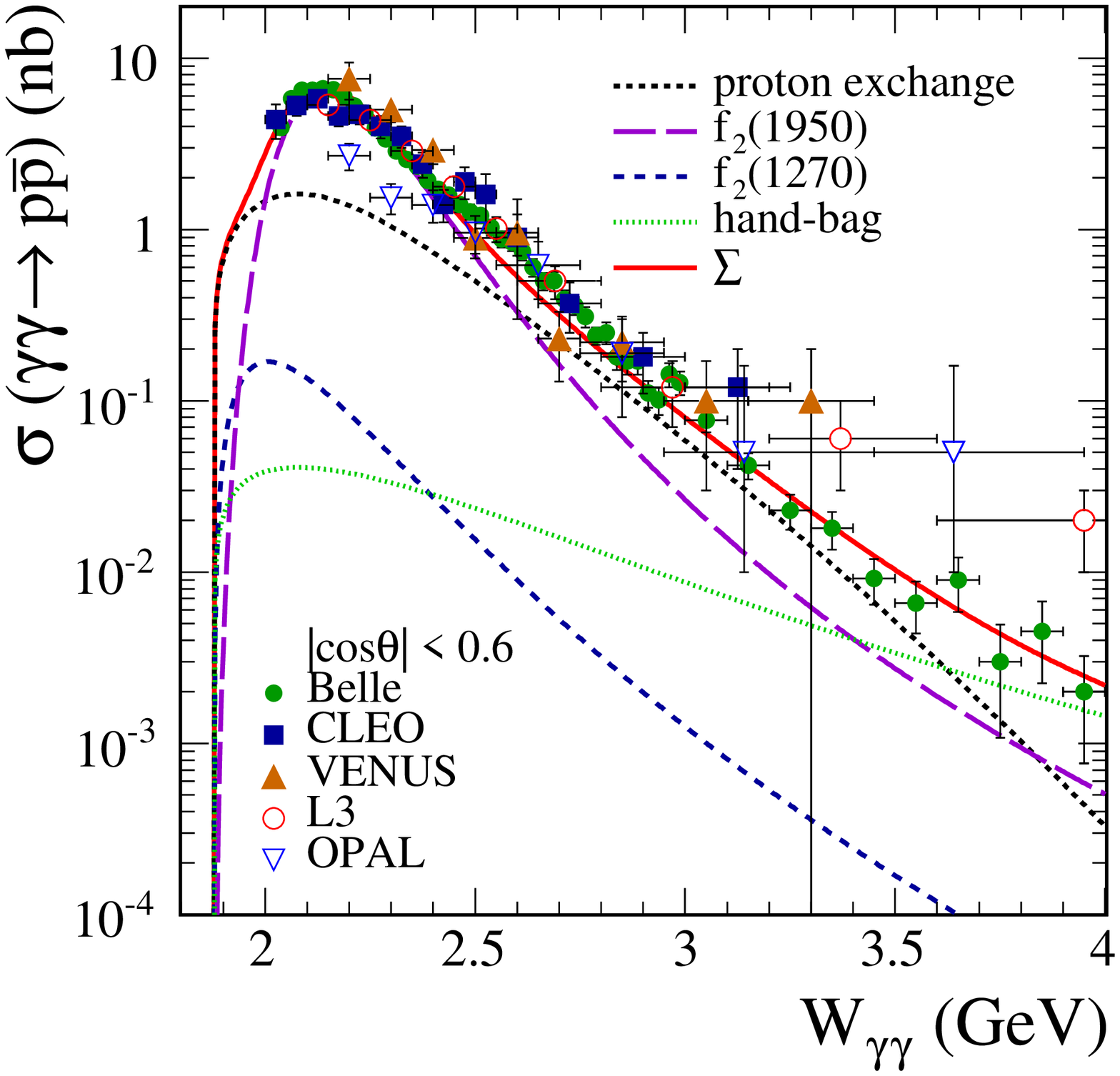}
(c)\includegraphics[width=0.47\textwidth]{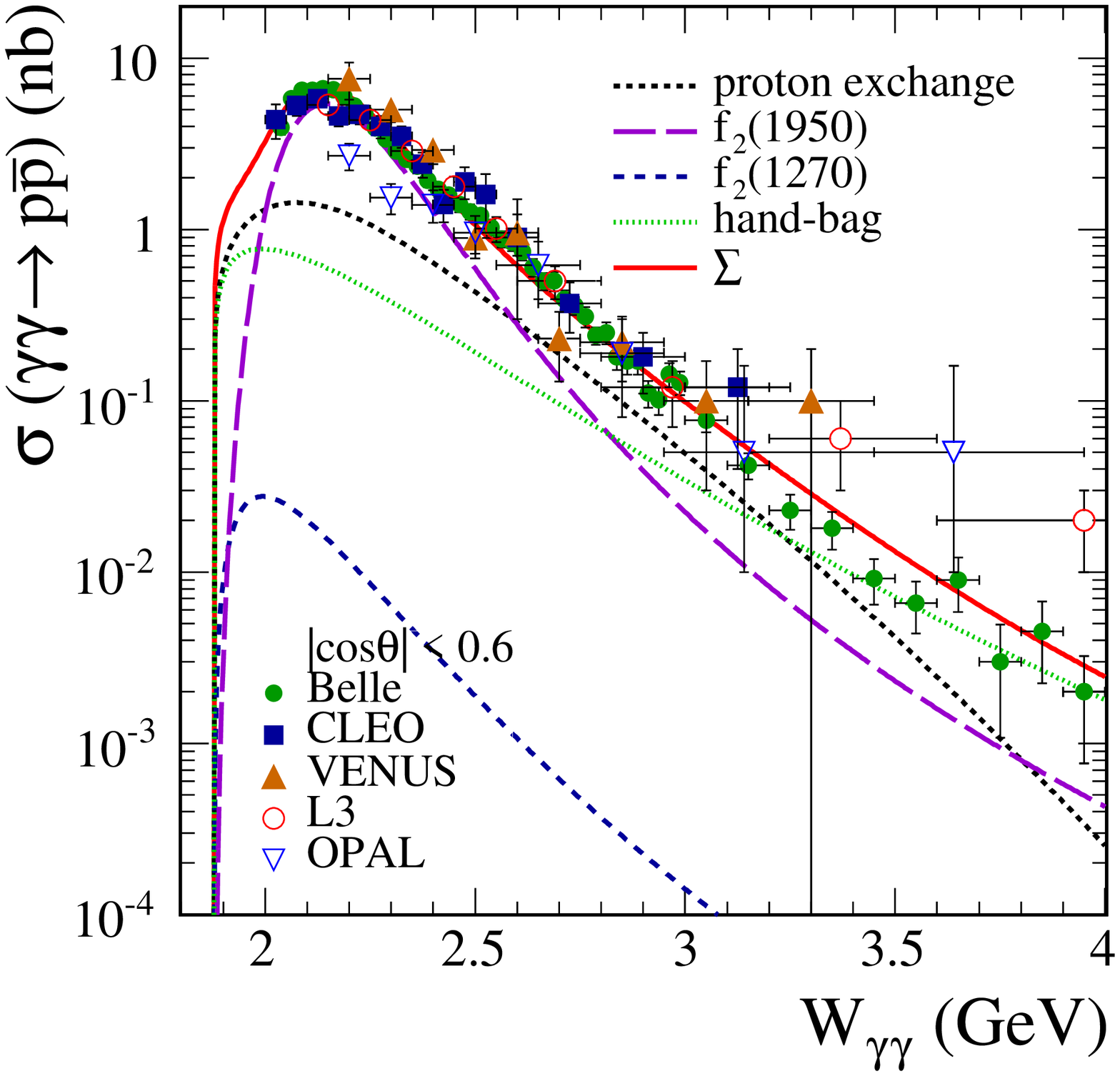}
\caption{\label{fig:sigma_W_3mech}
\small
Energy dependence of the total cross section for $\gamma \gamma \to p \bar{p}$
for $|\cos \theta|<0.6$. 
The experimental data are from the CLEO \cite{Artuso:1993xk},
VENUS \cite{Hamasaki:1997cy},
OPAL \cite{Abbiendi:2002bxa},
L3 \cite{Achard:2003jc}, 
and Belle \cite{Kuo:2005nr} experiments.
In the panel (a) we show the results
for the tensor meson exchanges and the proton-exchange contributions,
and their coherent sum (see the red solid line).
In the panels (b) and (c) we show the results including, in addition, the hand-bag contribution.
In the panels (a) and (b) we used the parameter set A
while in the panel (c) we used the parameter set B; see Table~\ref{table:parameters}.
}
\end{figure}

In Fig.~\ref{fig:sigma_W_3mech} we show the energy dependence of the cross section
for the $\gamma \gamma \to p \bar{p}$ reaction.
In the panel (a) we present results for the proton exchange and 
the $f_{2}(1270)$ and $f_{2}(1950)$ $s$-channel exchanges
together with the experimental data of the CLEO \cite{Artuso:1993xk},
VENUS \cite{Hamasaki:1997cy},
OPAL \cite{Abbiendi:2002bxa},
L3 \cite{Achard:2003jc}, 
and Belle \cite{Kuo:2005nr} experiments.
An agreement between the Belle experimental data \cite{Kuo:2005nr} 
and the earlier measurements \cite{Artuso:1993xk,Hamasaki:1997cy,Achard:2003jc}
with the exception of the OPAL experiment \cite{Abbiendi:2002bxa}
in the low mass region $W_{\gamma\gamma} = M_{p\bar{p}} < 3$~GeV can be observed
(within the quoted uncertainties); see also Fig.~\ref{fig:dsig_dz_W3to4} below.
For the $f_{2}(1270)$ contribution the coupling constants
$a_{f_{2} \gamma\gamma}$ and $b_{f_{2} \gamma\gamma}$ are relatively well known and 
taken from \cite{Ewerz:2013kda}.
We take into account only one $f_{2}(1270) p\bar{p}$ coupling 
($g^{(1)}_{f_{2}(1270) pp} = 11.04$)
and neglect the term with $g^{(2)}_{f_{2}(1270) pp}$.
For the $f_{2}(1950)$ contribution we take
only the term with $a_{f_{2}(1950) \gamma\gamma} g^{(2)}_{f_{2}(1950) p\bar{p}} = 
\frac{e^{2}}{4 \pi} \,13.05$~GeV$^{-3}$.
In the vertices for the meson exchange contributions we assume 
the same type of the form factors (\ref{formfactor_f2_secondform}) 
and $\Lambda_{f_2,pow}=1.15$~GeV;
see Eqs.~(\ref{formfactor_f2gamgam}) and (\ref{formfactor_f2pp}).
We take $\Lambda_{p}=1.08$ GeV for the proton-exchange contribution; see (\ref{ff_Poppe}). 
One can observe the dominance of the $f_2(1950)$ resonance term at low energies.
We slightly underestimate the Belle data from $\sqrt{s}=2.4$ to $2.9$~GeV.
The panels (b) and (c) show results including also the hand-bag contribution.
The hand-bag contribution is important at $W_{\gamma \gamma}>3$~GeV.
To illustrate uncertainties of our model we take 
in the calculation two sets of parameters.
For the convenience of the reader we collect in Table~\ref{table:parameters}
the parameters of our model and their numerical values
used here and in the following.

\begin{table}[!h]
\caption{Model parameters and their numerical values used.
The second column indicates the equation numbers where
the parameter is defined.
}
\begin{tabular}{|c|c|c|c|}
\hline
parameter for			& eq. 	& value (set A) & value (set B)\\ \hline \hline
nonresonant $p\bar{p}$	& &  &	\\ \hline 
$\kappa_{p}$ &	(\ref{F1_normalisation})\; et seq. 			& 1.7928 & 1.7928   \\
$\Lambda_{p}$ &	(\ref{ff_tus}), (\ref{ff_Poppe}) 		& 1.08~GeV & 1.07~GeV	\\ \hline \hline
$f_{2}(1270)$	& & &	\\ \hline
$a_{f_{2}\gamma\gamma}$ & (\ref{3.29new}); (3.40) of \cite{Ewerz:2013kda} & $\frac{e^{2}}{4 \pi} \,1.45$~GeV$^{-3}$    & $\frac{e^{2}}{4 \pi} \,1.45$~GeV$^{-3}$     \\
$b_{f_{2}\gamma\gamma}$ & (\ref{3.29new}); (3.40) of \cite{Ewerz:2013kda} & $\frac{e^{2}}{4 \pi} \,2.49$~GeV$^{-1}$  & $\frac{e^{2}}{4 \pi} \,2.49$~GeV$^{-1}$      \\
$M_{0}$ & (\ref{A01})\; et seq. & 1~GeV & 1~GeV\\
$g^{(1)}_{f_{2}pp}$ & (\ref{A01}), (\ref{A03}) & 11.04    & 11.04\\
$g^{(2)}_{f_{2}pp}$ & (\ref{A02}), (\ref{A04}) & 0   & 0  \\
$\Lambda_{f_{2},pow}$ &	(\ref{formfactor_f2_secondform}) 		& 1.15~GeV 	& 1~GeV \\ \hline \hline
$f_{2}(1950)$	& &  &	\\ \hline
$a_{f_{2}\gamma\gamma} g^{(2)}_{f_{2}pp}$ & 
(\ref{3.29new}), (\ref{A02}), (\ref{A04}) & $\frac{e^{2}}{4 \pi} \,13.05$~GeV$^{-3}$   & $\frac{e^{2}}{4 \pi} \,12$~GeV$^{-3}$ \\
$b_{f_{2}\gamma\gamma}$ & (\ref{3.29new}) & 0 & 0   \\
$g^{(1)}_{f_{2}pp}$ & (\ref{A01}), (\ref{A03}) & 0 & 0   \\
$\Lambda_{f_{2},pow}$ &	(\ref{formfactor_f2_secondform}) 		& 1.15~GeV & 1.15~GeV	\\ \hline \hline
hand-bag contribution	& & & 	\\ \hline 
$C_{A}$ &$R_{A}(s) = C_{A}/s$& 0.14~GeV$^{2}$ &  \\
$\tilde{C}_{A}$ &$R_{A}(s) = \tilde{C}_{A}/s^{2}$&  & 2.5~GeV$^{4}$ \\
$\Lambda_{hb}$ &	(\ref{corr_hb}) 		& 0.85~GeV & 0.85~GeV	\\ \hline 
\end{tabular}
\label{table:parameters}
\end{table}

\begin{figure}[!ht]
(a)\includegraphics[width=0.3\textwidth]{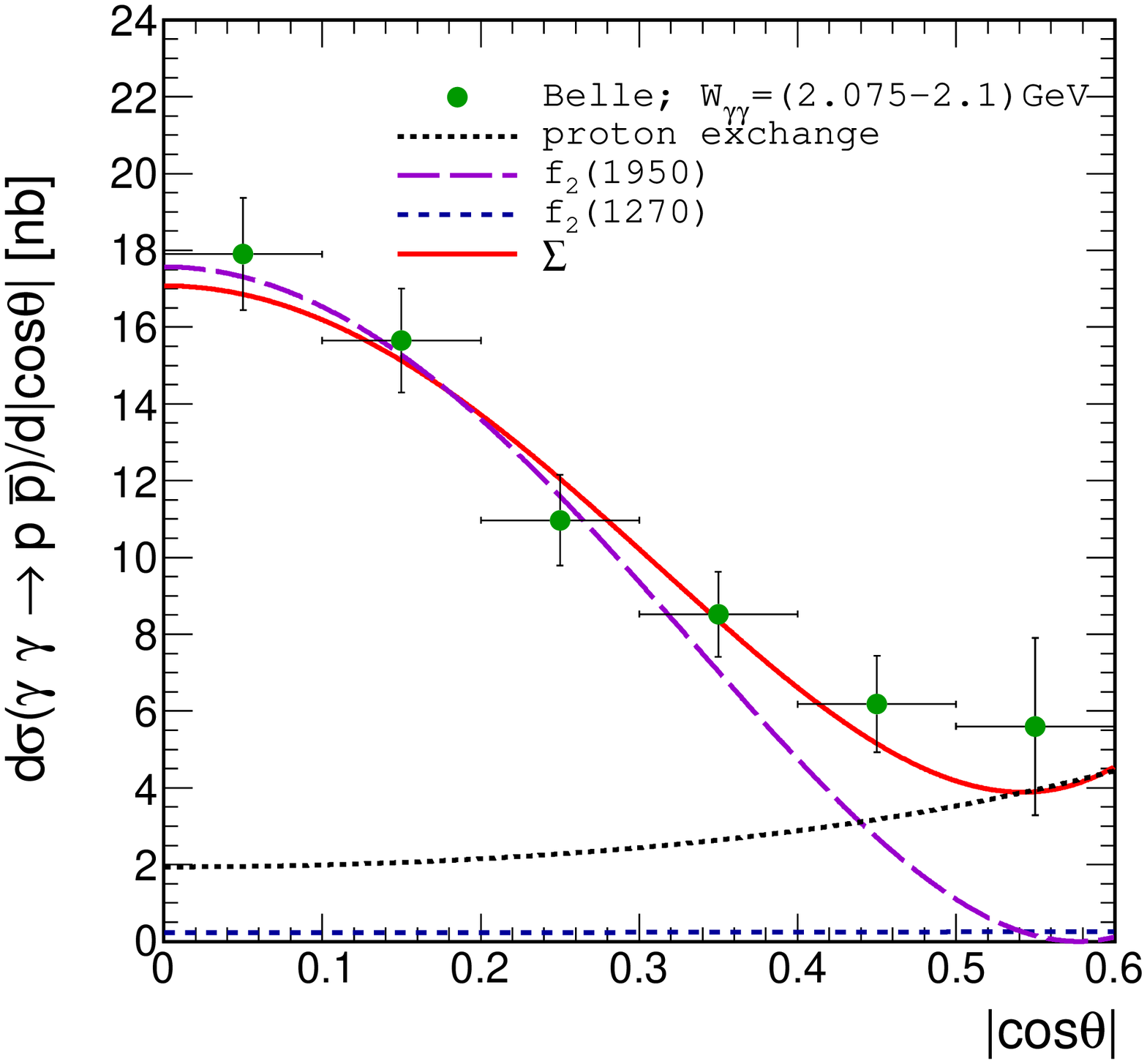}
(b)\includegraphics[width=0.3\textwidth]{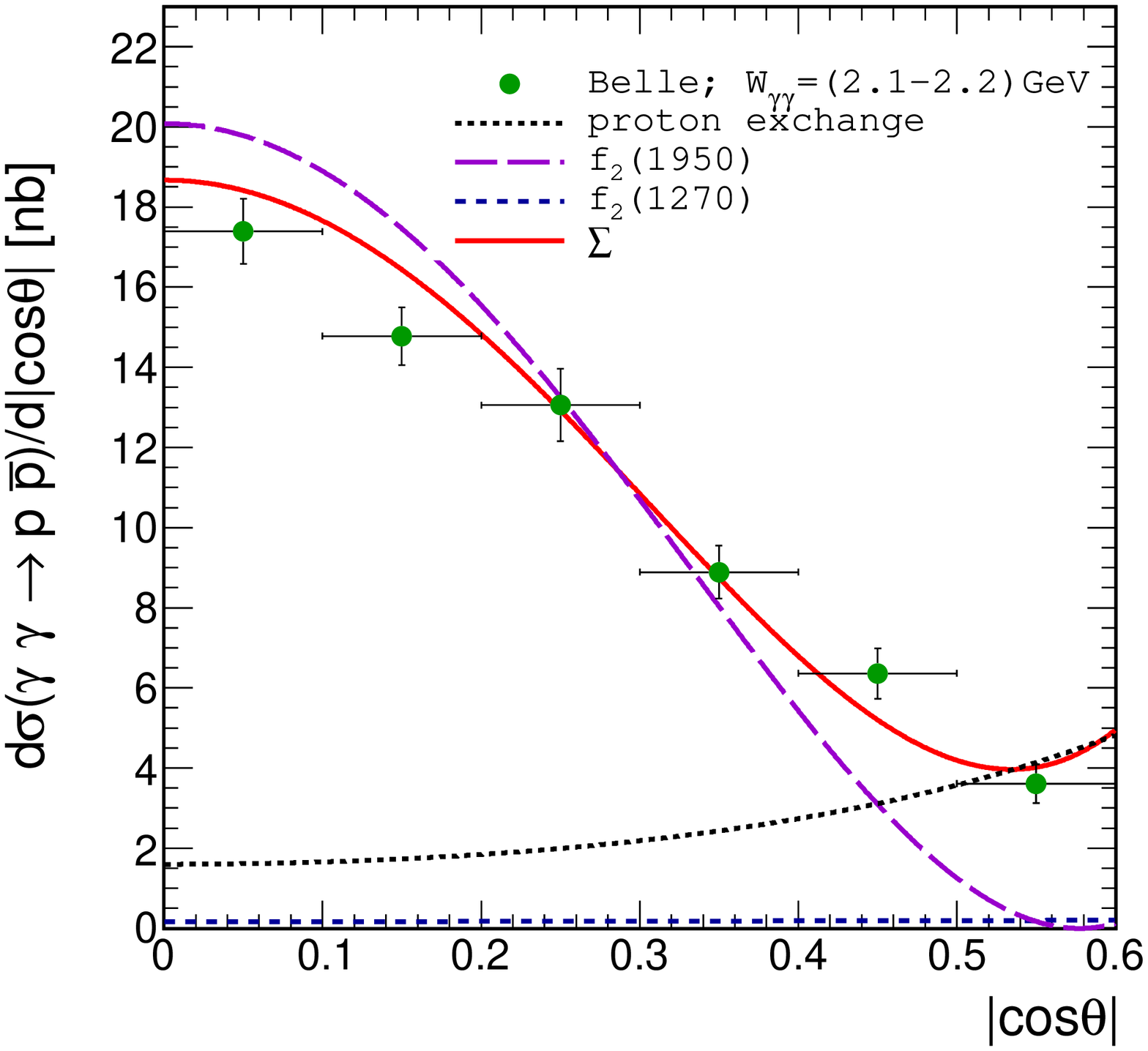}
(c)\includegraphics[width=0.3\textwidth]{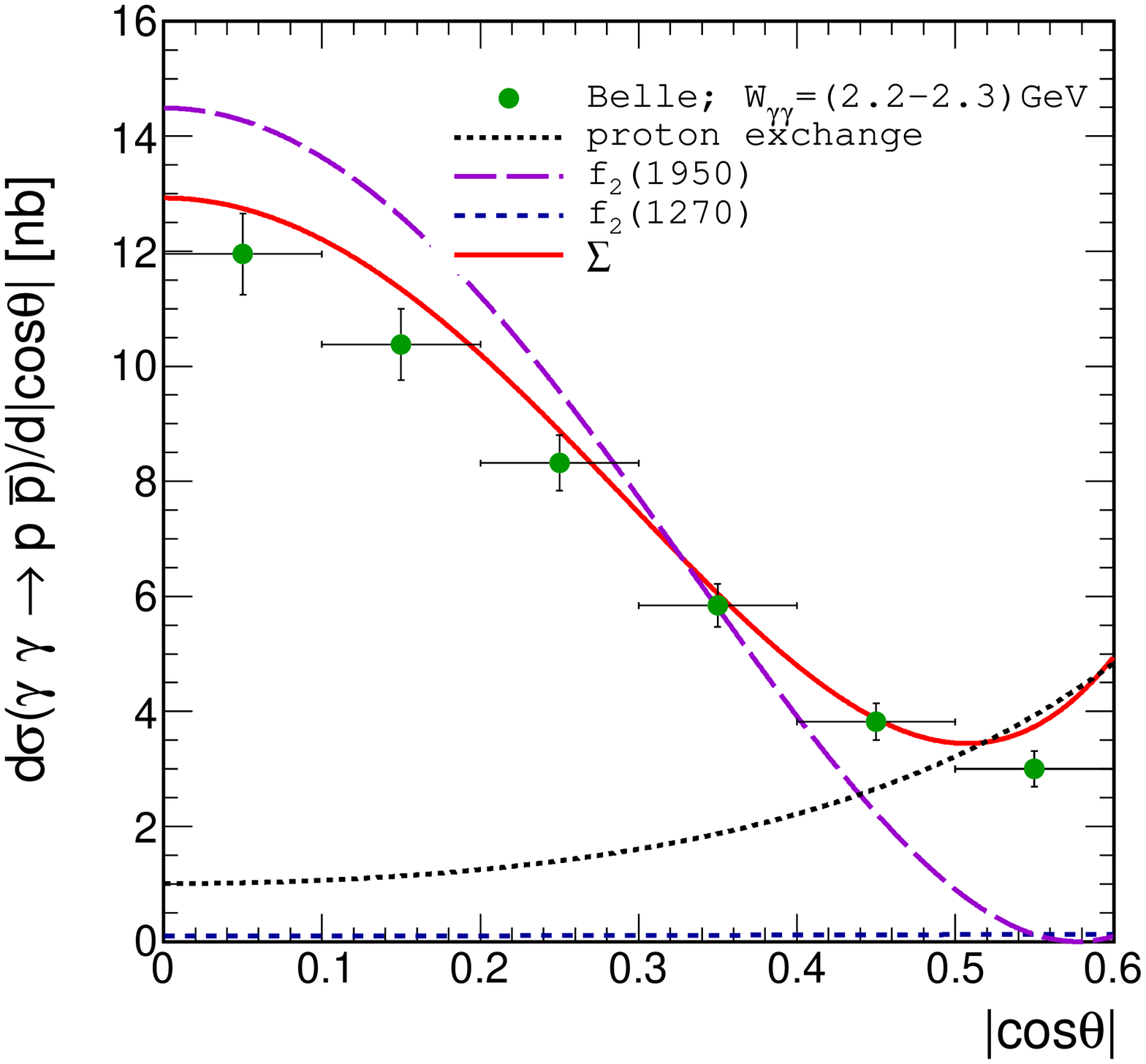}
(d)\includegraphics[width=0.3\textwidth]{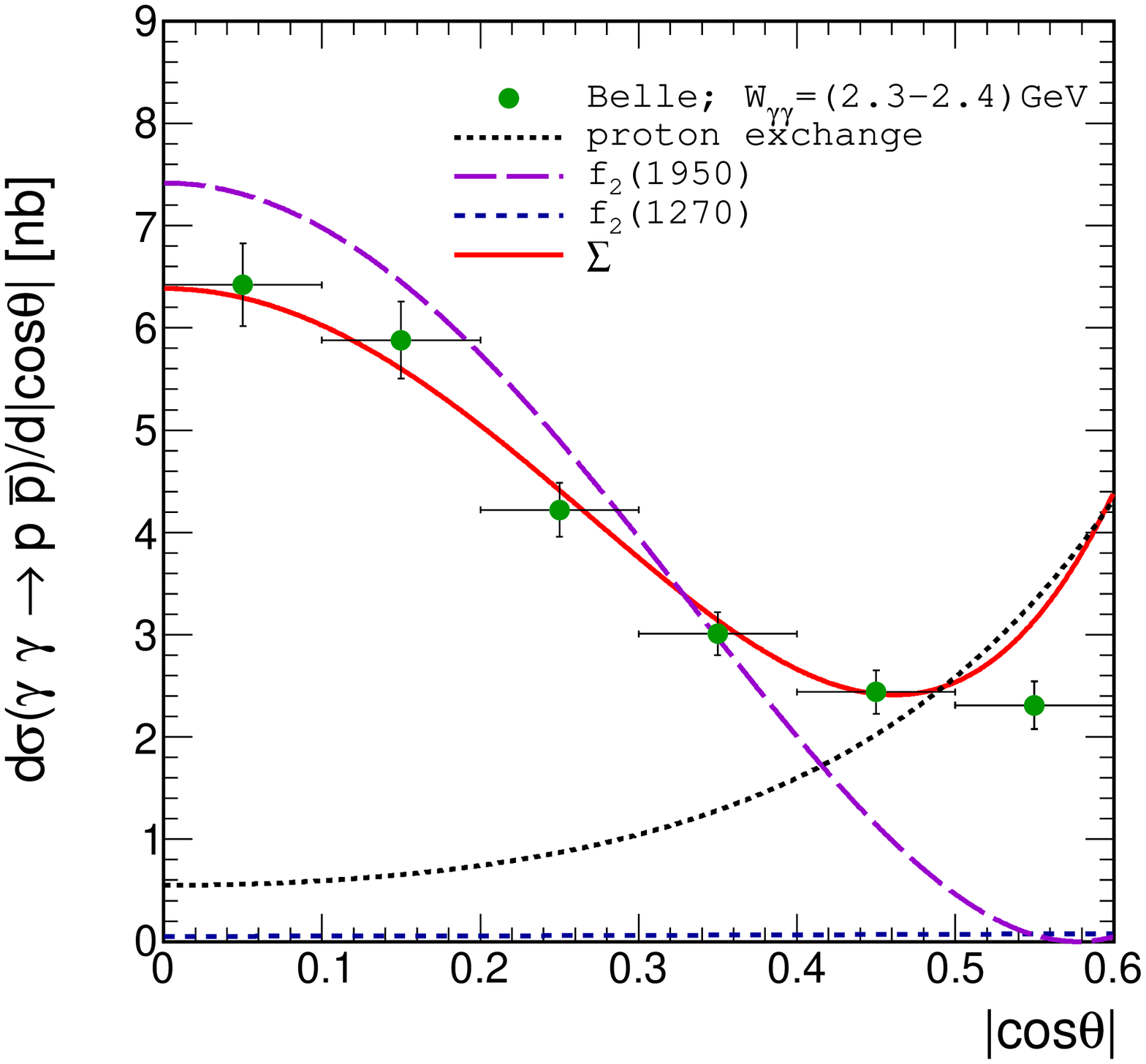}
(e)\includegraphics[width=0.3\textwidth]{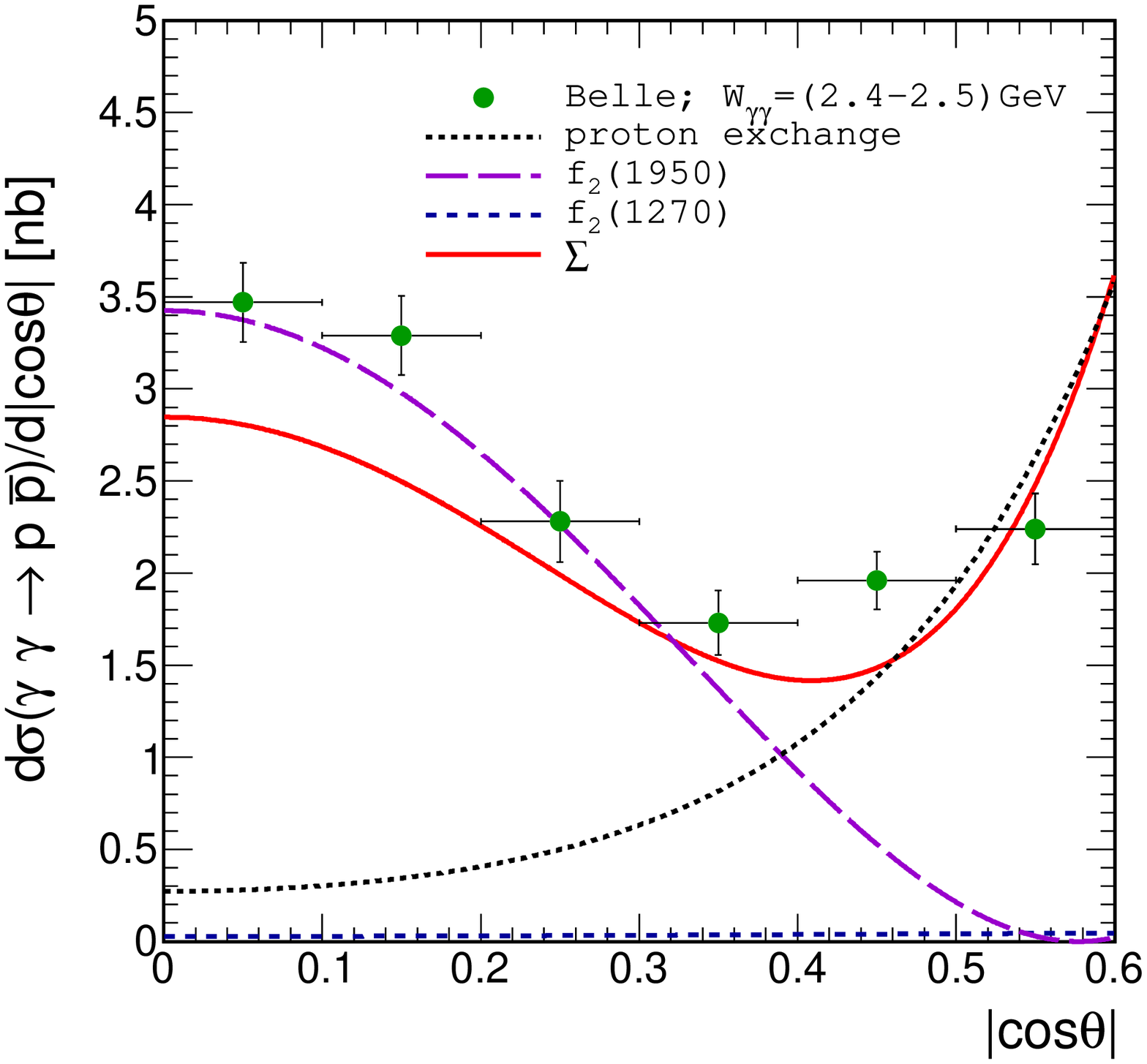}
(f)\includegraphics[width=0.3\textwidth]{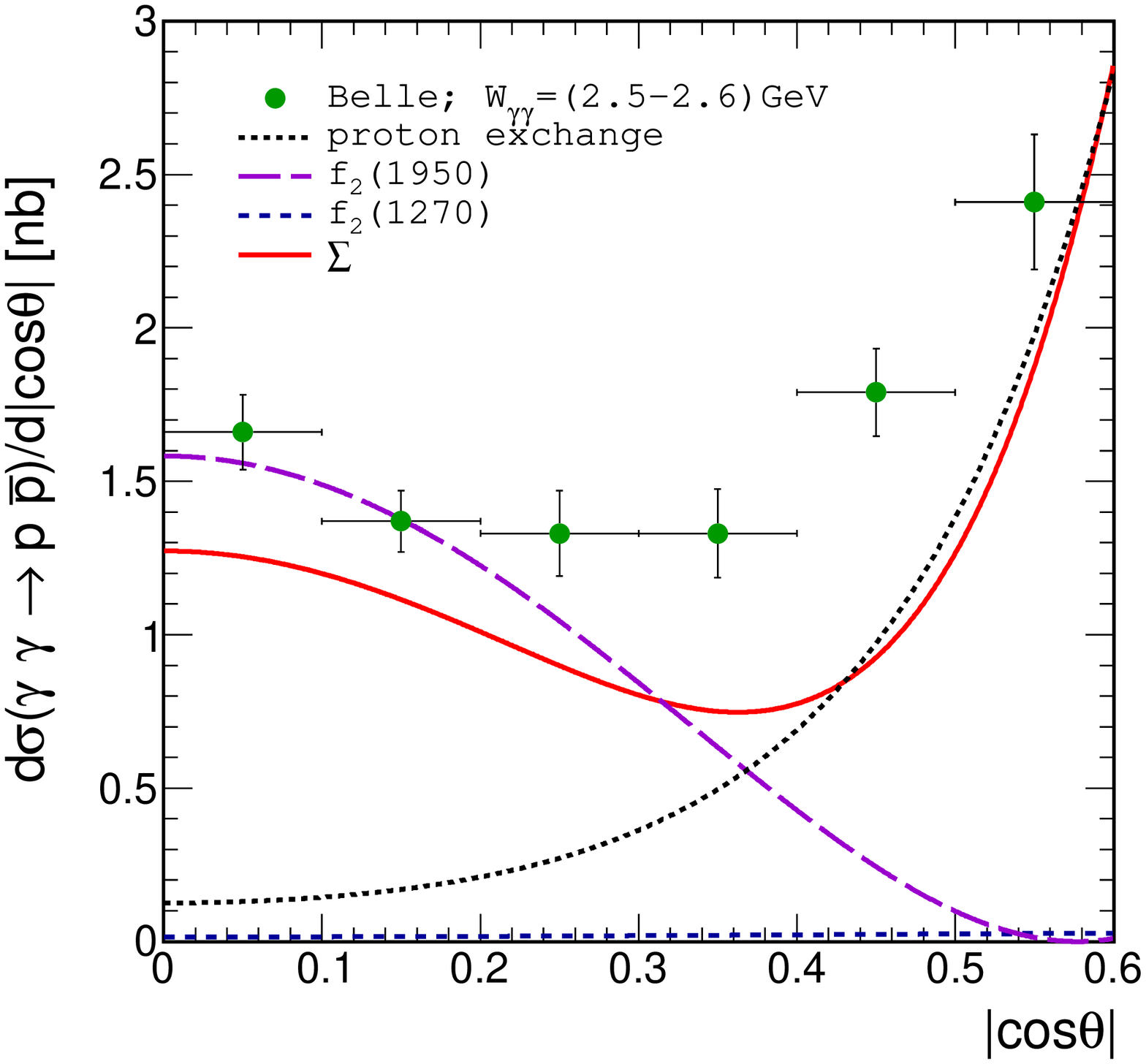}
(g)\includegraphics[width=0.3\textwidth]{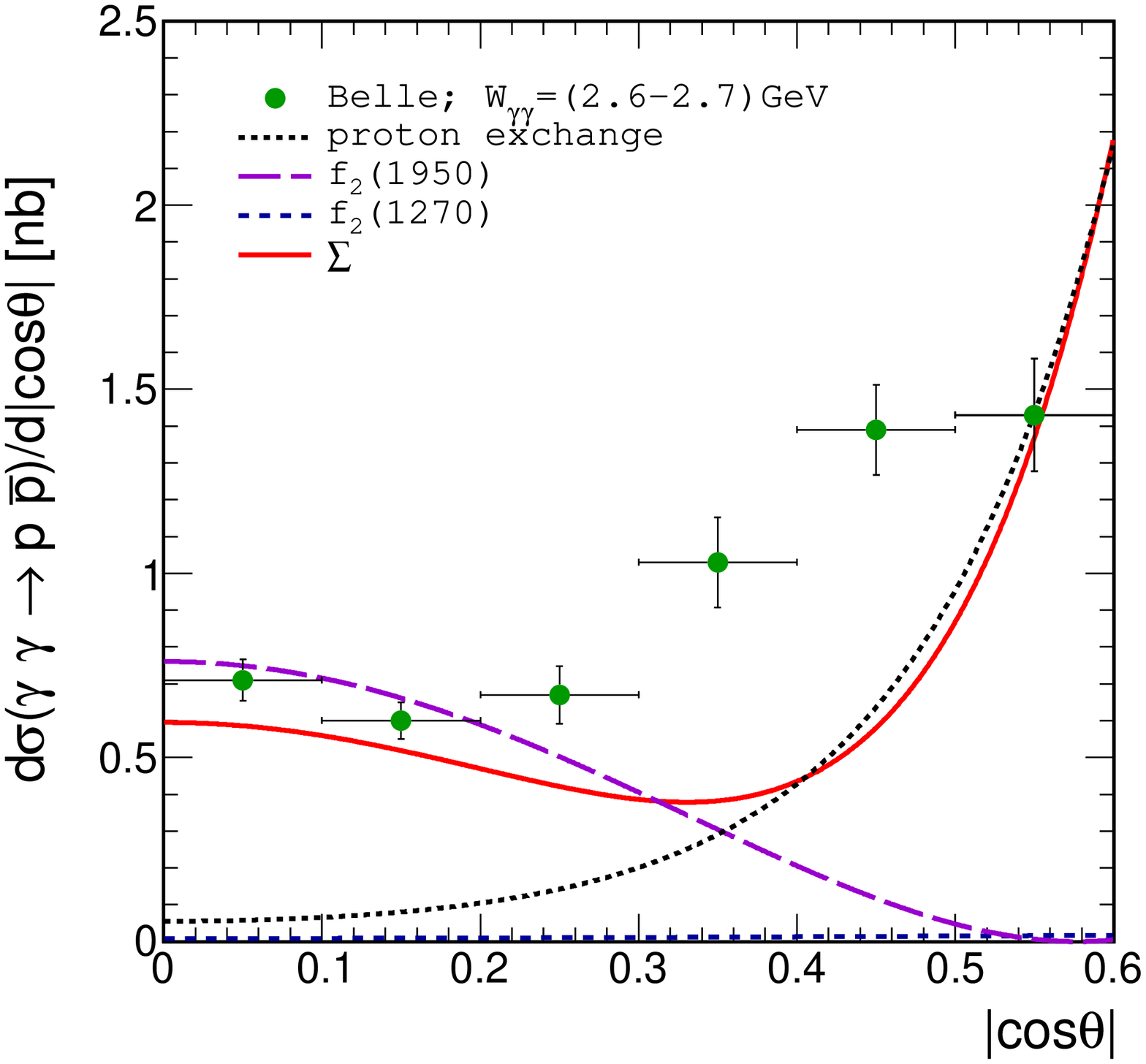}
(h)\includegraphics[width=0.3\textwidth]{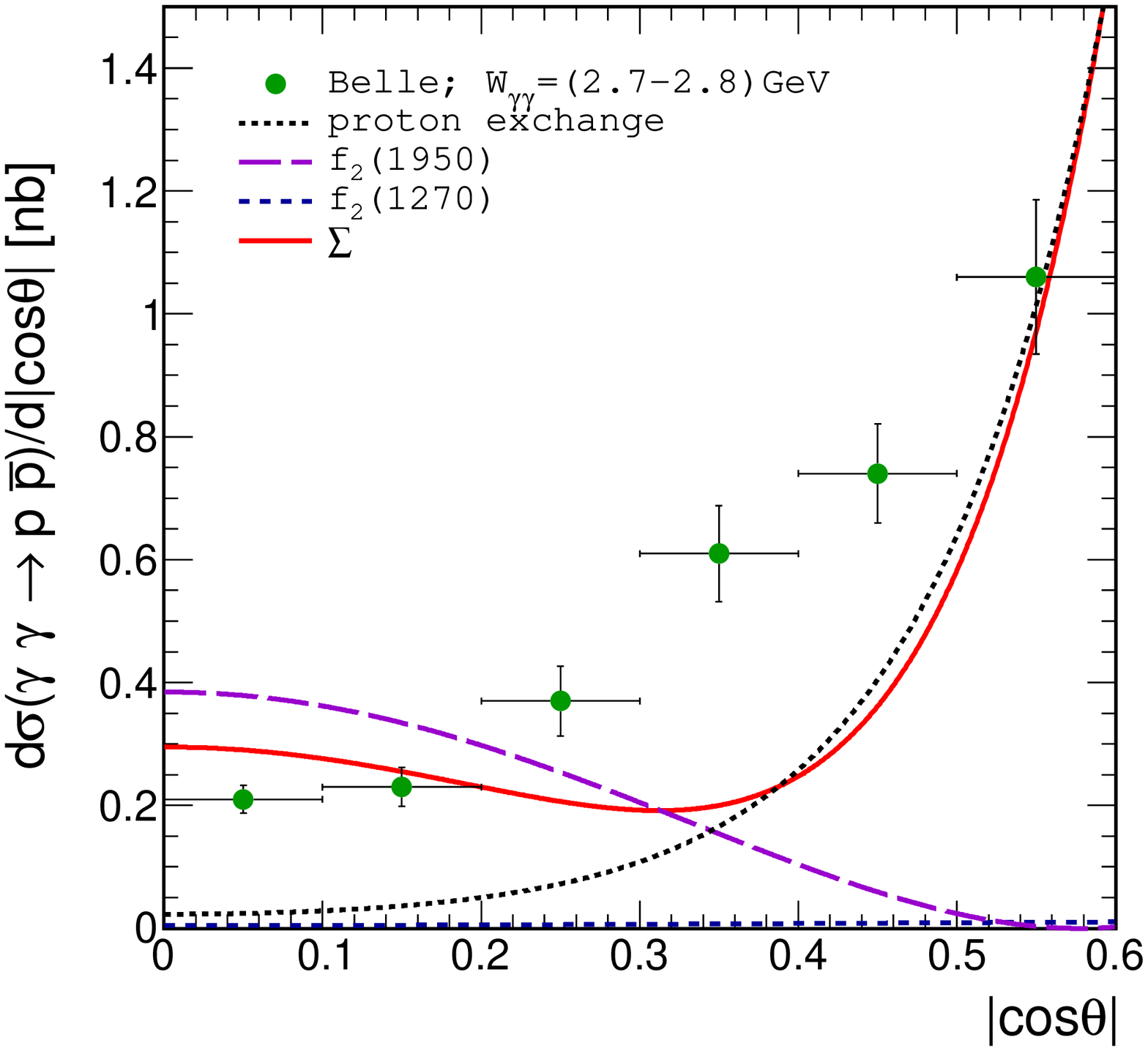}
(i)\includegraphics[width=0.3\textwidth]{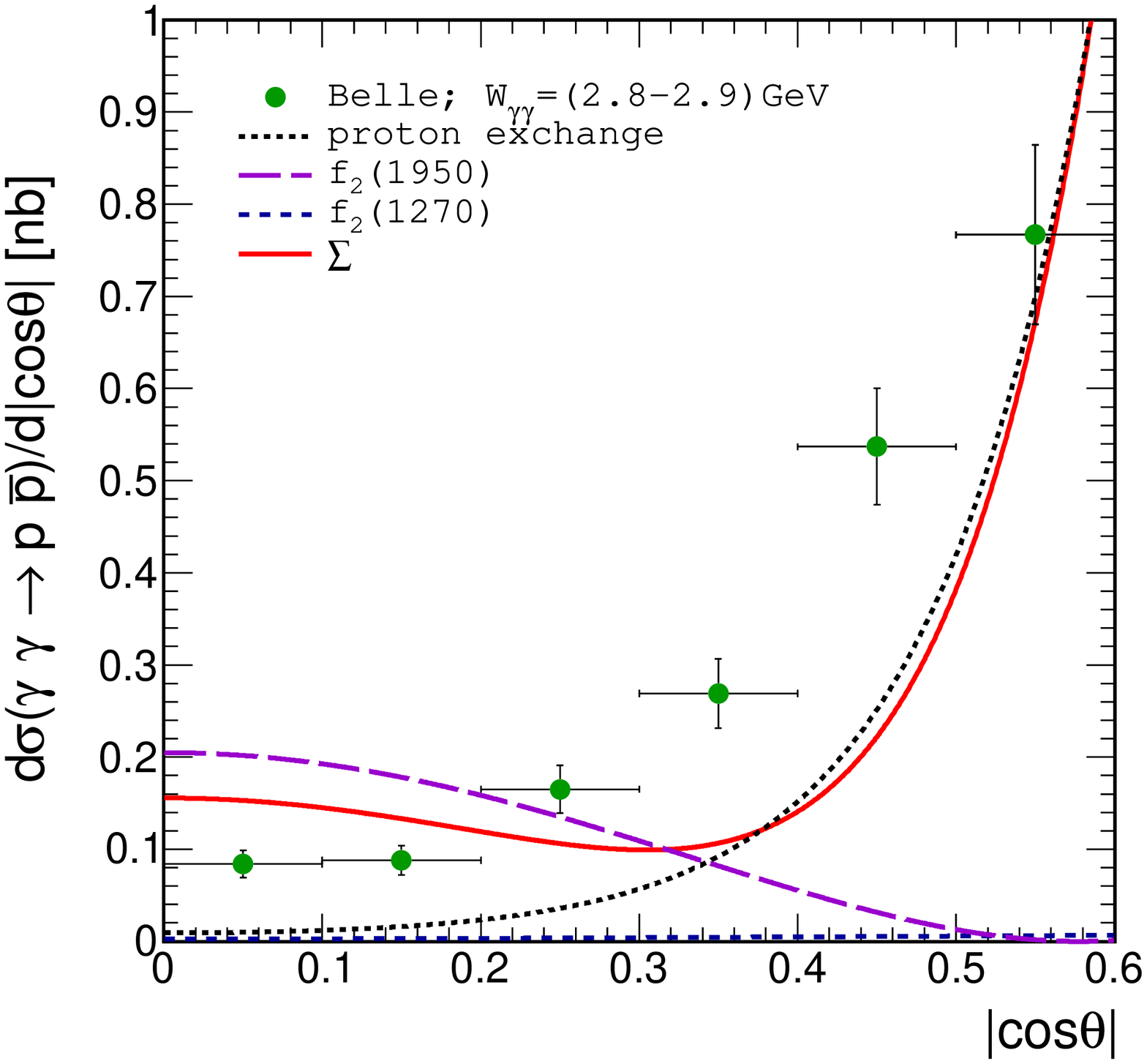}
(j)\includegraphics[width=0.3\textwidth]{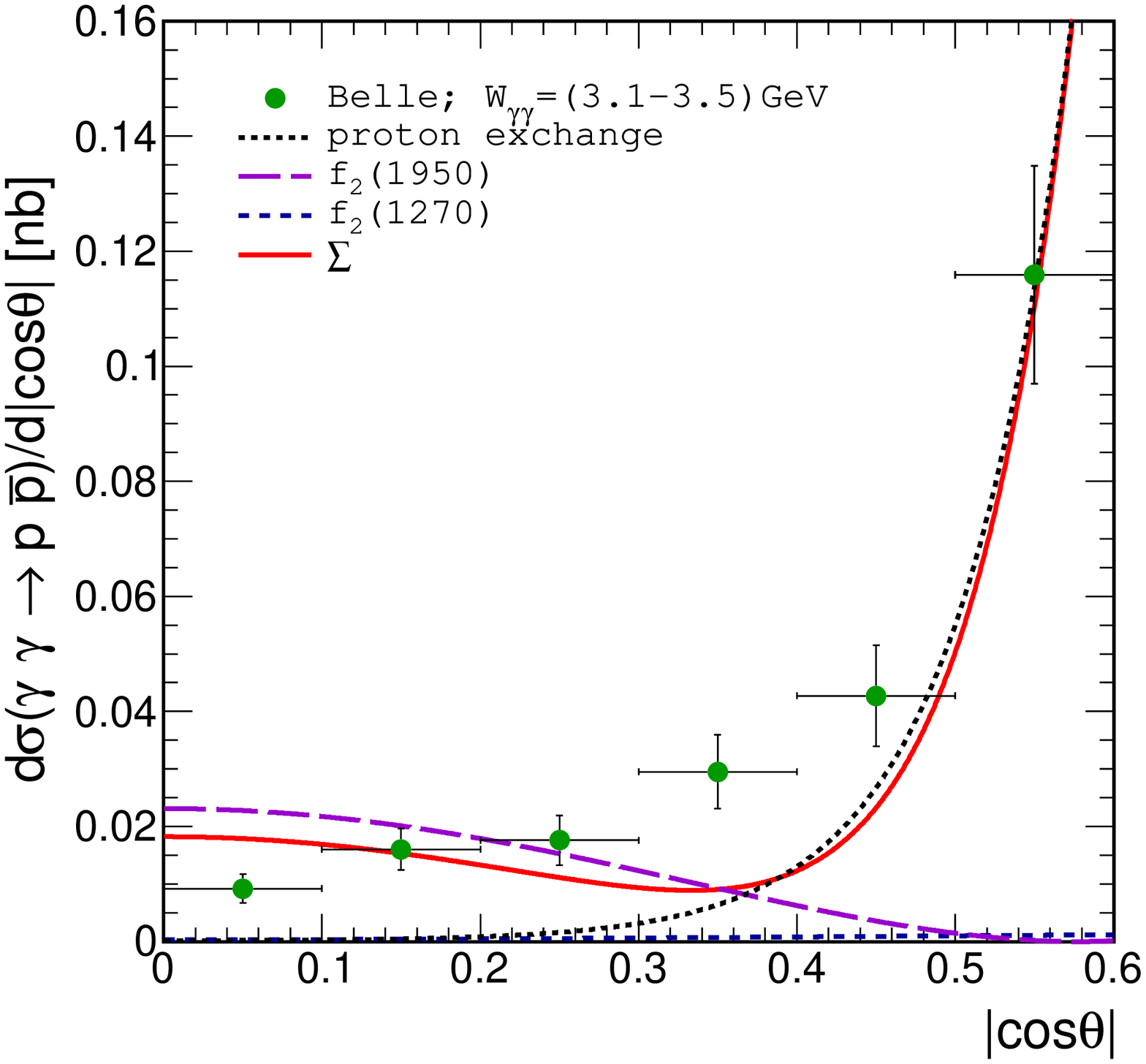}
(k)\includegraphics[width=0.3\textwidth]{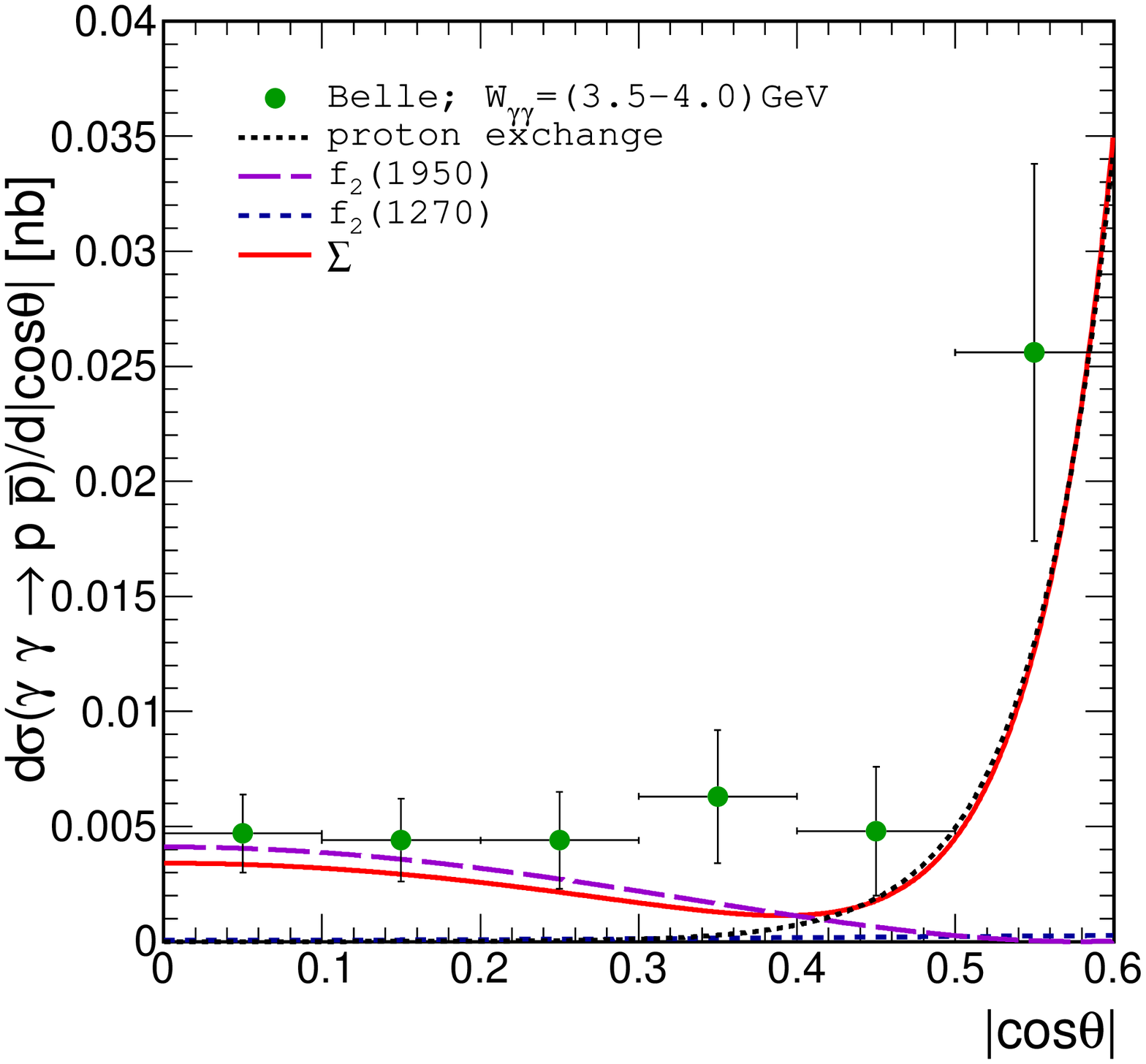}
  \caption{\label{fig:dsig_dz_3mech_nohb}
  \small
Differential cross sections for the $\gamma \gamma \to p \bar{p}$ reaction
as a function of $|\cos\theta|$ for different $W_{\gamma\gamma}$ ranges.
For the Belle data \cite{Kuo:2005nr} 
both statistical and systematic uncertainties are included.
Calculations were done with 
$\Lambda_{f_{2},pow} = 1.15$~GeV
in (\ref{formfactor_f2_secondform}),
and $\Lambda_{p} = 1.08$~GeV in (\ref{ff_Poppe}).
The hand-bag model contribution is not included here.
Here we used the parameter set~A from Table~\ref{table:parameters}.
}
\end{figure}

\begin{figure}[!ht]
(a)\includegraphics[width=0.3\textwidth]{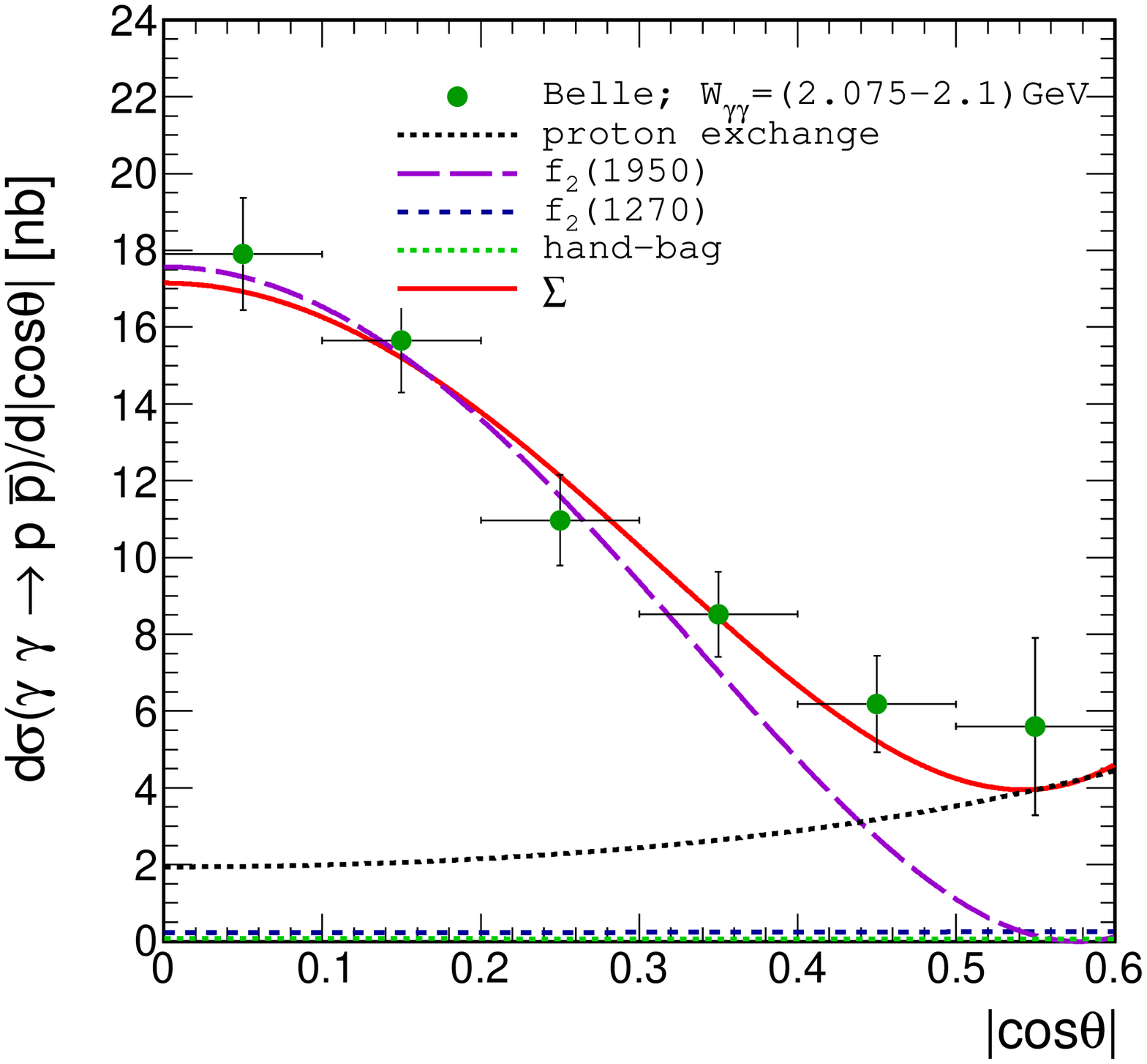}
(b)\includegraphics[width=0.3\textwidth]{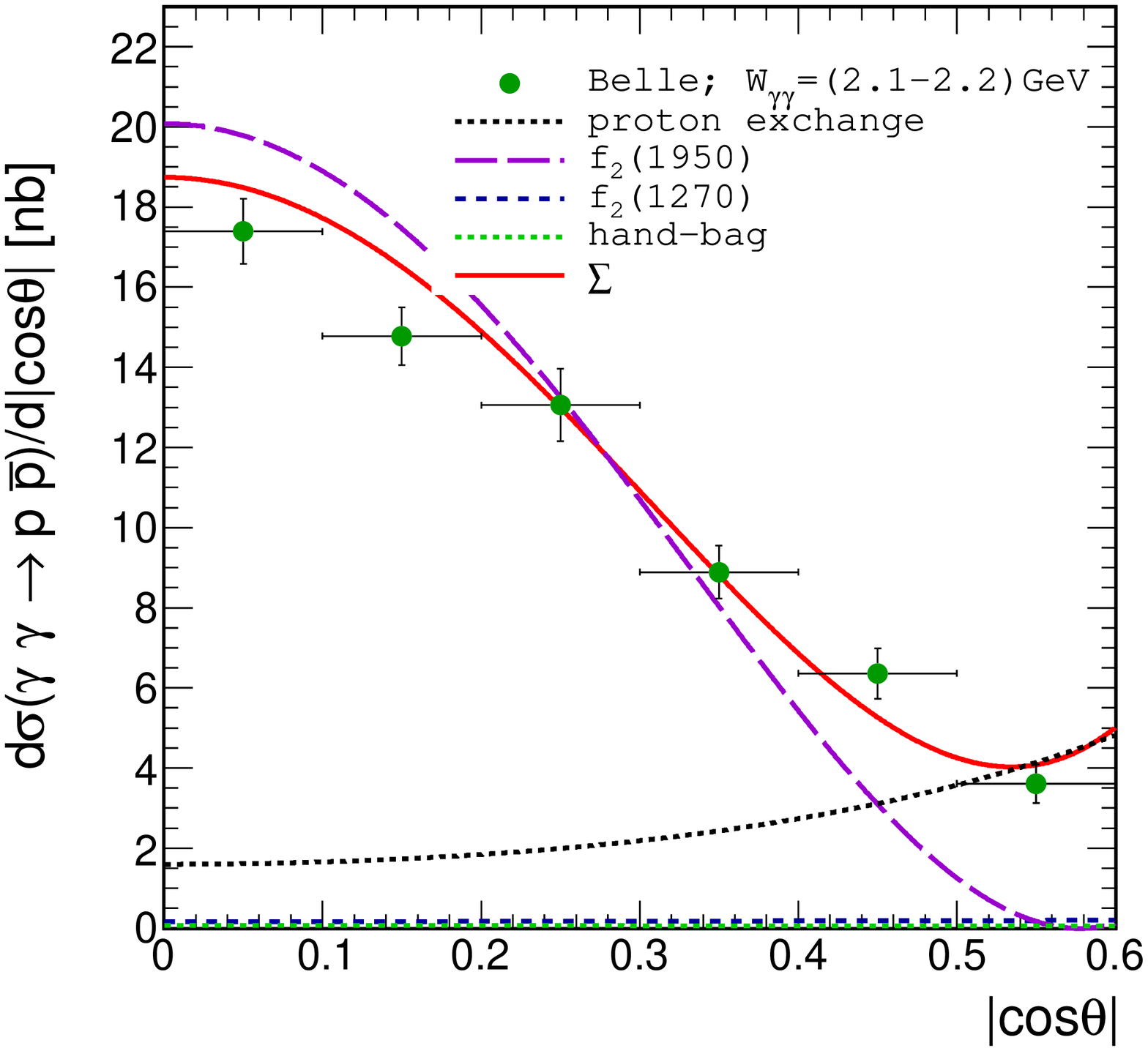}
(c)\includegraphics[width=0.3\textwidth]{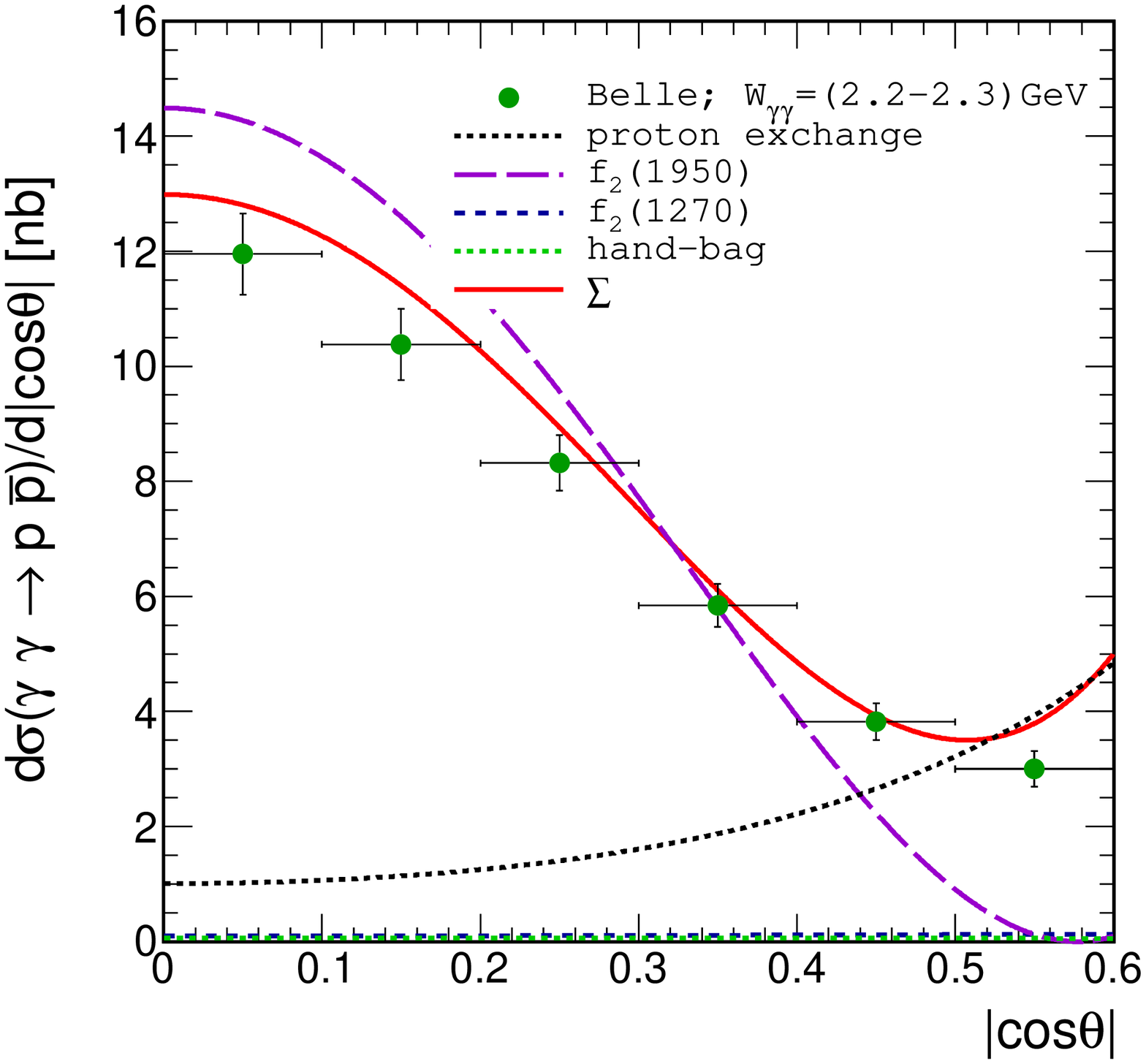}
(d)\includegraphics[width=0.3\textwidth]{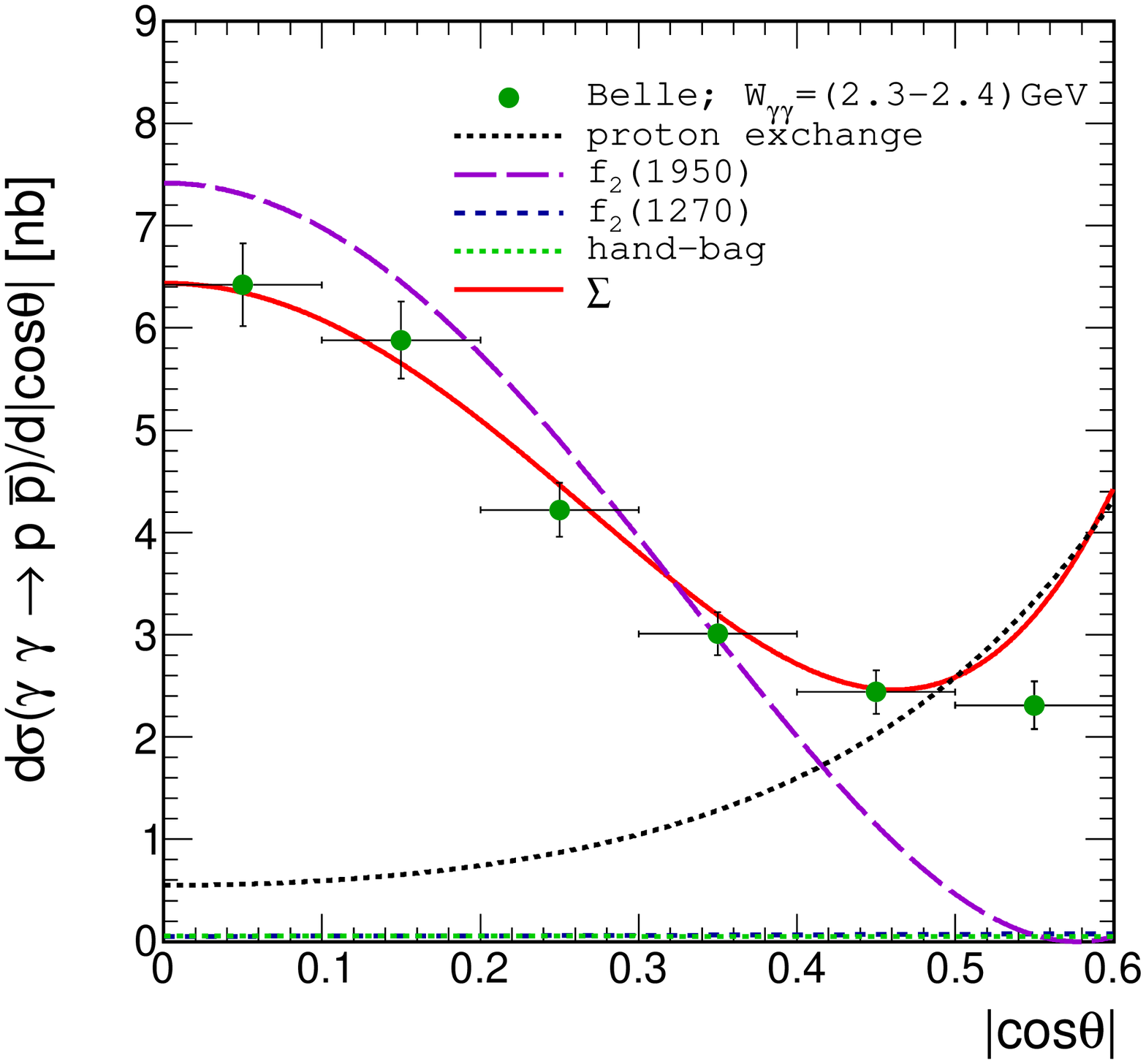}
(e)\includegraphics[width=0.3\textwidth]{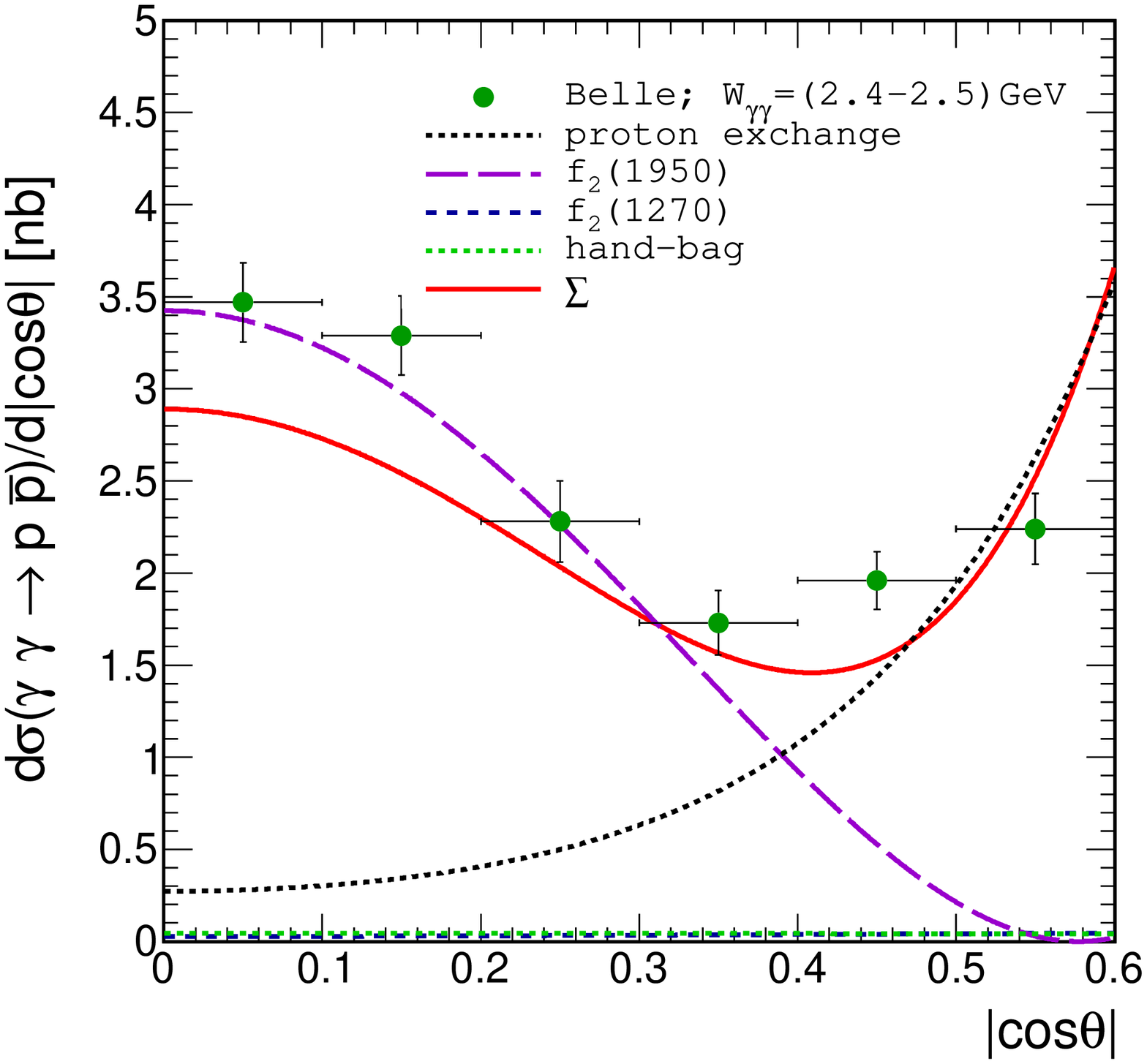}
(f)\includegraphics[width=0.3\textwidth]{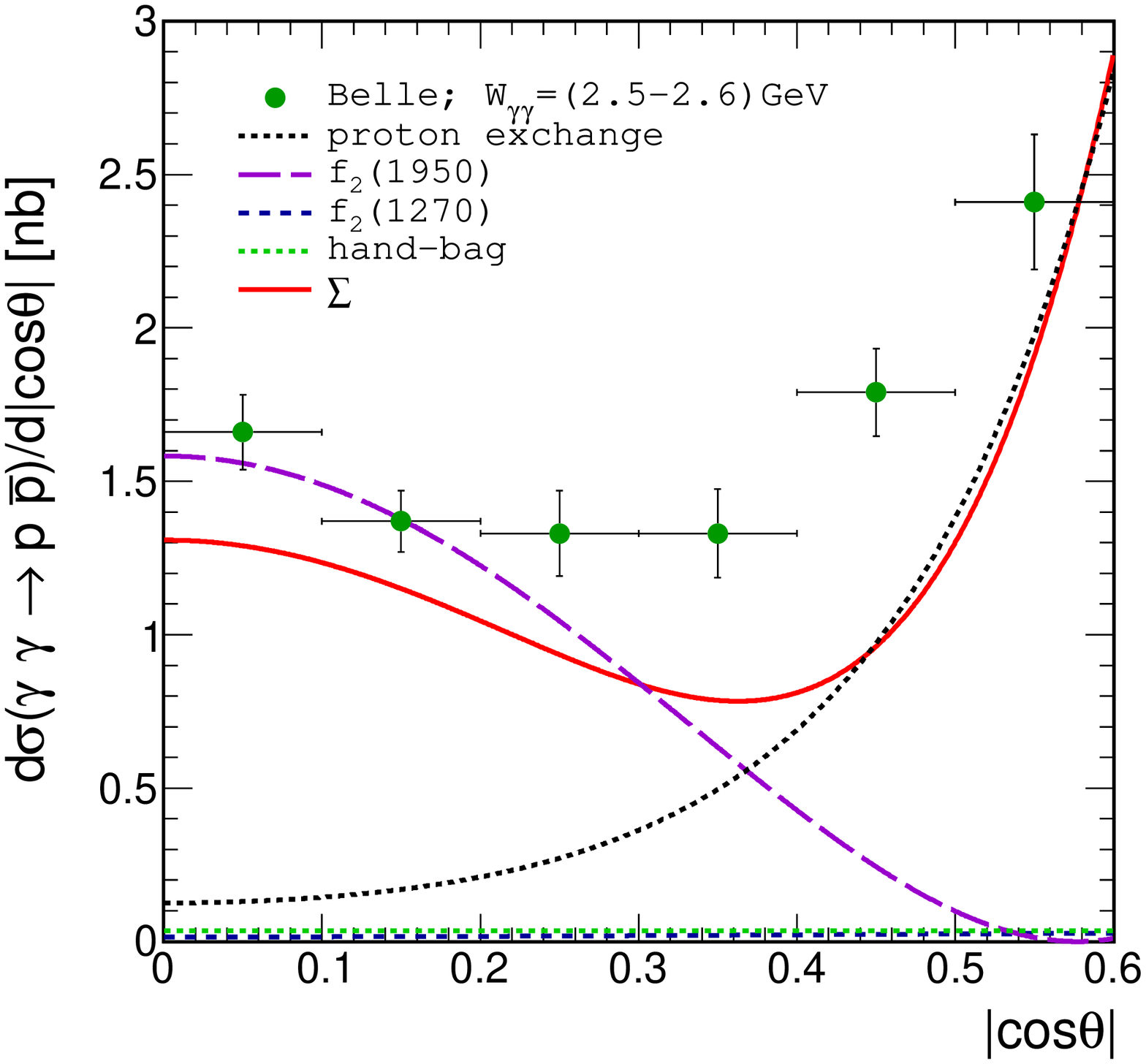}
(g)\includegraphics[width=0.3\textwidth]{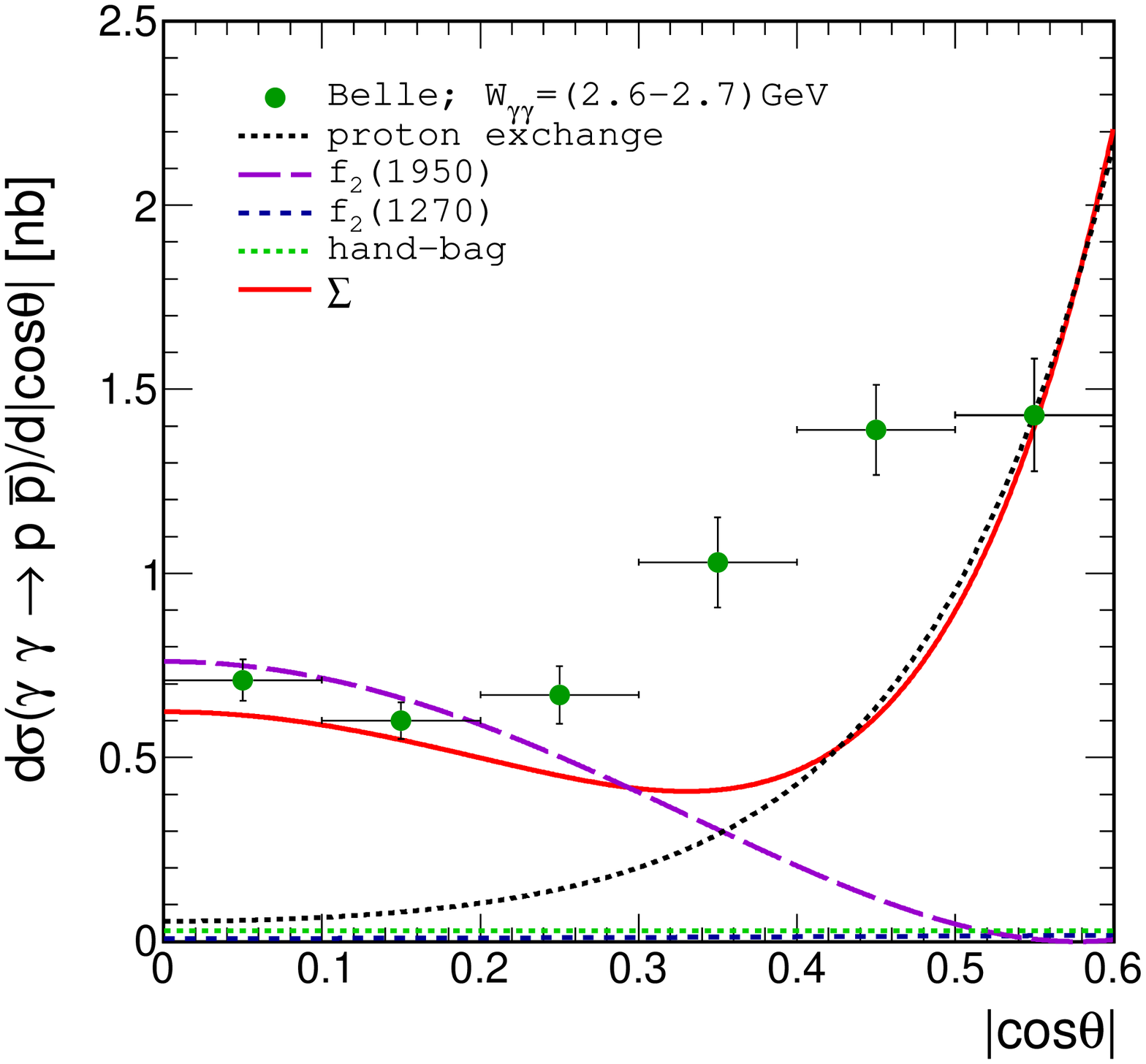}
(h)\includegraphics[width=0.3\textwidth]{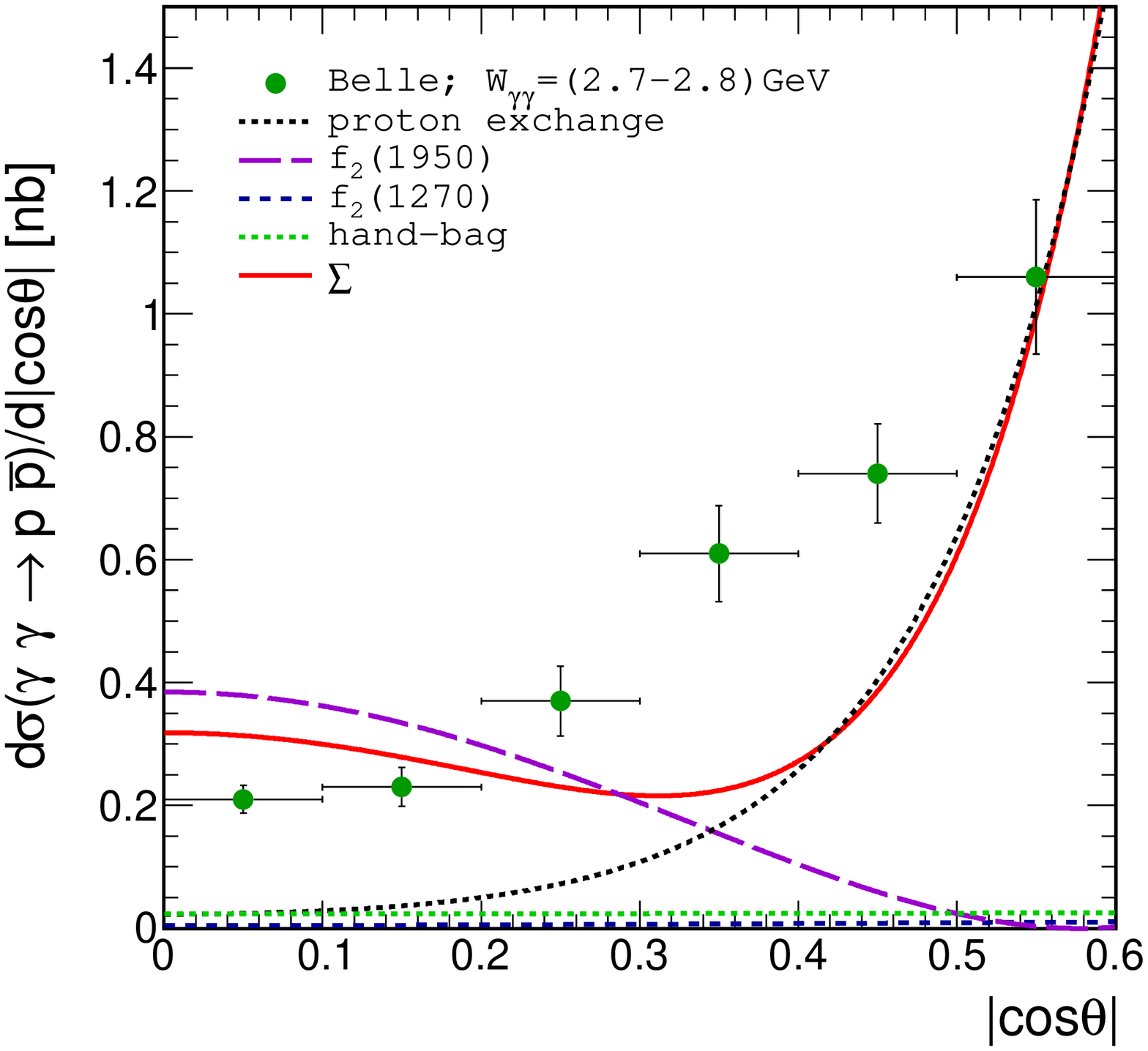}
(i)\includegraphics[width=0.3\textwidth]{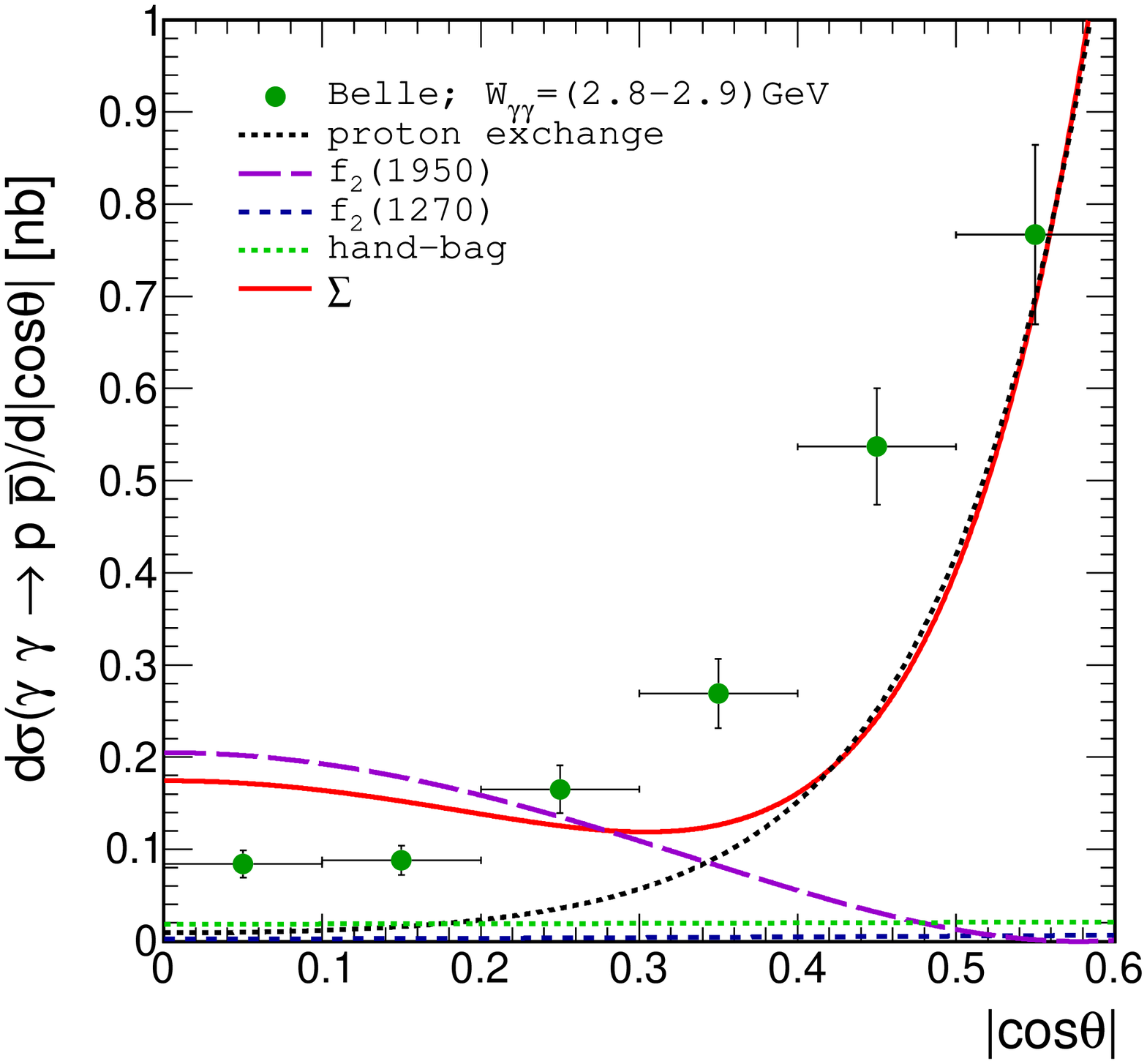}
(j)\includegraphics[width=0.3\textwidth]{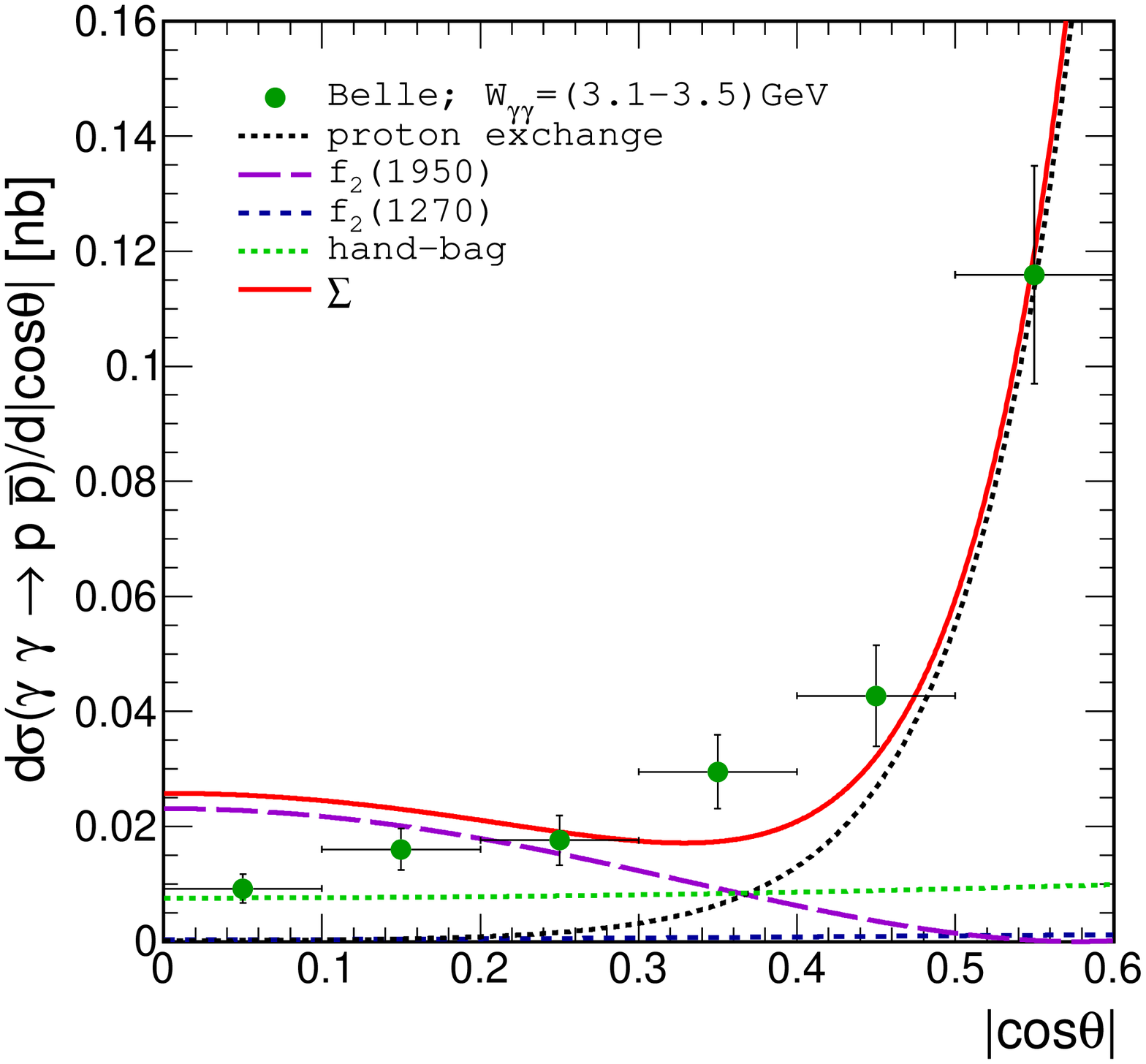}
(k)\includegraphics[width=0.3\textwidth]{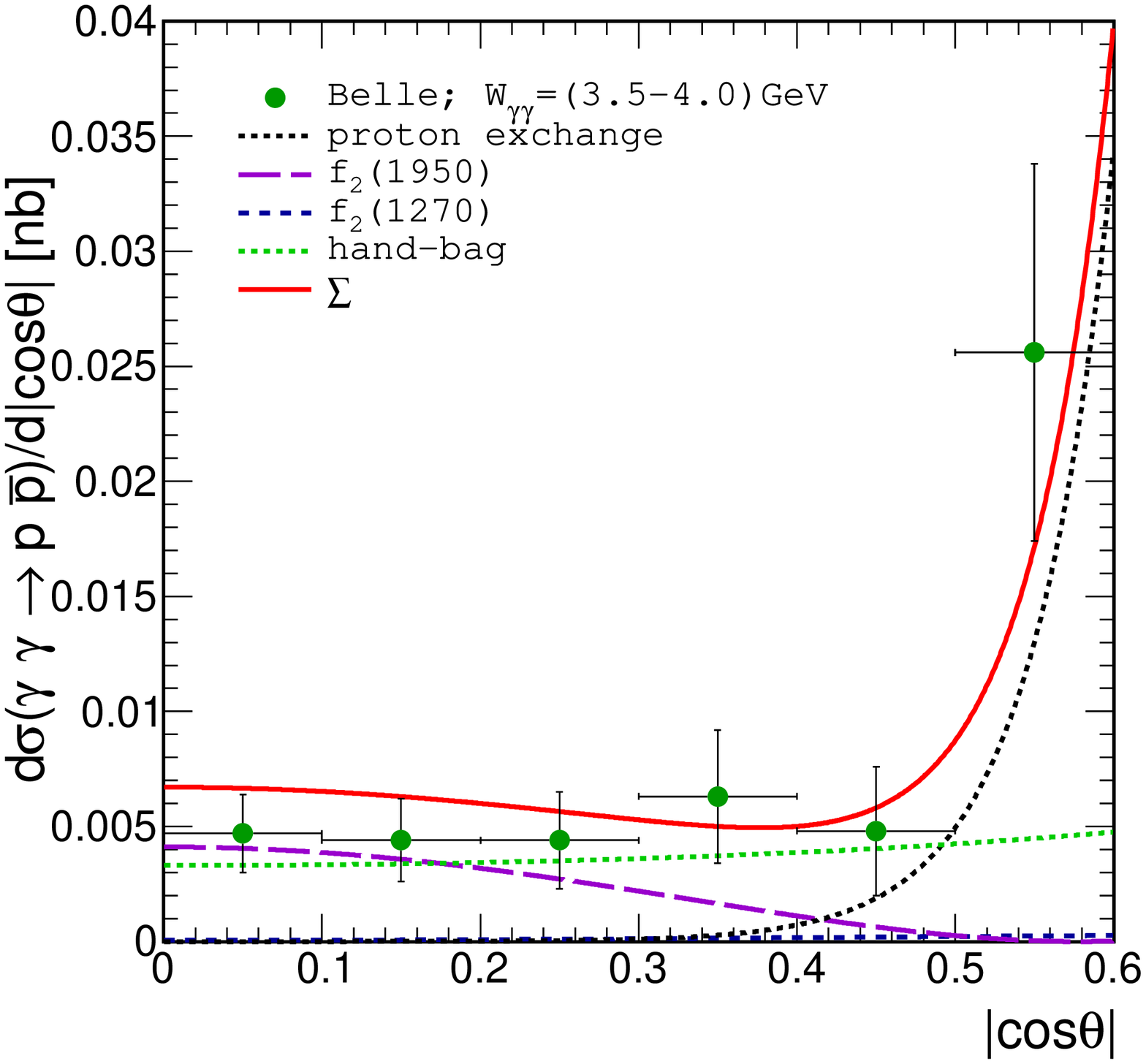}
  \caption{\label{fig:dsig_dz_3mech}
  \small
The same as in Fig.~\ref{fig:dsig_dz_3mech_nohb} but here
the hand-bag contribution is included. 
The green dotted line shows the contribution of the hand-bag mechanism.
Here we used the parameter set~A from Table~\ref{table:parameters}.
}
\end{figure}

\begin{figure}[!ht]
(a)\includegraphics[width=0.3\textwidth]{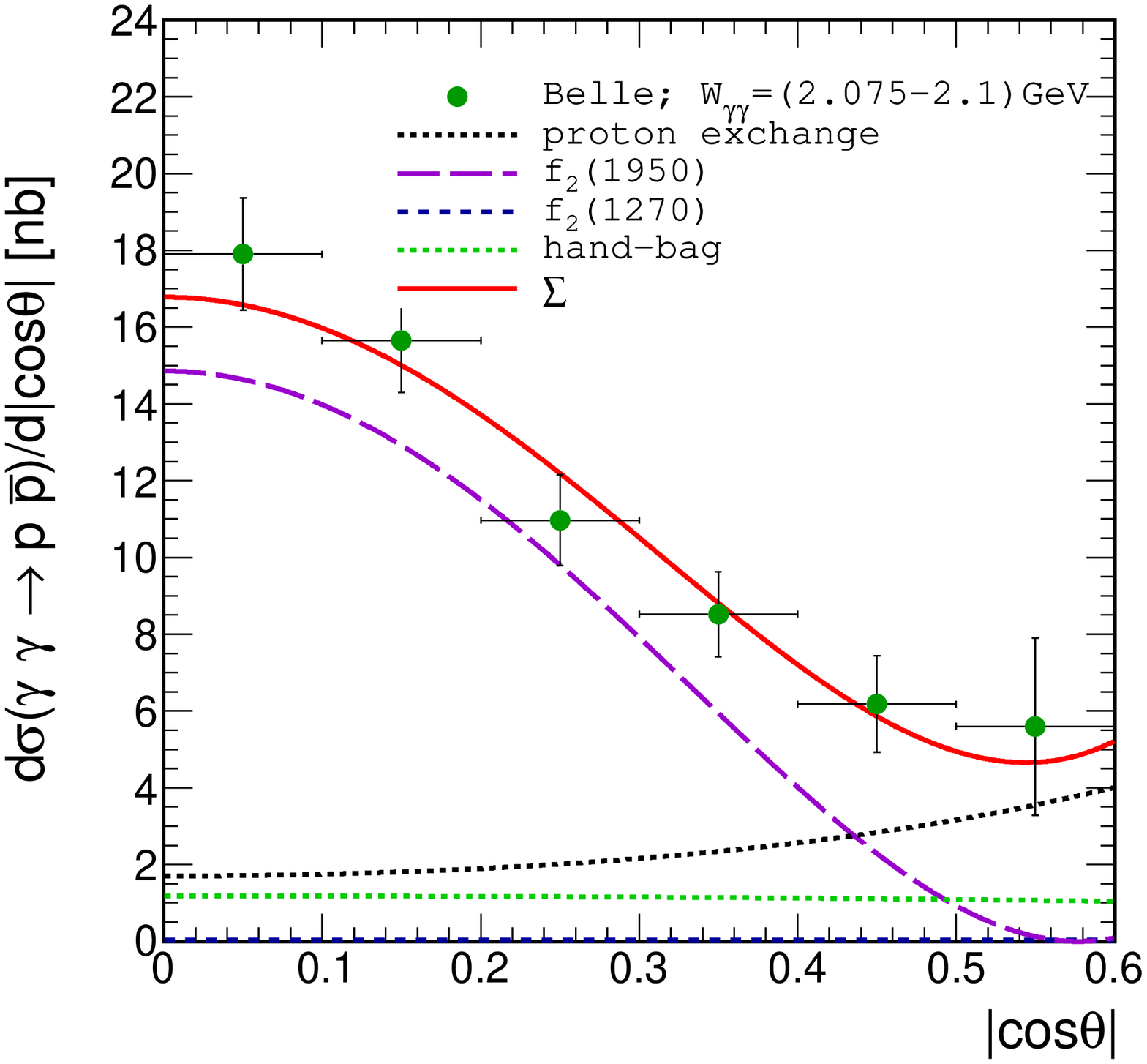}
(b)\includegraphics[width=0.3\textwidth]{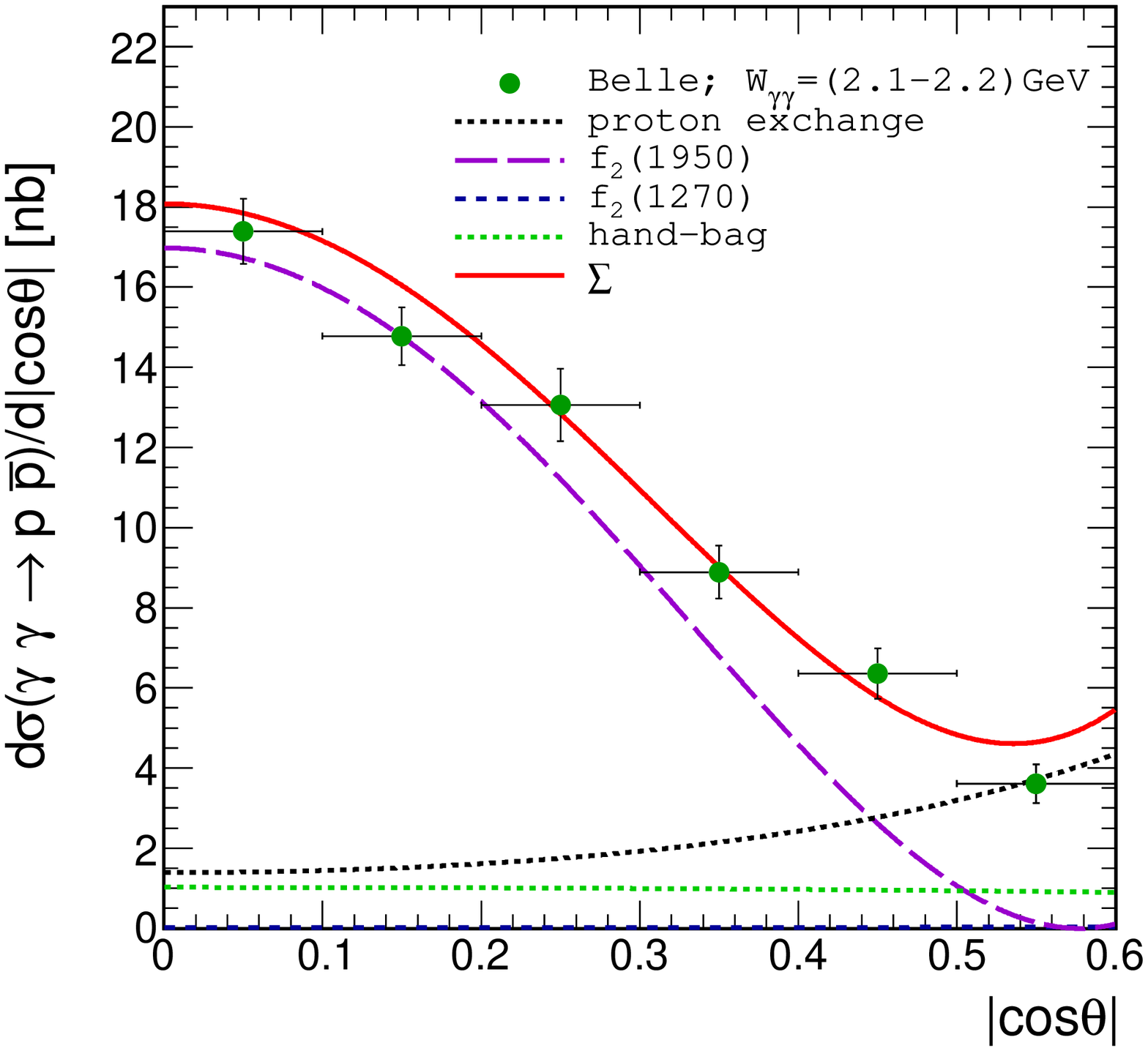}
(c)\includegraphics[width=0.3\textwidth]{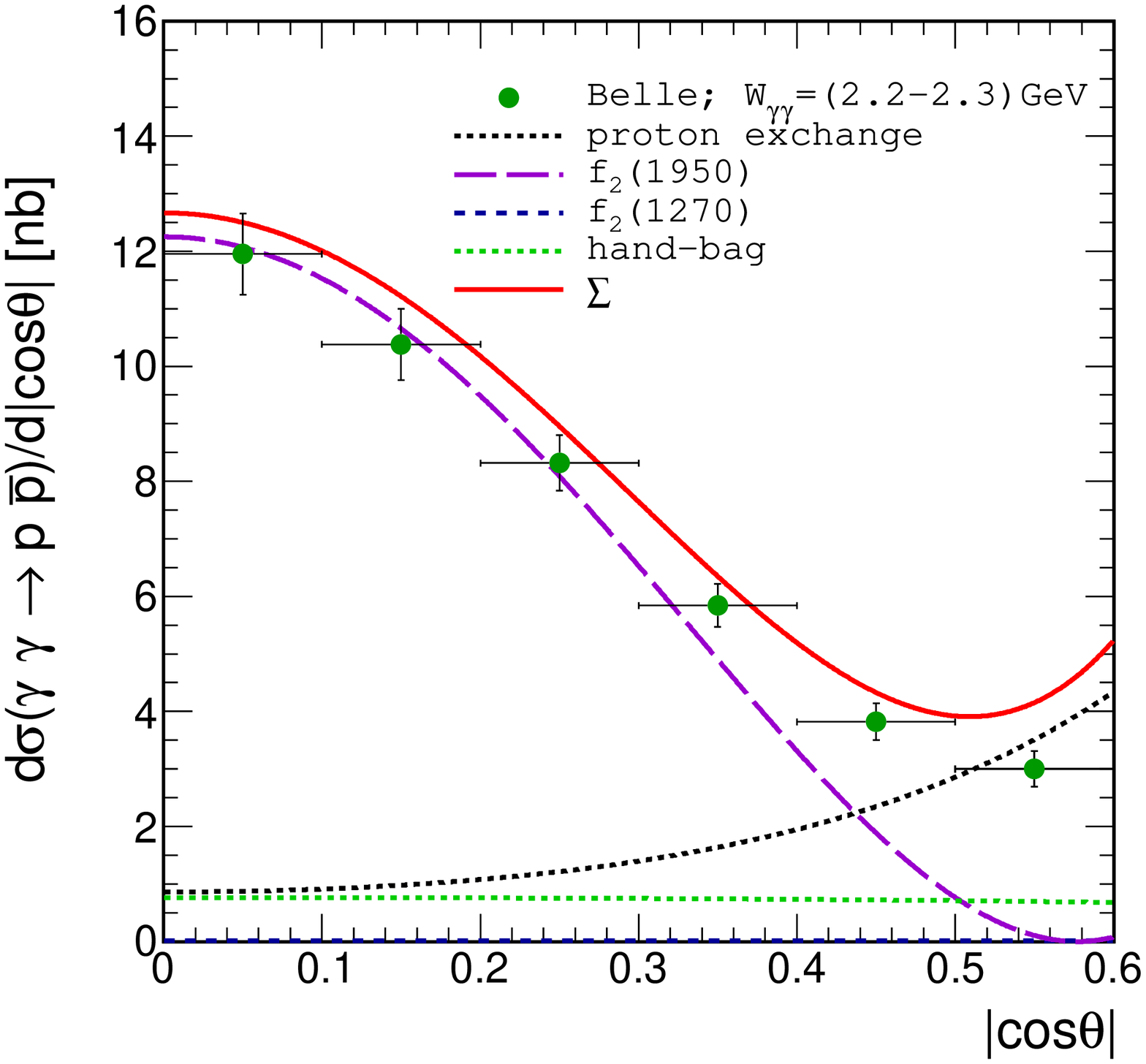}
(d)\includegraphics[width=0.3\textwidth]{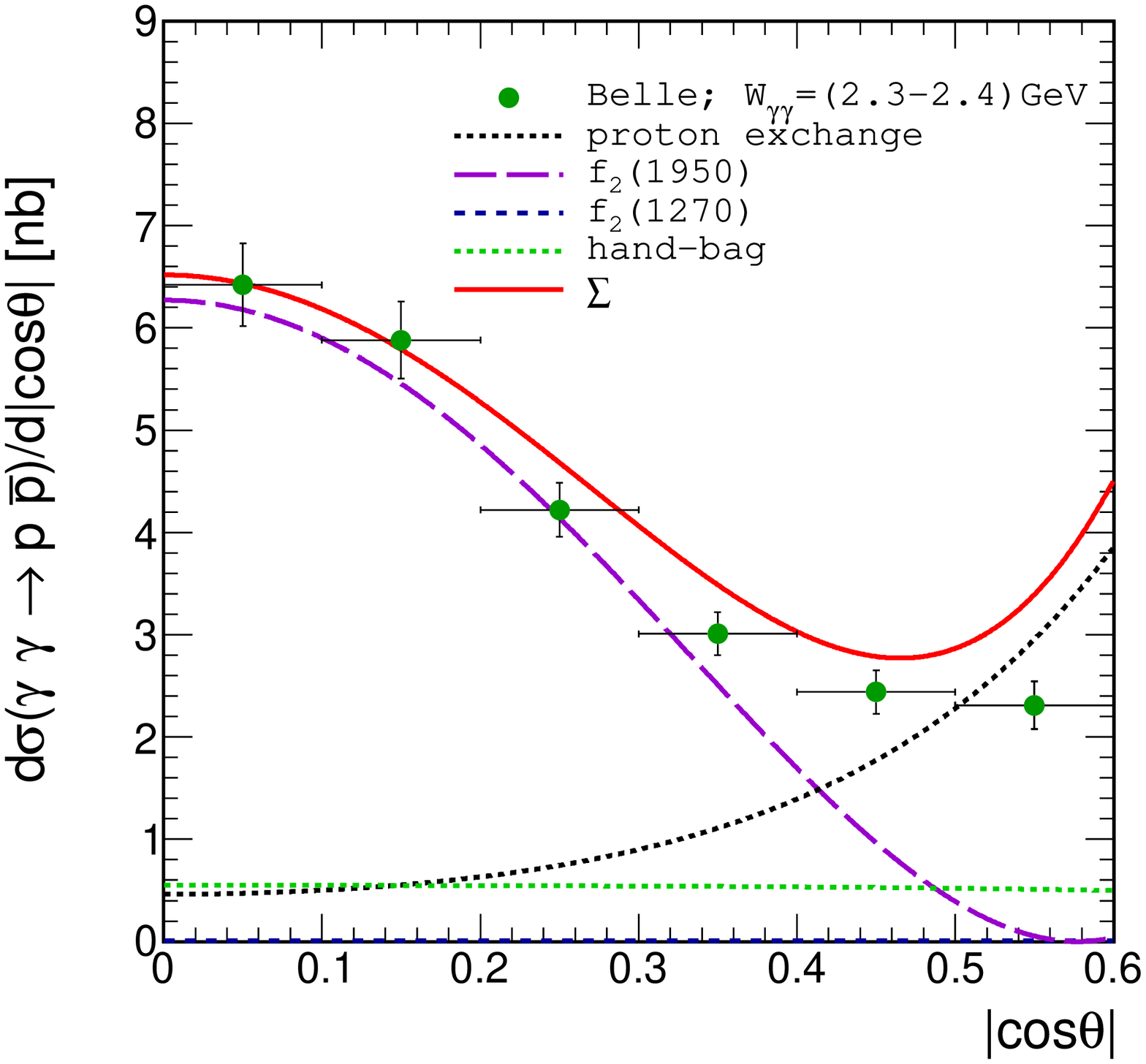}
(e)\includegraphics[width=0.3\textwidth]{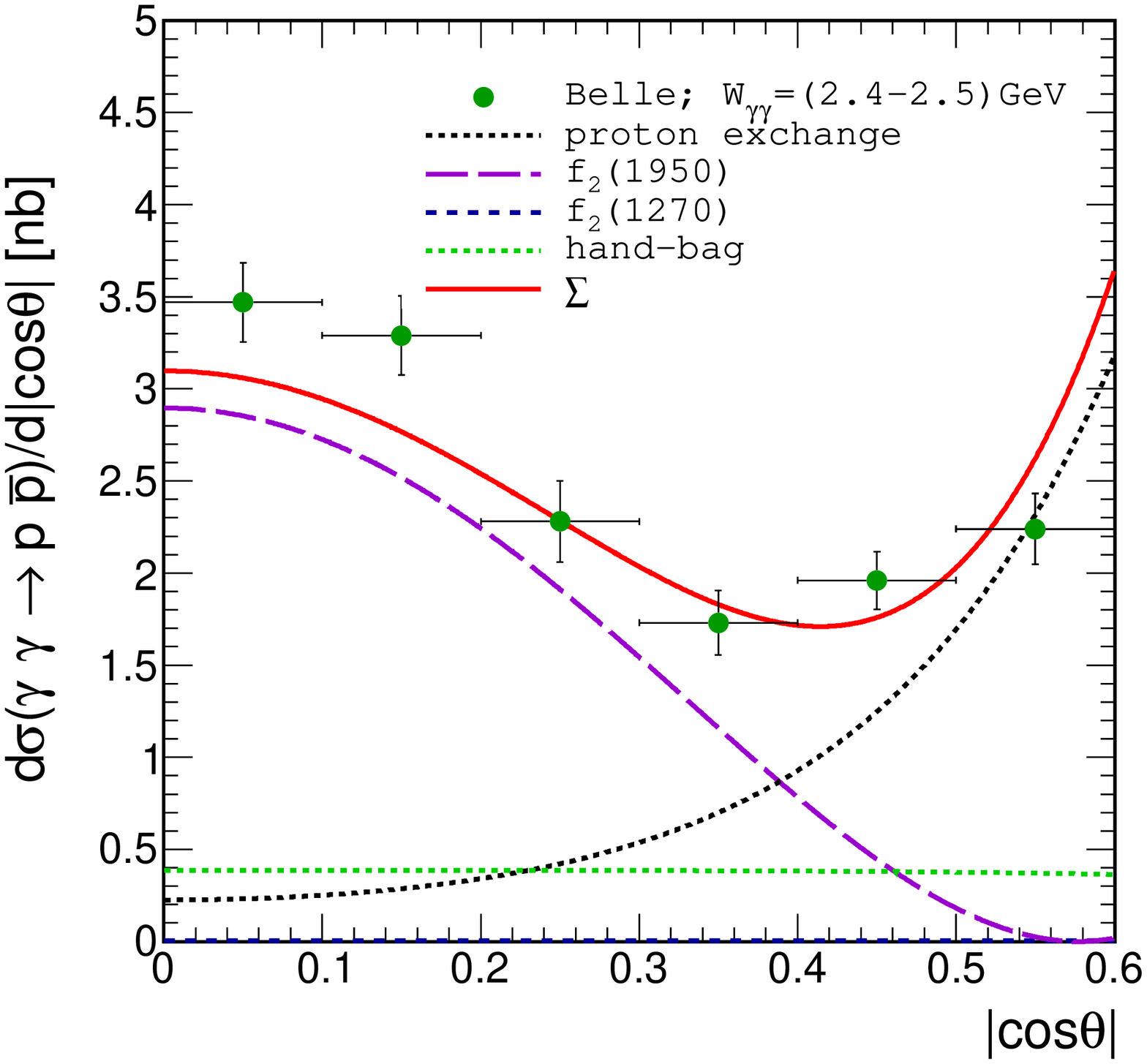}
(f)\includegraphics[width=0.3\textwidth]{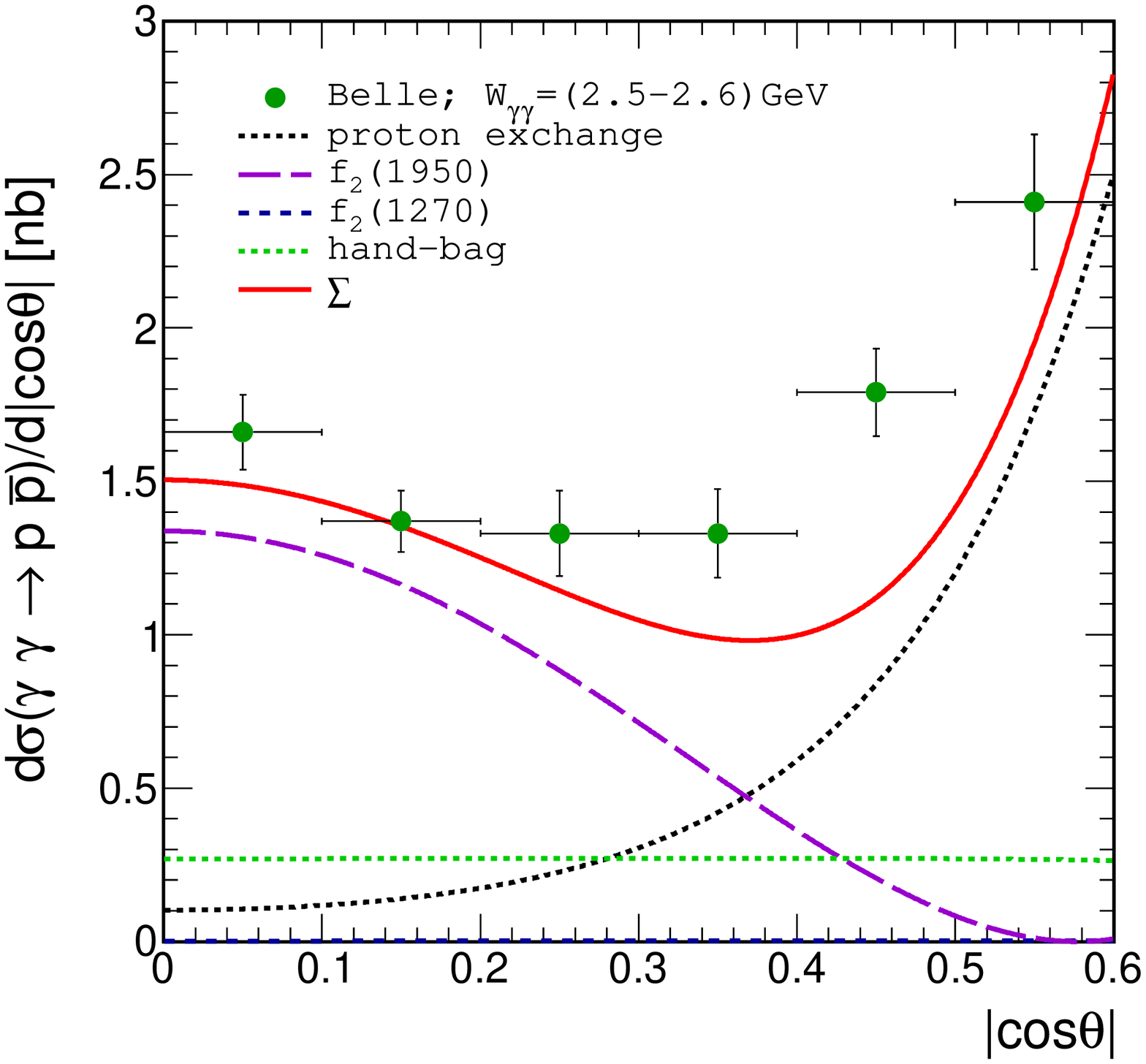}
(g)\includegraphics[width=0.3\textwidth]{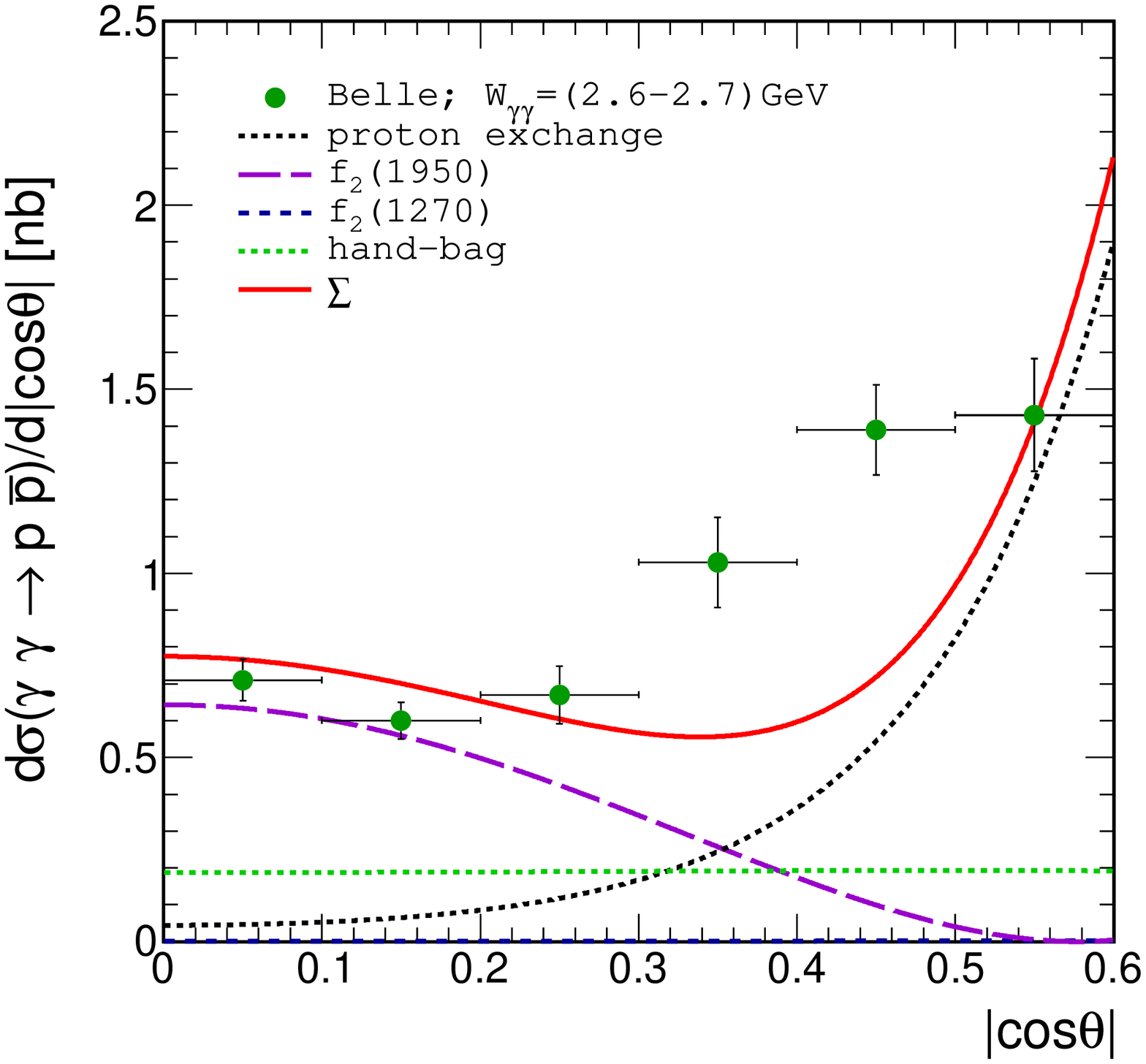}
(h)\includegraphics[width=0.3\textwidth]{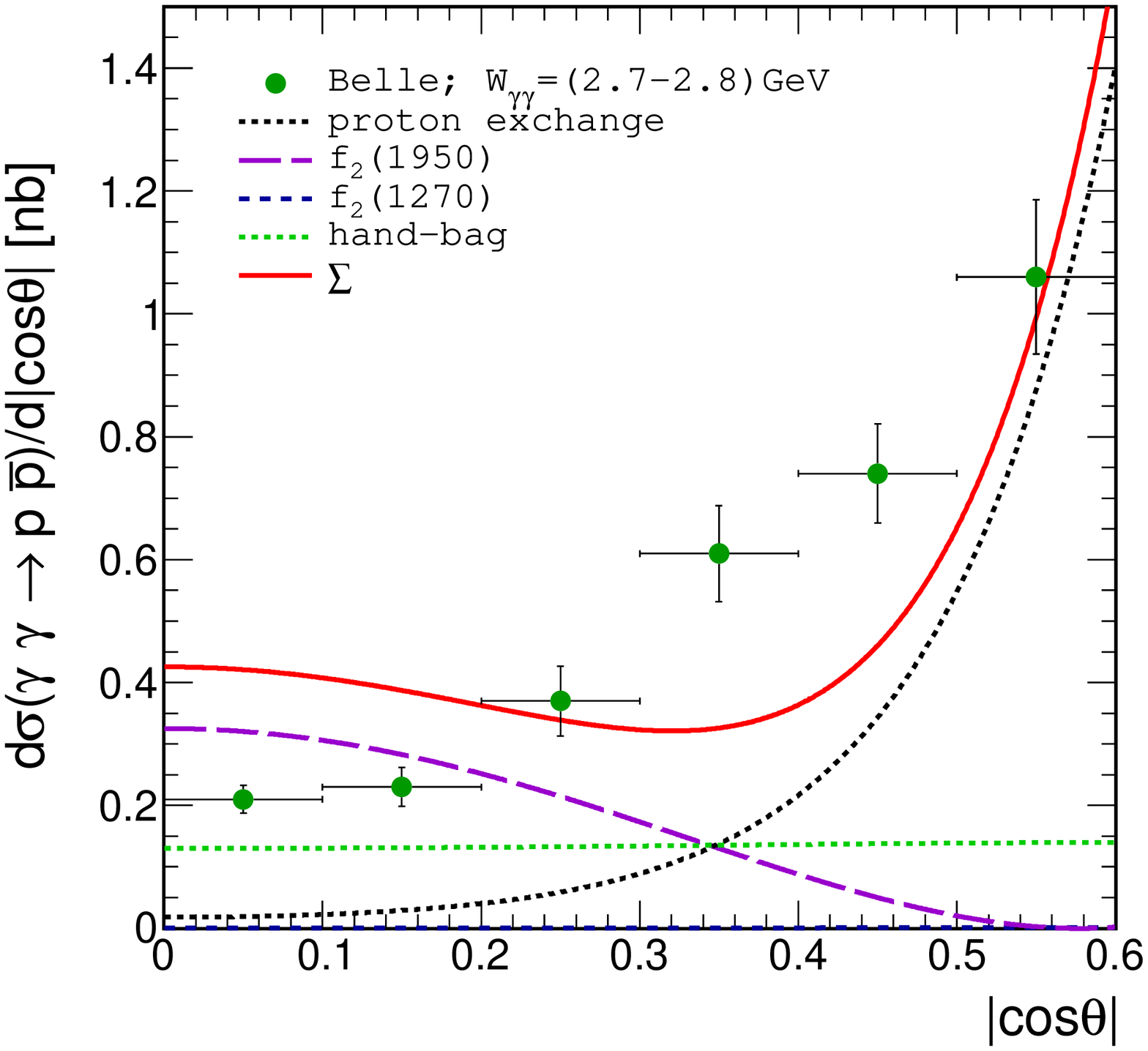}
(i)\includegraphics[width=0.3\textwidth]{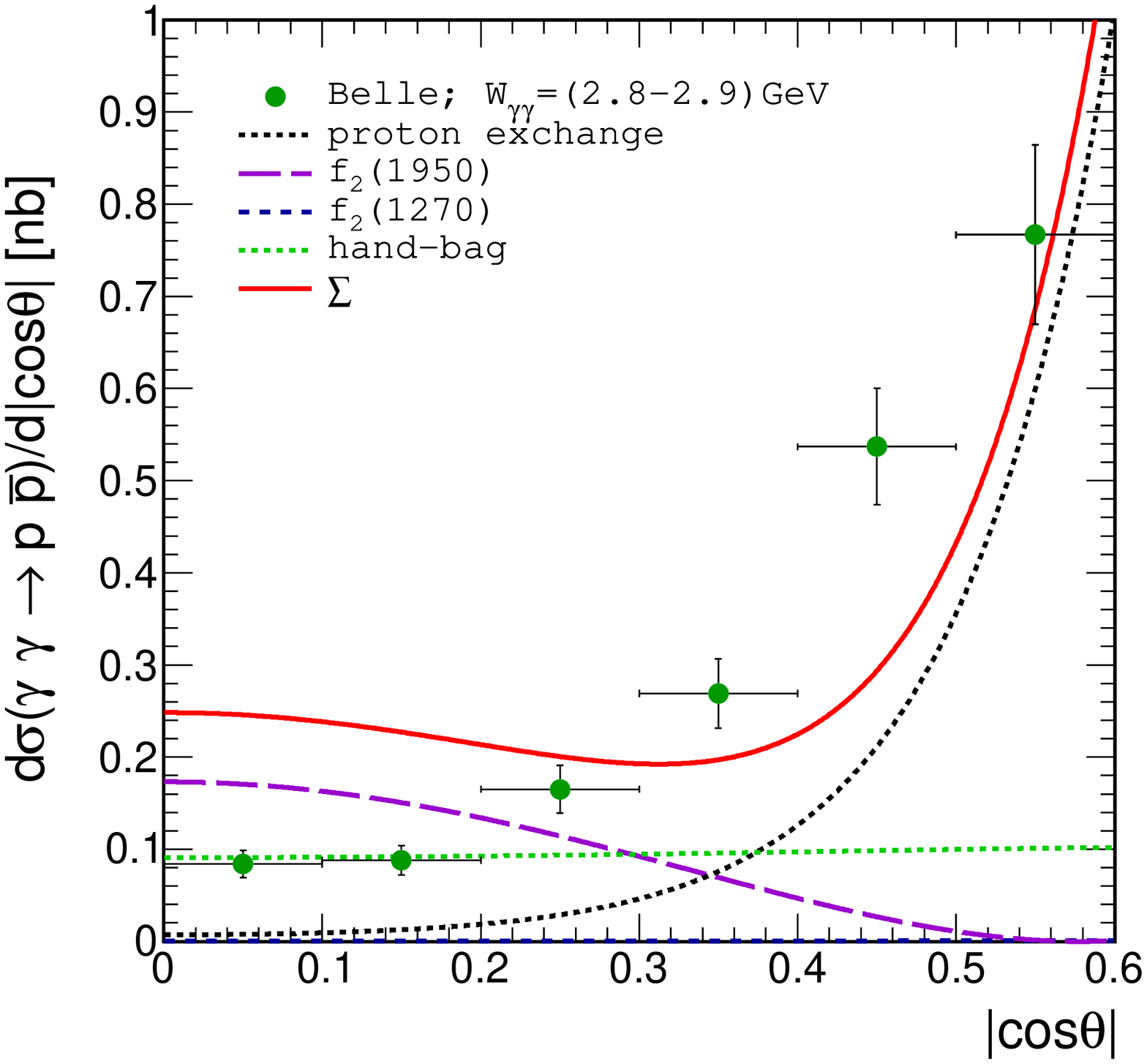}
(j)\includegraphics[width=0.3\textwidth]{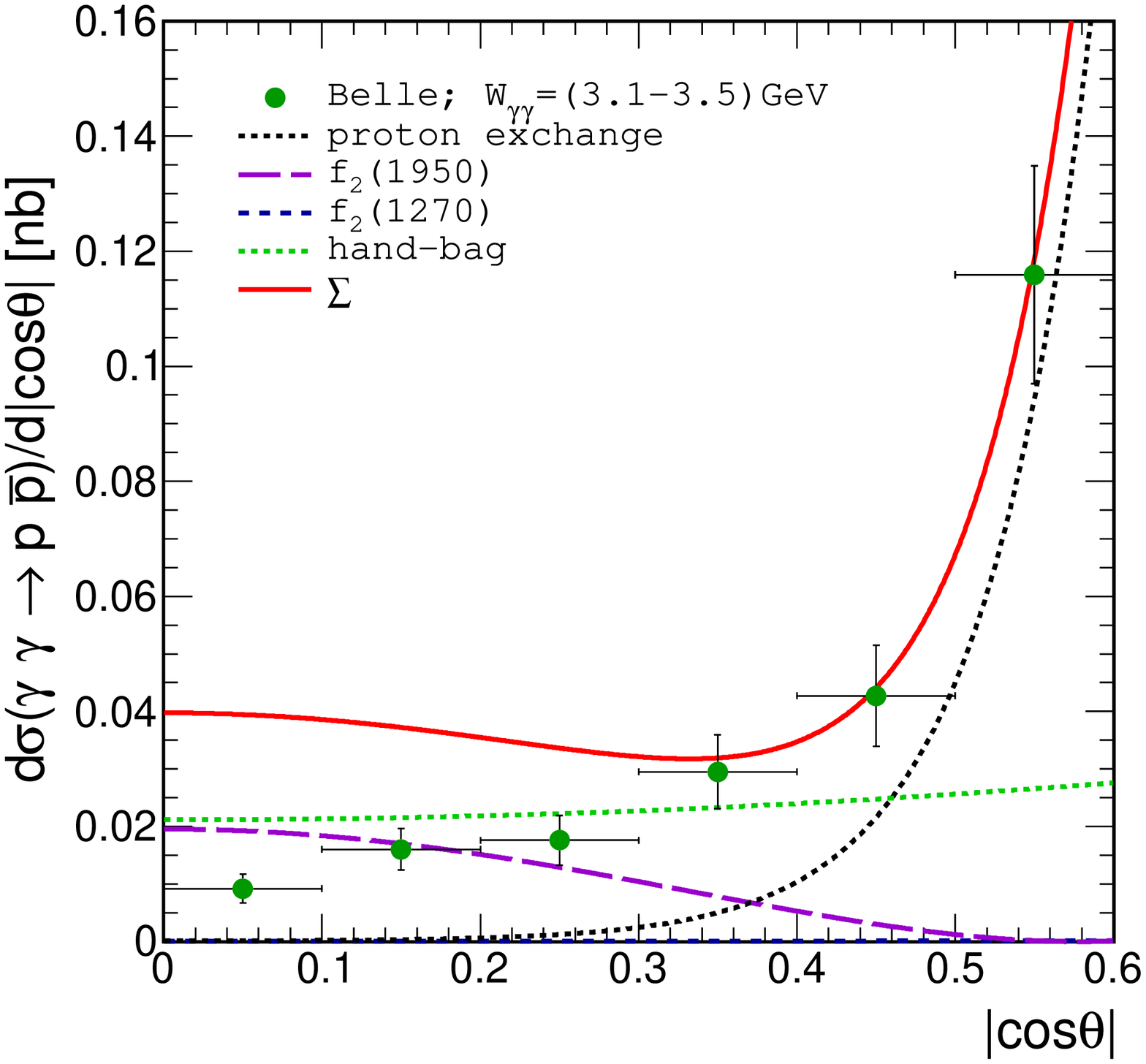}
(k)\includegraphics[width=0.3\textwidth]{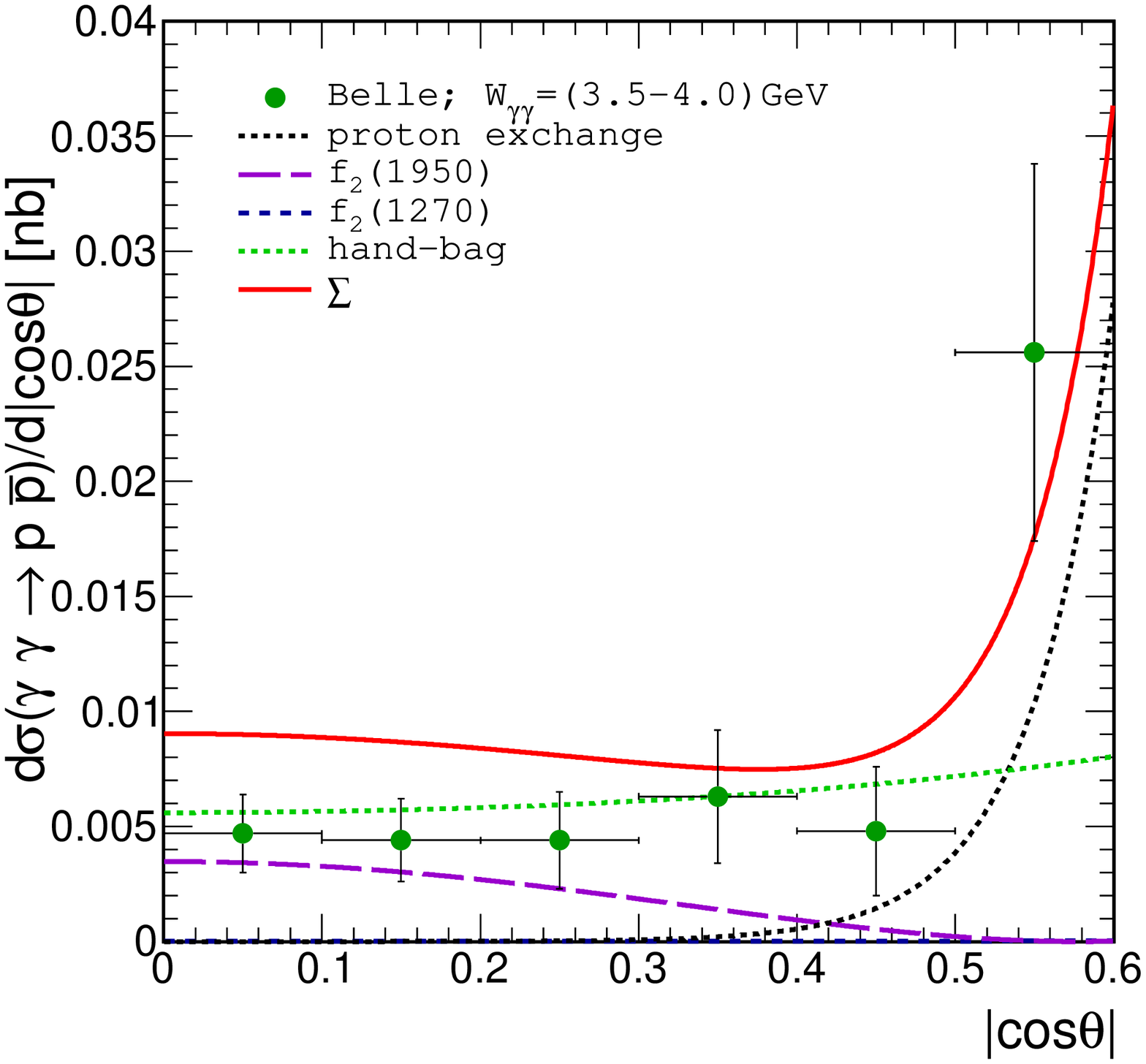}
  \caption{\label{fig:dsig_dz_3mech_setB}
  \small
The same as in Fig.~\ref{fig:dsig_dz_3mech} 
but here we used the parameter set~B from Table~\ref{table:parameters}.}
\end{figure}

In Figs.~\ref{fig:dsig_dz_3mech_nohb} and \ref{fig:dsig_dz_3mech},
we show our fits to the Belle angular distributions
\footnote{The cross section $d\sigma/d|z|$, $z=\cos\theta$, was calculated 
for the Belle angular range of $-0.6 < z < 0.6$, 
but plotted for $0 < z < 0.6$ after multiplication by a factor 2.}.
Here we use the same parametrization as in Fig.~\ref{fig:sigma_W_3mech} (a) 
(see set~A of Table~\ref{table:parameters}).
In Fig.~\ref{fig:dsig_dz_3mech_nohb} we present results
for the $f_{2}(1270)$, $f_{2}(1950)$ and proton-exchange contributions separately,
as well as their coherent sum.
At large angles, $\cos\theta \approx 0$,
the inclusion of the $f_{2}(1270)$ contribution
lowers the cross section compared to the case when 
only the $f_{2}(1950)$ and proton-exchange are taken into account.
In Fig.~\ref{fig:dsig_dz_3mech} we show results including the hand-bag contribution.
The $C_{A}$ parameter obtained from the fit is $C_{A} = 0.14$~GeV$^{2}$.
In Fig.~\ref{fig:dsig_dz_3mech_setB} 
we use, as in Fig.~\ref{fig:sigma_W_3mech} (c), the parameter set~B of Table~\ref{table:parameters}.
The $\tilde{C}_{A}$ parameter obtained from the fit is $\tilde{C}_{A} = 2.5$~GeV$^{4}$.
In Ref.~\cite{Diehl:2002yh} $\tilde{C}_{A}$ was estimated 
to be in the range $4.9 \div 8.0$~GeV$^{4}$
which is the same order of magnitude as we find.

Experimentally the angular distributions were
averaged over rather large intervals of (sub)process energies.
For a better comparison with the experimental data
we use the formula, with $z \equiv \cos\theta$,
\begin{equation}
\Big{\langle \frac{d \sigma}{dz}(W_{\gamma\gamma}) \Big\rangle}_{\Delta W_{\gamma\gamma}} = 
\frac{1}{\Delta W_{\gamma\gamma}}
\int_{W_{\gamma\gamma}-\frac{\Delta W_{\gamma\gamma}}{2}}^{W_{\gamma\gamma}+\frac{\Delta W_{\gamma\gamma}}{2}} \frac{d\sigma}{dz}(W_{\gamma\gamma}) dW_{\gamma\gamma} \,,
\label{averaging_dsigma_dz}
\end{equation}
instead of $\frac{d\sigma}{dz}(W_{\gamma\gamma} = \frac{W_{\gamma\gamma,min} + W_{\gamma\gamma,max}}{2})$.


In Fig.~\ref{fig:dsig_dz_W3to4} we compare the Belle data \cite{Kuo:2005nr} 
and the earlier OPAL and L3 data \cite{Abbiendi:2002bxa,Achard:2003jc} 
with our model results.
Due to the large error bars of the OPAL and L3 data
only the comparison of the model results with the Belle data 
gives significant information.
\begin{figure}[!ht]
\includegraphics[width=0.45\textwidth]{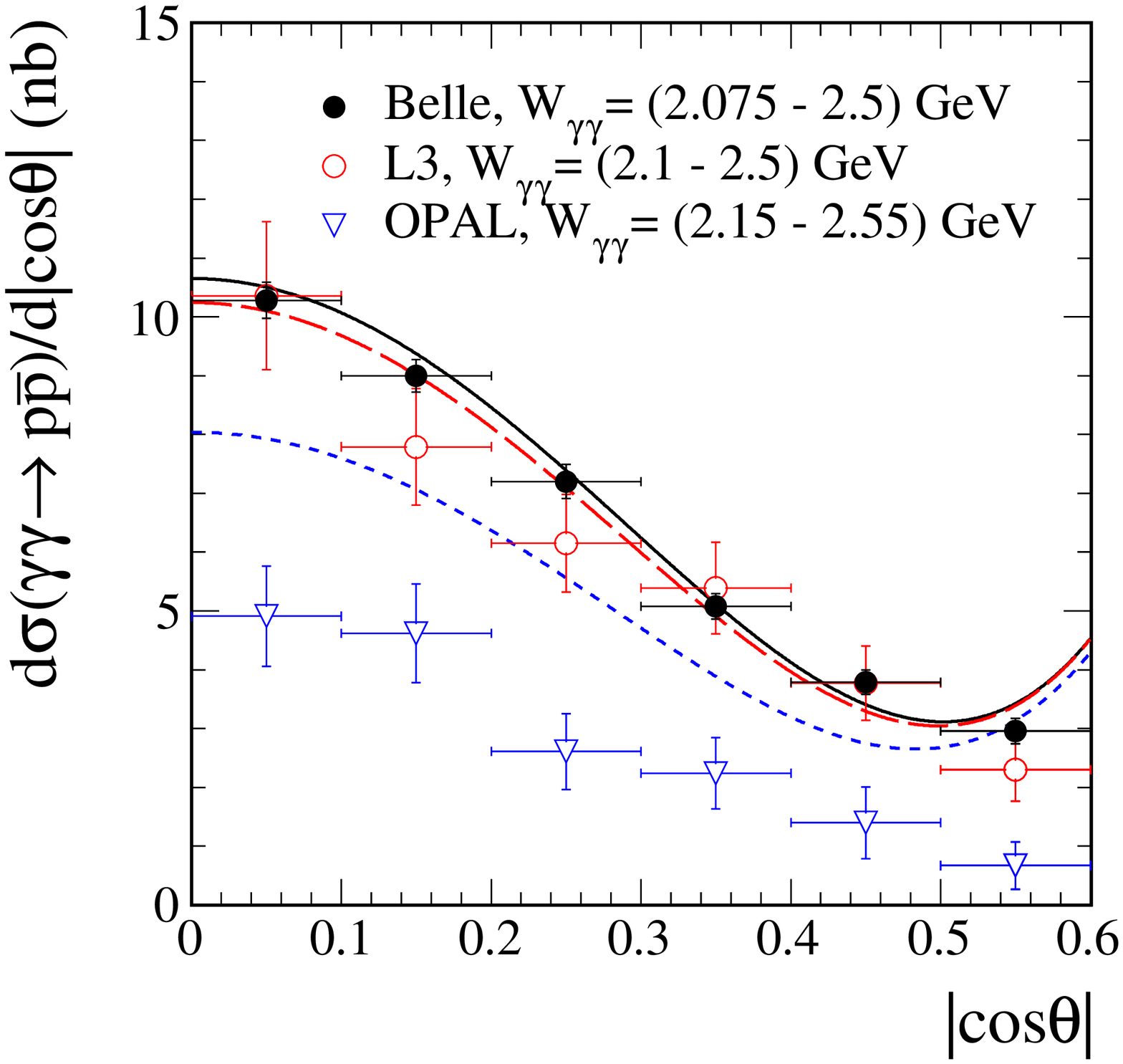}
\includegraphics[width=0.45\textwidth]{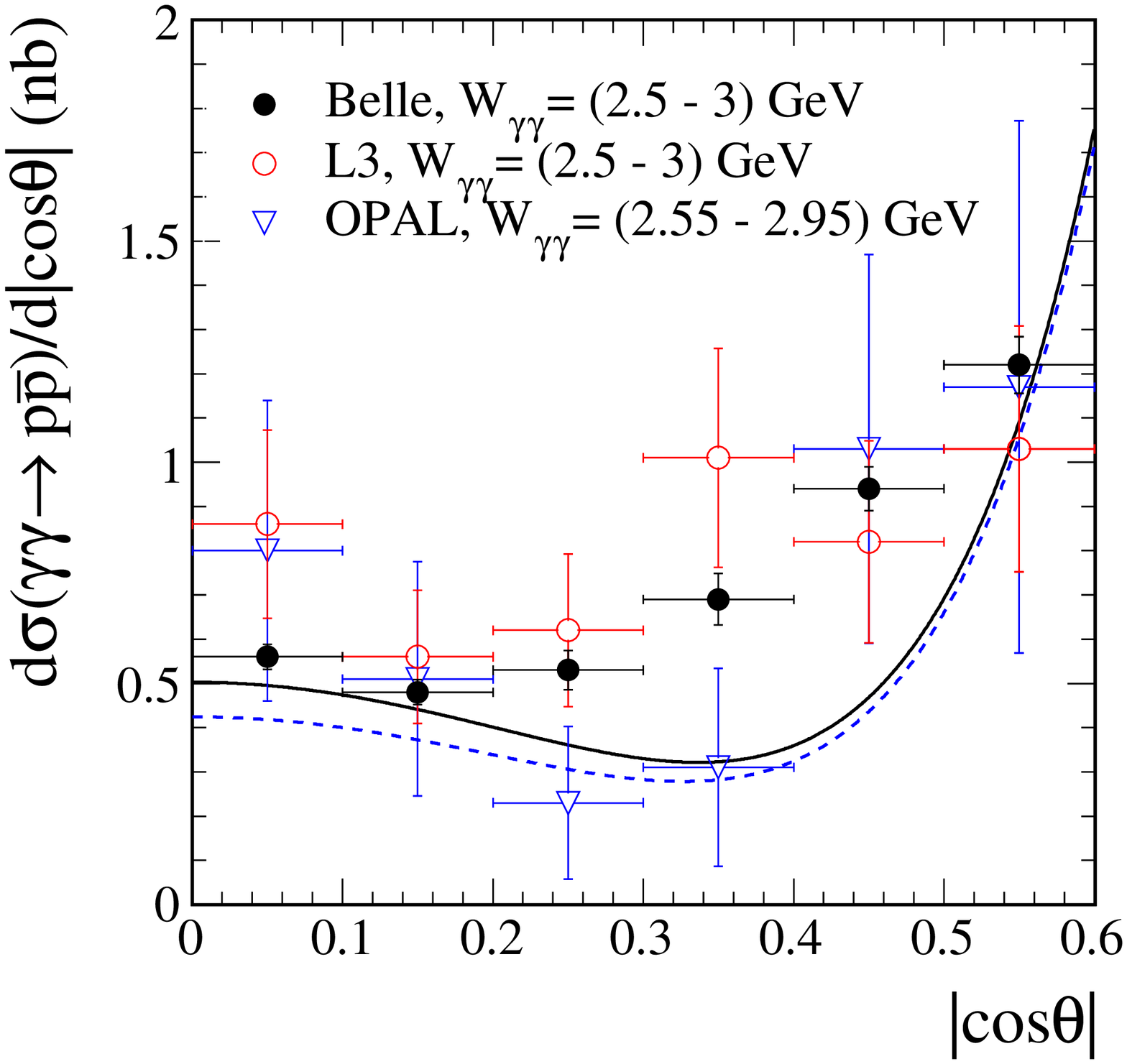}
\includegraphics[width=0.45\textwidth]{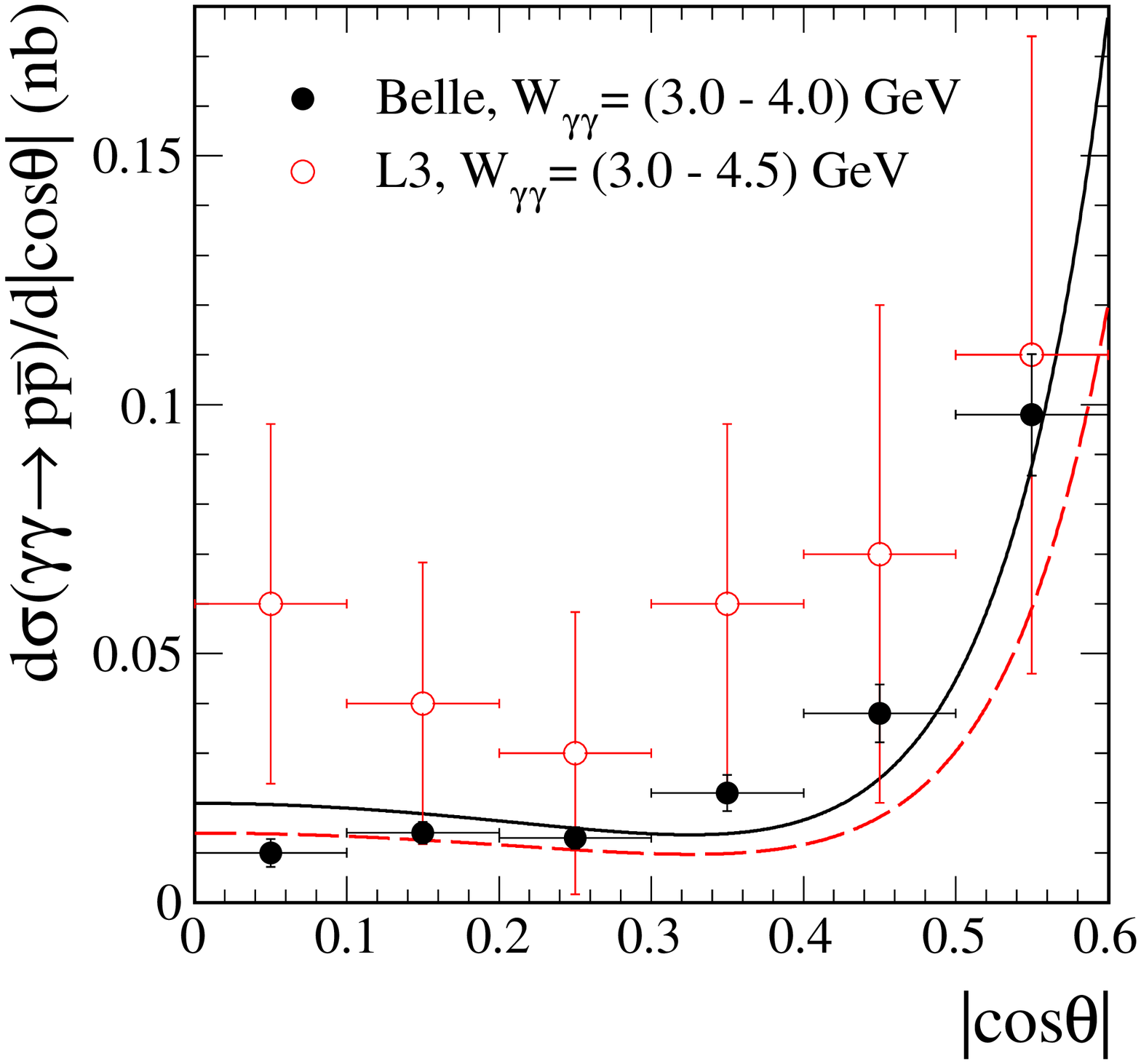}
  \caption{\label{fig:dsig_dz_W3to4}
  \small
Differential cross sections for the $\gamma \gamma \to p \bar{p}$ reaction
as a function of $|\cos\theta|$
for different $W_{\gamma \gamma}$ ranges.
We compare our total model results (including the hand-bag contribution)
with the Belle data \cite{Kuo:2005nr},
the L3 data \cite{Achard:2003jc},
and the OPAL data \cite{Abbiendi:2002bxa};
see the black solid line, the red long-dashed line, 
and the blue short-dashed line, respectively.
Here we used the parameter set~A from Table~\ref{table:parameters}.
}
\end{figure}

Heaving shown that the results of our approach,
including three mechanisms, describe the Belle experimental data reasonably well
we shall present our predictions for the nuclear reaction (\ref{nuclear_reaction})
in the next section.

\section{Predictions for the nuclear ultraperipheral collisions}

Having described the Belle angular distributions
we go to the predictions for the nuclear collisions.
In this section we show the integrated cross sections 
and several differential distributions 
for the nuclear process (\ref{nuclear_reaction}) 
calculated as described in Sec.~\ref{sec:nuclear_theory}
including three mechanisms 
discussed in Secs.~\ref{sec:gamgam_ppbar} and \ref{sec:results_ppbar}.
In the calculations below we used the parameter set~A 
from Table~\ref{table:parameters}.

\begin{figure}[!ht]
\includegraphics[width=0.45\textwidth]{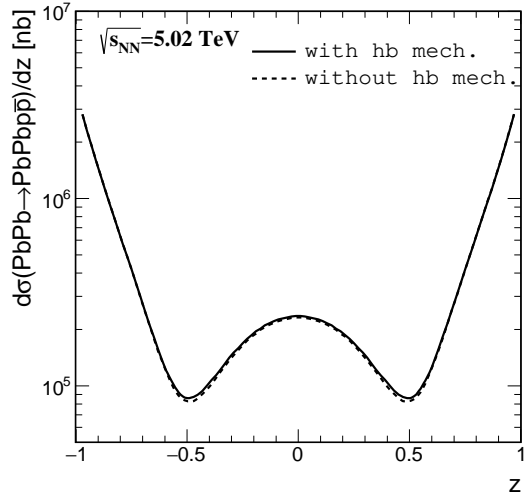}
  \caption{\label{fig:nuclear_dsig_dz}
  \small
The distribution in $z = \cos\theta$, integrating over
$2 m_{p} < W_{\gamma \gamma} < 4$~GeV,
for the $PbPb \to Pb Pb p \bar{p}$ reaction
at the $PbPb$ collision energy $\sqrt{s_{NN}} = 5.02$~TeV.
}
\end{figure}
\begin{figure}[!ht]
\includegraphics[width=0.45\textwidth]{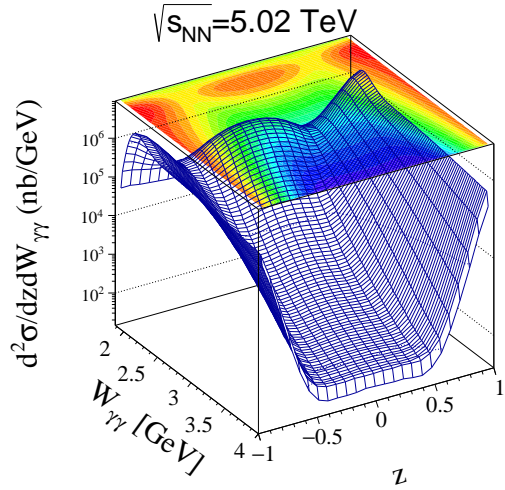}
  \caption{\label{fig:nuclear_dsig_dzdW}
  \small
Distribution in ($z,W_{\gamma\gamma}$)
for the $PbPb \to Pb Pb p \bar{p}$ reaction (\ref{nuclear_reaction}) 
at the LHC energy $\sqrt{s_{NN}}=5.02$ TeV.
}
\end{figure}

In Fig.~\ref{fig:nuclear_dsig_dz} we present
the angular distribution $d\sigma/dz$ 
($z = \cos\theta$ in the $\gamma \gamma$ c.m. system)
at the $PbPb$ collision energy $\sqrt{s_{NN}} = 5.02$~TeV.
Here we show the nuclear results when the hand-bag mechanism 
is included (solid line) and excluded (dotted line).
One can conclude that the hand-bag contribution does not play an important role
in the $p\bar{p}$ angular distribution.
We wish to emphasize that the enhancements at $z= \pm 1$ are the consequence
of our model presented in Sec.~\ref{sec:gamgam_ppbar}.
One can better visualize this behavior with the help of 
the two dimensional distribution $d^2 \sigma/dzdW_{\gamma\gamma}$.
From Fig.~\ref{fig:nuclear_dsig_dzdW} we clearly see that
the result for the nuclear reaction corresponds to 
that for elementary $\gamma\gamma \to p \bar{p}$ reaction
discussed in the previous section.
The $f_2(1950)$ contribution dominates at smaller $W_{\gamma\gamma}$ and
at $z \approx 0$ and $z \approx \pm 1$. 
This coincides with the result which was presented 
in Fig.~\ref{fig:dsig_dz_helicity_structure} (left panel, solid line).
In contrast to the resonant contribution, 
the proton-exchange one is concentrated mostly at larger invariant masses
and around $z = \pm 1$.

\begin{figure}[!ht]
(a)\includegraphics[width=0.45\textwidth]{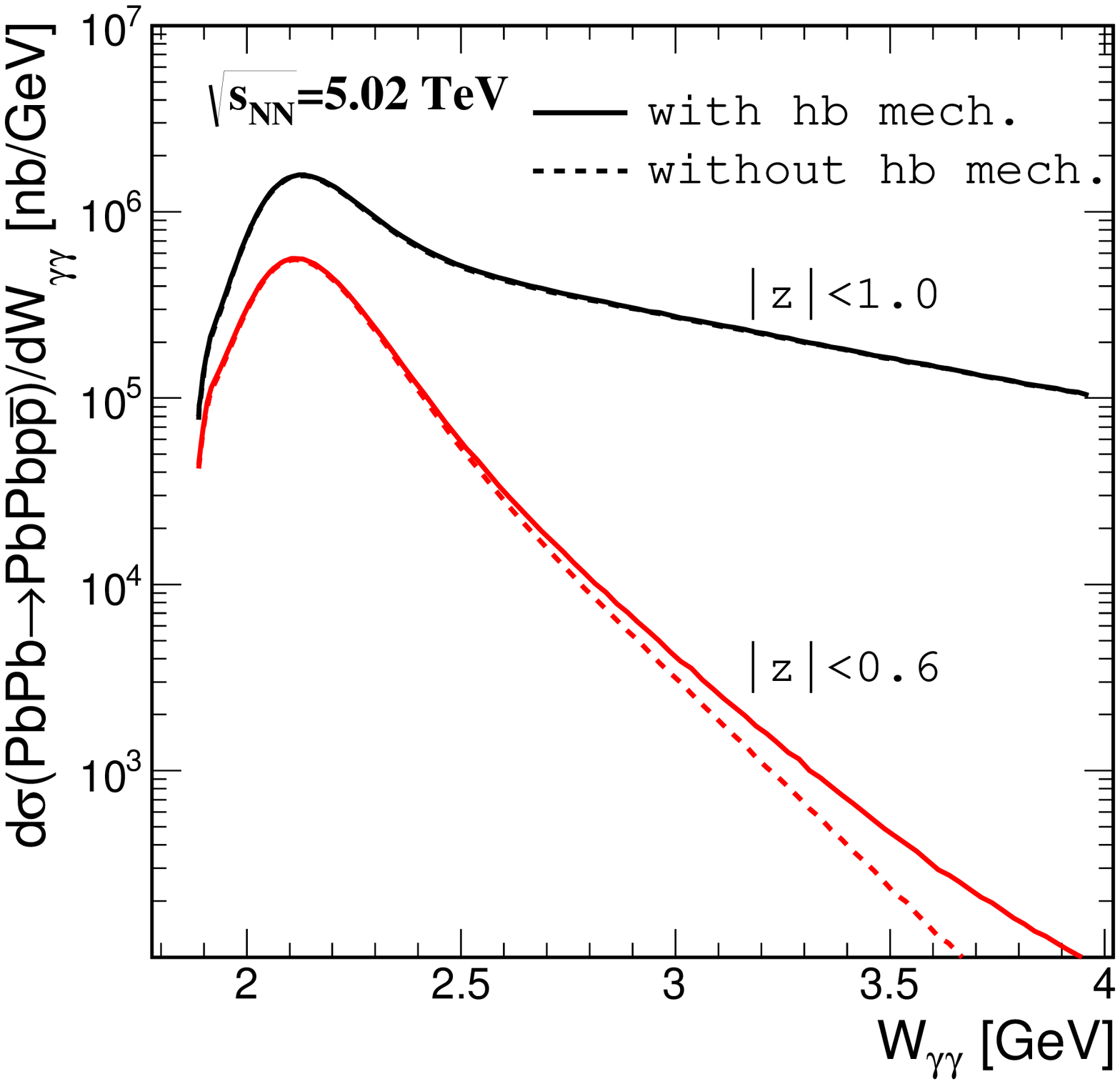}
(b)\includegraphics[width=0.45\textwidth]{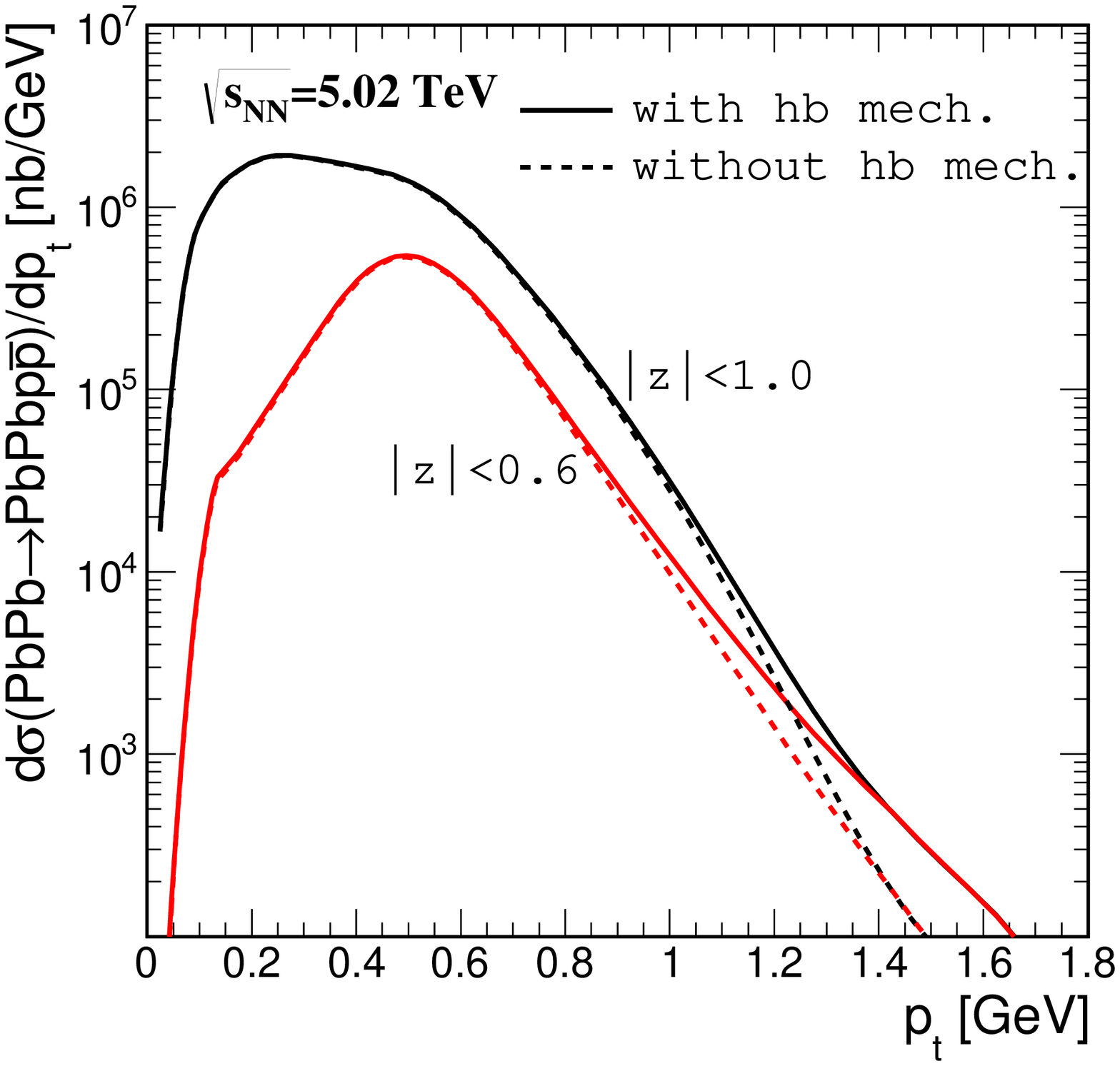}\\
(c)\includegraphics[width=0.45\textwidth]{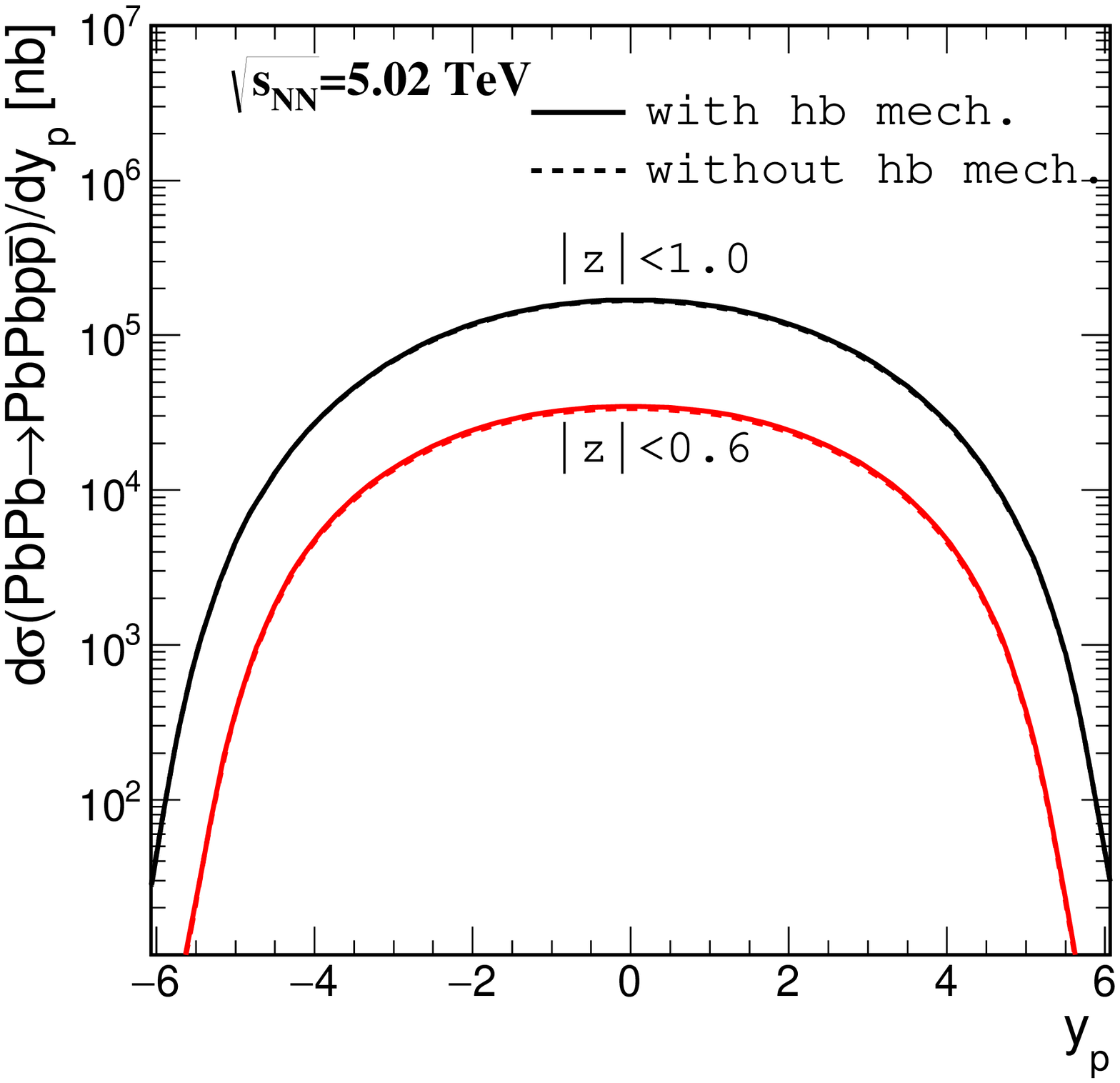}
(d)\includegraphics[width=0.45\textwidth]{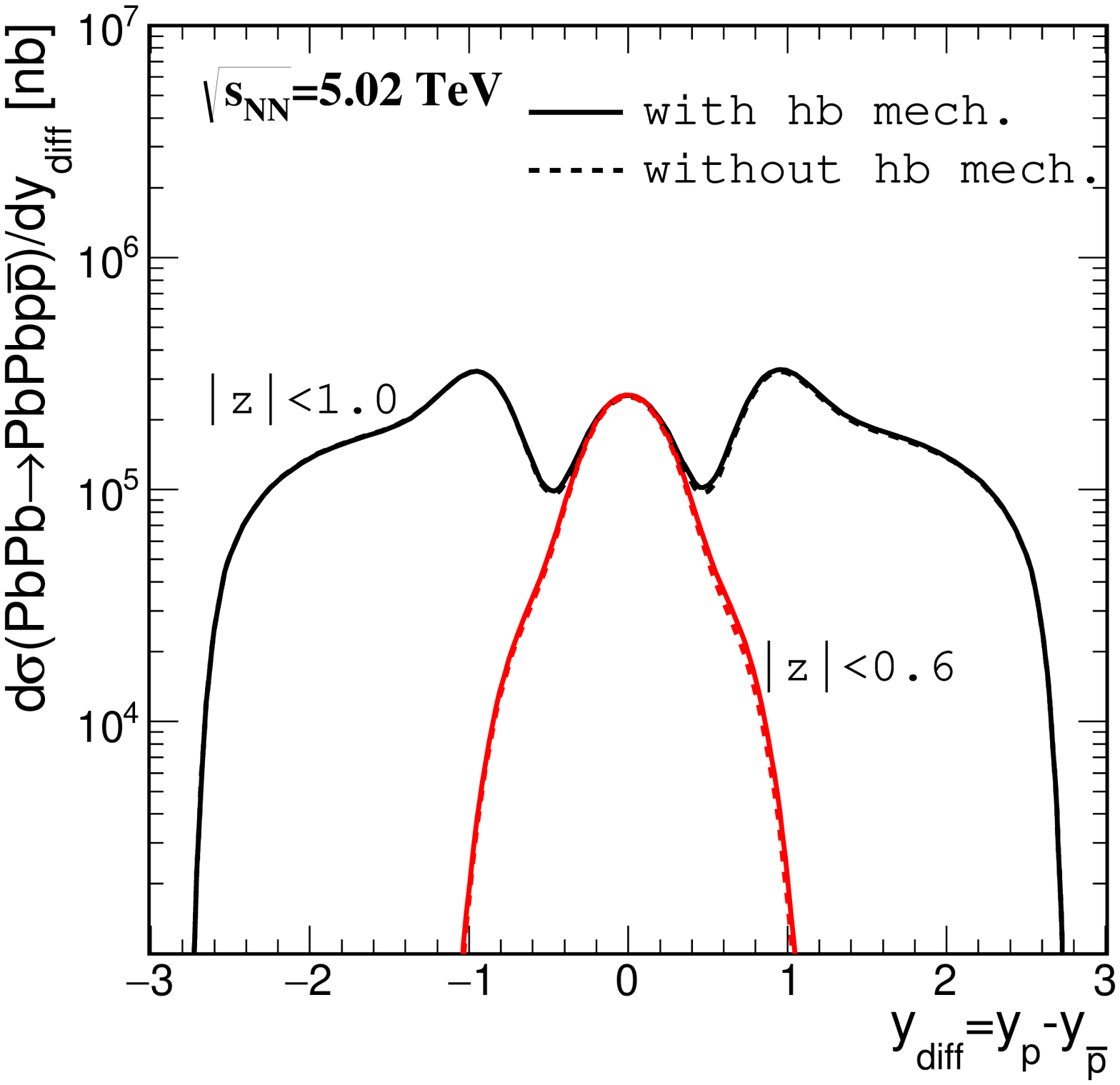}
  \caption{\label{fig:nuclear}
  \small
The differential nuclear cross sections 
for the $Pb Pb \to Pb Pb p \bar{p}$ reaction (\ref{nuclear_reaction})
at $\sqrt{s_{NN}}=5.02$ TeV.
Results for the full range of $z$ (the black lines)
and for $|z|<0.6$ (the red lines) are presented.
In panels (b) - (d) we integrate for $2 m_{p} \leqslant W_{\gamma \gamma} < 4$~GeV.
No other cuts have been imposed here.
}
\end{figure}
In Fig.~\ref{fig:nuclear} we present
the nuclear differential cross sections for two ranges of $z$:
the red lines are for $|z| < 0.6$, as in the Belle measurement,
the black lines are for $|z| \leqslant 1$~(full range).
Panel (a) shows the distribution in proton-antiproton invariant mass
($M_{p \bar{p}} \equiv W_{\gamma\gamma}$).
The $M_{p \bar{p}}$ distribution for the full $z$-range extends 
to much larger invariant masses
while for the Belle $z$-range it falls steeply down. 
Similar as for the elementary cross section (Fig.~\ref{fig:sigma_W_3mech}),
the hand-bag mechanism contributes significantly at $M_{p \bar{p}} > 3$~GeV.
Simultaneously, the difference between the results with (solid lines) 
and without (dotted lines) hand-bag contribution appears 
more pronounced for the case
when the angular phase space is narrowed.
In the present calculations we integrate 
for $2 m_{p} \leqslant W_{\gamma \gamma} < 4$~GeV.
The transverse momentum distributions of protons and antiprotons
shown in panel (b) are identical. Therefore we label them by $p_t$.
For large $p_t$ the distributions fall steeply.
The limitation on the phase space ($|z| < 0.6$) 
has a significant impact for smaller values of $p_t$
and has no influence for $p_t > 1.4$ GeV.
In the panel (c) we show distributions 
in rapidity of the proton or antiproton (which are identical).
Here we see only a difference in the normalization, 
and not in the shape for the two different ranges of $z$.
Finally, in the panel (d) we show the distribution in rapidity
distance between proton and antiproton $\mathrm{y}_{diff}=\mathrm{y}_p-\mathrm{y}_{\bar{p}}$. 
The larger the range of phase space
the broader is the distribution in $\mathrm{y}_{diff}$.
There are three maxima when no extra cuts are imposed.
The broad peak at $\mathrm{y}_{diff} \approx 0$ corresponds to the region $|z| < 0.6$.
It seems that observation of the broader $\mathrm{y}_{diff}$ distribution,
in particular identification of the outer maxima, could be a good test of our model.
As we see from Fig.~\ref{fig:nuclear_dsig_dz} the cross section
decreases quickly with $W_{\gamma \gamma} = M_{p \bar{p}}$
for $|z| < 0.6$, but stays large for $|z| \to 1$.
Thus, extending the integration to $W_{\gamma \gamma} > 4$~GeV
should not change the distributions of Fig.~\ref{fig:nuclear} (b) - (d)
for $|z| < 0.6$ but could have a sizeable influence on
those for $|z| \leqslant 1$.

\begin{figure}[!ht]
\includegraphics[width=0.45\textwidth]{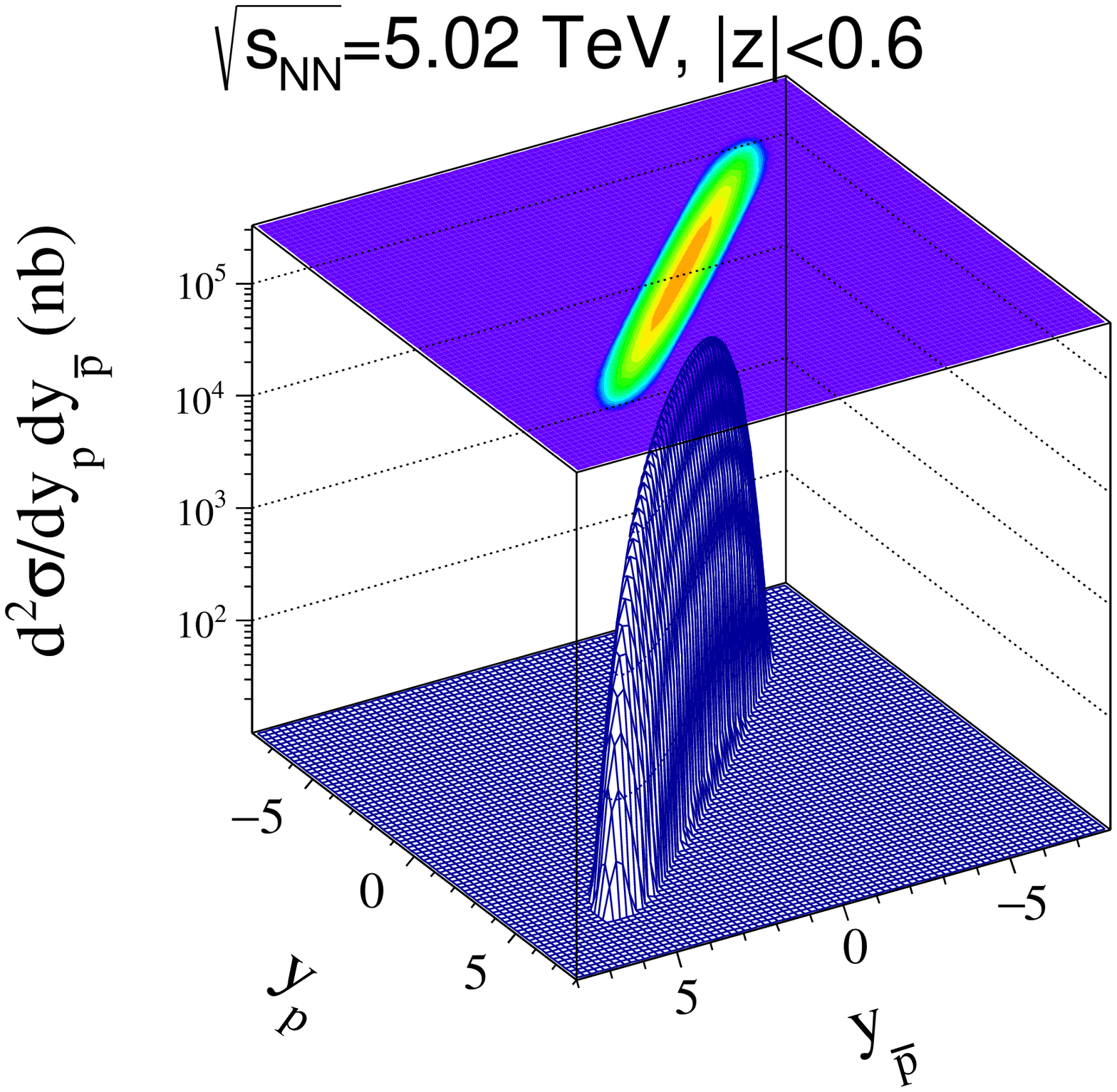}
\includegraphics[width=0.45\textwidth]{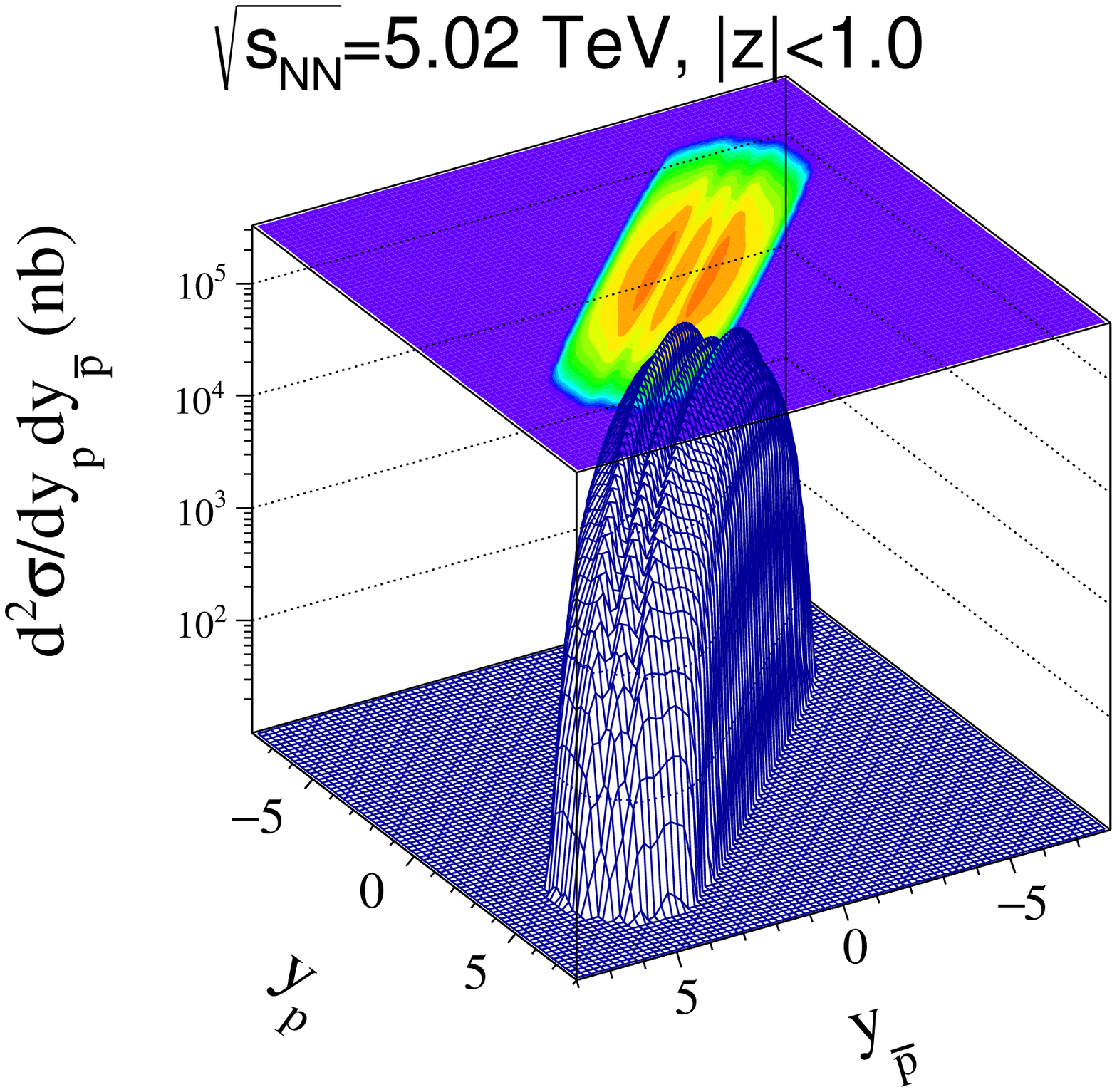}
  \caption{\label{fig:nuclear:y1y2}
  \small
The two-dimensional distributions in proton and antiproton rapidities
for the reaction (\ref{nuclear_reaction}) at $\sqrt{s_{NN}}=5.02$~TeV
for two different $z$-ranges of outgoing nucleons. 
The results include the hand-bag contribution.
The results are integrated for $2 m_{p} < W_{\gamma \gamma} < 4$~GeV.
}
\end{figure}
In Fig.~\ref{fig:nuclear:y1y2} we show
the two-dimensional distributions in (${\rm y}_{p}, {\rm y}_{\bar{p}}$) 
again for two ranges of $z$ (left panel relates to the Belle angle limitation 
and right panel is for full phase space). 
The cross section is concentrated along the diagonal ${\rm y}_p \simeq {\rm y}_{\bar{p}}$.

The ALICE Collaboration can measure $p \bar{p}$ in $Pb$-$Pb$ collisions
for $|{\rm y}| < 0.9$; 
see \cite{Kryshen:2017jfz} where the $J/\psi \to p \bar{p}$ decay was observed.
\footnote{We thank E.~L. Kryshen for some information on the recent ALICE measurement.}
We predict 46 events for $|{\rm y}| < 0.9$ and $p_t > 1$~GeV
for our $\gamma \gamma \to p \bar{p}$ contribution,
including three mechanisms, 
for ALICE integrated luminosity $L_{int} = 95$~$\mu$b$^{-1}$ \cite{Kryshen:2017jfz}. 
On the other hand the coherent $J/\psi$ photoproduction \cite{Klusek-Gawenda:2015hja}
in the $p \bar{p}$ channel gives 583 events
assuming approximately isotropic decay of $J/\psi \to p \bar{p}$.
This strongly suggests dominance of the coherent photoproduction mechanism of $J/\psi$
over the $\gamma \gamma$ contribution.
With such a transverse momentum cut 
as for the ALICE preliminary result 
a lot of the $\gamma \gamma \to p \bar{p}$ contribution 
is lost (with respect to the full phase space)
but considerably less of coherent $J/\psi \to p \bar{p}$ contribution,
where the maximum of the $p \bar{p}$ emission occurs
at $p_t = \frac{m_{J/\psi}}{2} \approx 1.5$~GeV
(sharp Jacobian peak associated with the fact that
transverse momentum of the coherent $J/\psi$ is very small).
Generally, the range covered by the ATLAS and CMS detectors 
for $p \bar{p}$ pairs in UPC is somewhat larger, $|{\rm y}|<2.5$.
The LHCb Collaboration can measure $p \bar{p}$ production in nuclear collisions
for $2<\eta<4.5$ and $p_{t}>0.2$~GeV.
\footnote{We thank R. McNulty and T. Shears for some information on the recent LHCb measurement.}

\begin{figure}[!ht]
\includegraphics[width=0.48\textwidth]{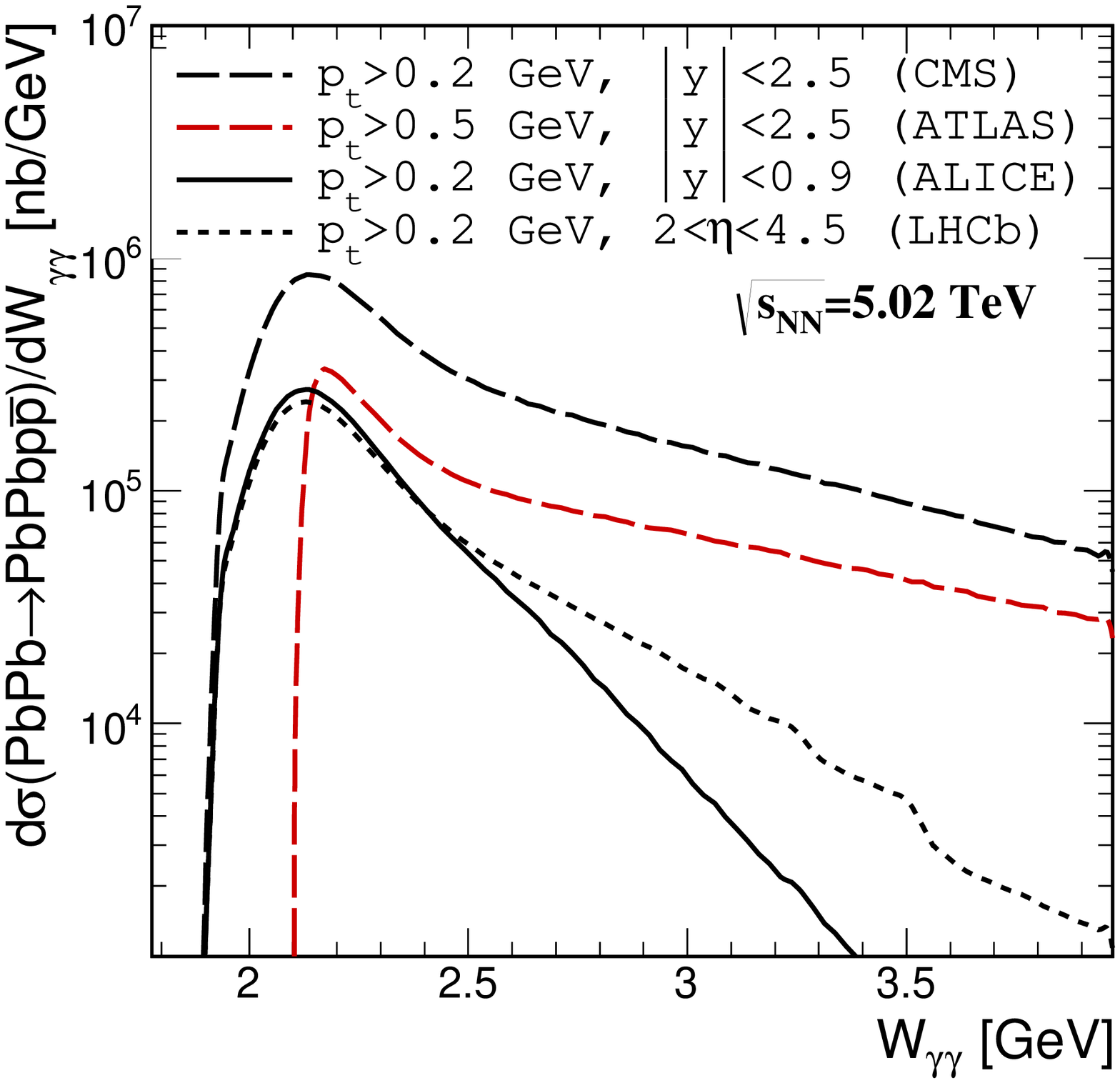}
\includegraphics[width=0.48\textwidth]{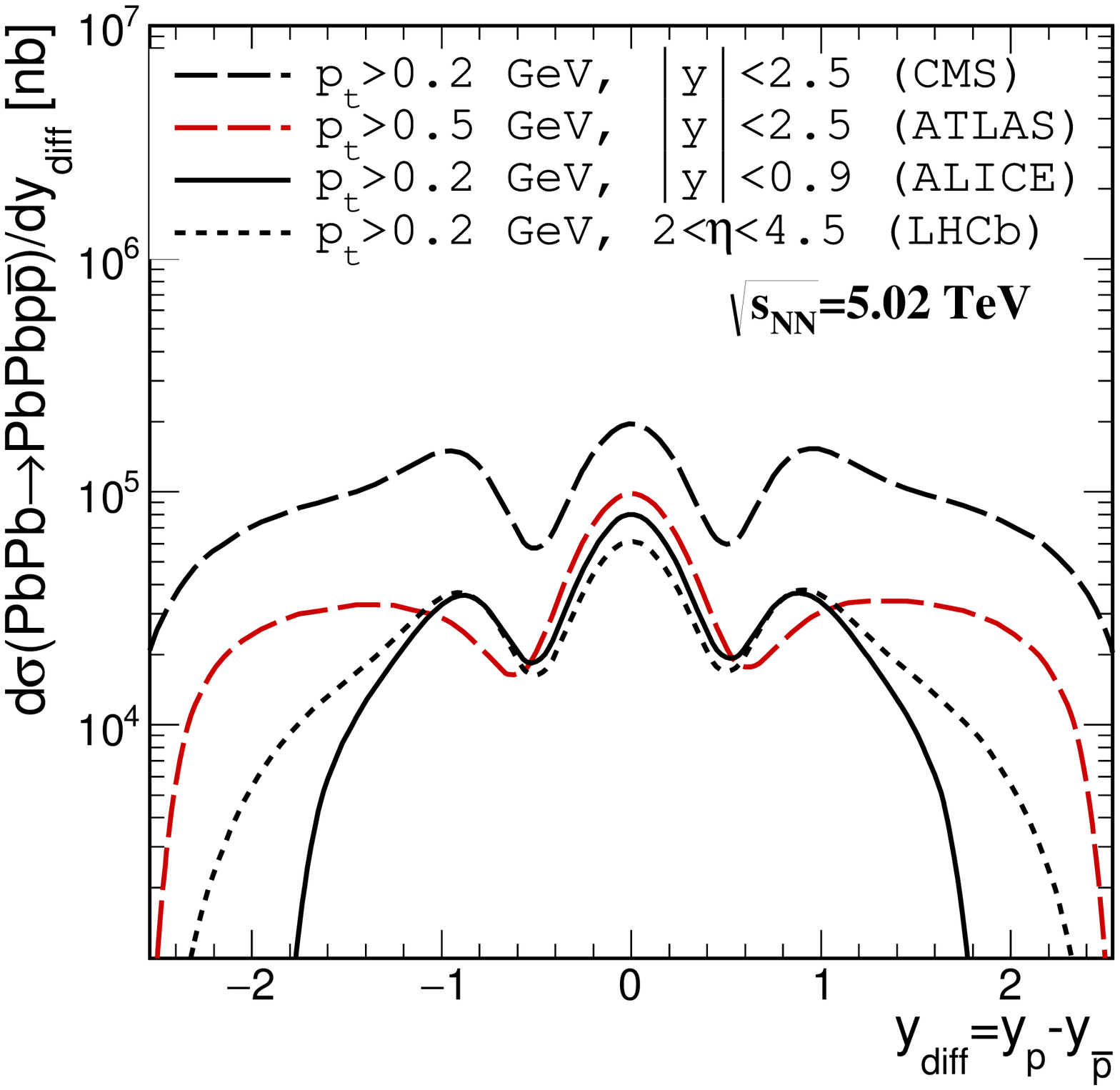}
  \caption{\label{fig:nuclear_ALICE}
  \small
The differential nuclear cross sections as a function of $p \bar{p}$ invariant mass
(the left panel) and ${\rm y}_{diff}={\rm y}_p-{\rm y}_{\bar{p}}$ (the right panel) 
for the $Pb Pb \to Pb Pb p \bar{p}$ reaction (\ref{nuclear_reaction}).
The results for different experimental cuts are presented.
}
\end{figure}
In Fig.~\ref{fig:nuclear_ALICE} we present distributions in 
$W_{\gamma \gamma} \equiv M_{p \bar{p}}$ (the left panel) and 
$\mathrm{y}_{diff} = \mathrm{y}_{p} - \mathrm{y}_{\bar{p}}$ (the right panel) 
imposing cuts on rapidities and transverse momenta of outgoing baryons.
From the left panel, we can observe that
the dependence on invariant mass of the $p \bar{p}$ pair 
is sensitive to the (pseudo)rapidity cut imposed.
Note that due to the cut on $p_t > 0.5$~GeV
the $W_{\gamma \gamma}$ distribution begins with a larger value of 2.1~GeV
(compare also with Fig.~\ref{fig:nuclear}~(a)).
The distribution in the difference of proton and antiproton rapidities is interesting.
Again (comparing with Fig.~\ref{fig:nuclear}~(d), $|z|<1.0$)
the $\mathrm{y}_{diff}$-distributions show three maxima.
The experimental cuts imposed on $p_{t}$ do not remove the external
maxima predicted by our model.
Such characteristic features can be checked by future experiments.

For completeness, we give the cross sections 
for the $Pb Pb \to Pb Pb \,p \bar{p}$
reaction for the $\gamma \gamma$ contribution
for various experimental cuts on proton and antiproton (pseudo)rapidities
and transverse momenta at $\sqrt{s_{NN}}=5.02$~TeV.
We find the cross section of 100~$\mu$b taking into account the ALICE cuts 
($|{\rm y}|<0.9$, $p_{t}>0.2$~GeV),
160~$\mu$b for the ATLAS cuts ($|{\rm y}|<2.5$, $p_{t}>0.5$~GeV),
500~$\mu$b for the CMS cuts ($|{\rm y}|<2.5$, $p_{t}>0.2$~GeV),
and 104~$\mu$b for the LHCb cuts ($2<\eta<4.5$, $p_{t}>0.2$~GeV).

\section{Conclusions}

We have discussed in detail the production of proton-antiproton pairs
in photon-photon collisions. Previous theoretical papers on the subject 
tried to pick up only one simple mechanism out of many in principle possible ones. 
In our work we have tried to incorporate the known mechanisms, 
such as proton exchange, $s$-channel resonance exchange and the hand-bag contribution.

In our calculation of the nonresonant proton exchange
we have included both Dirac- and Pauli-type couplings of the photon to the nucleon
and form factors for the exchanged off-shell protons.
We have found that the Pauli-type coupling is very important, 
enhances the cross section considerably, and cannot therefore be neglected.

We have shown that the Belle data \cite{Kuo:2005nr} for low photon-photon energies
can be nicely described by including in addition to the proton exchange 
the $s$-channel exchange of the $f_2(1950)$ resonance,
which was observed to decay into the $\gamma\gamma$ and $p \bar{p}$ channels \cite{Olive:2016xmw}. 
We include in the calculation also the $s$-channel $f_2(1270)$ meson exchange contribution.
These two tensor mesons were also needed to describe the Belle data 
for the $\gamma\gamma \to \pi^+\pi^-$ and $\gamma\gamma \to \pi^0\pi^0$
processes \cite{Uehara:2009cka, Klusek-Gawenda:2013rtu}.
Our simple model has a few parameters; see Table~\ref{table:parameters}.
Adjusting the parameters of the vertex form factors for the proton exchange,
of the tensor meson $s$-channel exchanges,
and of the form factor (\ref{corr_hb}) in the hand-bag contribution 
we have managed to describe both total cross section
and differential angular distributions of the Belle Collaboration 
with significantly better agreement with the data than in all previous trials.

Having described the Belle data we have used the
$\gamma \gamma \to p \bar p$ cross section to calculate 
the integrated cross section and differential distributions 
for production of $p \bar{p}$ pairs in 
ultraperipheral, ultrarelativistic, collisions (UPC) of heavy ions
at $\sqrt{s_{NN}} = 5.02$~TeV. 
We have presented distributions in rapidity 
and transverse momentum of protons and antiprotons, 
invariant mass of the $p \bar p$ system as well as
in the difference of rapidities for protons and antiprotons.
We have presented results for the full angular range of $z = \cos\theta$
as well as for the Belle range $|z|<0.6$.
The integrated cross section for the full phase space
is by a factor 5 larger than the one corresponding to the Belle angular coverage.
The larger the range of phase space
the broader is the distribution in $\mathrm{y}_{diff}$,
the rapidity difference between proton and antiproton.

We have also made predictions for $Pb$-$Pb$ collisions at 
$\sqrt{s_{NN}} = 5.02$~TeV and experimental cuts for
the ALICE, ATLAS, CMS, and LHCb experiments.
Corresponding total cross sections and differential distributions have been presented.
The UPC of heavy ions may provide 
new information compared to the presently available data from
$e^+ e^-$ collisions, in particular, if the structures
of the $\mathrm{y}_{diff}$ distributions shown in Figs.~\ref{fig:nuclear}
and \ref{fig:nuclear_ALICE} can be observed.
  

\acknowledgments
The authors are grateful to Markus Diehl for correspondence and
comments and to Carlo Ewerz for discussions.
This research was partially supported by
the Polish National Science Centre Grant No. DEC-2014/15/B/ST2/02528 (OPUS),
the MNiSW Grant No. IP2014 025173 (Iuventus Plus),
and by the Center for Innovation and Transfer of Natural Sciences 
and Engineering Knowledge in Rzesz\'ow.

\appendix

\section{Helicity states for protons and antiprotons and helicity amplitudes}
\label{section:Appendix1}
The general theory of helicity amplitudes
for collisions of particles with spin was developed in \cite{Jacob:1959at}.
To make our article selfcontained and to fix the phases
of our states we discuss in the following the construction of helicity states 
for protons and antiprotons as we found convenient for our purposes. 
These states are then used to determine 
the independent helicity amplitudes
for the reaction $\gamma \gamma \to p \bar{p}$ (\ref{2to2_reaction}).

\begin{figure}[!ht]
\includegraphics[width=0.42\textwidth]{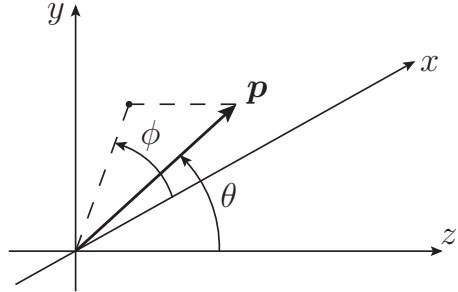}
\caption{\label{fig:coordinate_system}
\small
Coordinate system and momentum vector $\bp$.
}
\end{figure}

We consider protons and antiprotons in a fixed reference frame;
see Fig.~\ref{fig:coordinate_system}.
Let $\bp$ be the 3-momentum of the proton and
\begin{equation}
\begin{split}
& \bhp = \frac{\bp}{|\bp|}=
\left( \begin{array}{c}
\sin\theta \,\cos\phi  \\
\sin\theta \,\sin\phi  \\
\cos\theta  \\
\end{array} \right), \\
& 0 \leqslant \theta \leqslant \pi \,,\quad
  0 \leqslant \phi < 2\pi\,.
\end{split}
\label{B_01}
\end{equation}
We use throughout our paper a boldface notation for 3-vectors, $\bp$, $\bec$ etc.

For $\bhp \cdot \bec \neq -1$ ($\theta \neq \pi$)
we define the spinors of definite helicity of type $a$ as
\begin{equation}
\begin{split}
& u_{s}^{(h,a)}(p) =
\sqrt{p^{0} + m_{p}}
\left( \begin{array}{c}
\chi_{s}^{(a)}(\bhp)\\\\
2 s \dfrac{|\bp|}{p^{0} + m_{p}}\, \chi_{s}^{(a)}(\bhp) \\
\end{array} \right),\\
& s \in \{+1/2,-1/2 \}\,,
\end{split}
\label{B_02}
\end{equation}
where
\begin{equation}
\begin{split}
& \chi_{s}^{(a)}(\bhp) = \dfrac{1 + 2s \,(\bsigma \cdot \bhp)}
{\sqrt{2(1+\bhp \cdot \bec)}}\;
\chi_{s}^{(1)} \,,\\
& \chi_{1/2}^{(1)} = 
\left( \begin{array}{c}
1 \\
0 \\
\end{array} \right)\,, \quad
\chi_{-1/2}^{(1)} = 
\left( \begin{array}{c}
0 \\
1 \\
\end{array} \right)\,.
\end{split}
\label{B_03}
\end{equation}
This gives
\begin{equation}
\begin{split}
& \chi_{1/2}^{(a)}(\bhp) = \dfrac{1}{\sqrt{2}}
\left( \begin{array}{c}
\sqrt{1 + \hat{p}_{z}} \\\\
\dfrac{\hat{p}_{x} + i \hat{p}_{y}}{\sqrt{1 + \hat{p}_{z}}} \\
\end{array} \right) =
\left( \begin{array}{c}
\cos\frac{\theta}{2} \\\\
\sin\frac{\theta}{2}\; e^{i \phi} \\
\end{array} \right), \\
& \chi_{-1/2}^{(a)}(\bhp) = \dfrac{1}{\sqrt{2}}
\left( \begin{array}{c}
-\dfrac{\hat{p}_{x} - i \hat{p}_{y}}{\sqrt{1 + \hat{p}_{z}}} \\\\
\sqrt{1 + \hat{p}_{z}} \\
\end{array} \right) =
\left( \begin{array}{c}
-\sin\frac{\theta}{2} \;e^{-i \phi} \\\\
\cos\frac{\theta}{2} \\
\end{array} \right), 
\end{split}
\label{B_04}
\end{equation}
\begin{equation}
\begin{split}
\bar{u}_{r}^{(h,a)}(p)\,u_{s}^{(h,a)}(p) = 2 m_{p} \,\delta_{rs}\,.
\end{split}
\label{B_05}
\end{equation}

Let us denote the usual spinors with spin in $\pm z$ direction as
\begin{equation}
\begin{split}
& u_{r}(p) =
\sqrt{p^{0} + m_{p}}
\left( \begin{array}{c}
\chi_{r}^{(1)}\\\\
\dfrac{\bsigma \cdot \bp}{p^{0} + m_{p}}\, \chi_{r}^{(1)} \\
\end{array} \right),\\
& r \in \{+1/2,-1/2 \}\,;
\end{split}
\label{B_06}
\end{equation}
see for instance \cite{Nachtmann:1990ta}.

We get then
\begin{equation}
\begin{split}
&\left(\bar{u}_{r}(p)\,u_{s}^{(h,a)}(p)\right) = 2 m_{p} \left(B^{(a)}_{rs}(\bhp)\right)\,,\\
&B^{(a)}(\bhp) = \left(B^{(a)}_{rs}(\bhp)\right) = \frac{1}{\sqrt{2}}
\left( \begin{array}{cc}
\sqrt{1 + \hat{p}_{z}} & \;\, -\dfrac{\hat{p}_{x}-i\hat{p}_{y}}{\sqrt{1 + \hat{p}_{z}}} \\ 
\dfrac{\hat{p}_{x}+i\hat{p}_{y}}{\sqrt{1 + \hat{p}_{z}}} & \;\, \sqrt{1 + \hat{p}_{z}} \\
\end{array} \right)\\
&\qquad \qquad \qquad \qquad \quad=
\left( \begin{array}{cc}
\cos\frac{\theta}{2} & \;\, -\sin\frac{\theta}{2}\; e^{-i \phi}\\ 
\sin\frac{\theta}{2}\; e^{i \phi} & \;\, \cos\frac{\theta}{2} \\
\end{array} \right).
\end{split}
\label{B_07}
\end{equation}
Furthermore we define the creation operators for a proton
in the helicity state $s$ of type $a$ by
\begin{equation}
\begin{split}
a^{\dagger}_{h,a}(\bp,s) = a^{\dagger}_{r}(\bp) \,B^{(a)}_{rs}(\bhp)\,,
\end{split}
\label{B_08}
\end{equation}
where $a^{\dagger}_{r}(\bp)$ are the usual creation operators
corresponding to the spinors (\ref{B_06}).
We have then
\begin{equation}
\begin{split}
&u_{s}^{(h,a)}(p)\, \bar{u}_{s}^{(h,a)}(p) = \slash{p} + m_{p}\,,\\
&a^{\dagger}_{h,a}(\bp,s)\, \bar{u}_{s}^{(h,a)}(p) = 
 a^{\dagger}_{r}(\bp)\, \bar{u}_{r}(p)\,.
\end{split}
\label{B_09}
\end{equation}

For $\bhp \cdot \bec \neq 1$ ($\theta \neq 0$)
we can define helicity spinors of type $b$ as follows
\begin{equation}
\begin{split}
& u_{s}^{(h,b)}(p) =
\sqrt{p^{0} + m_{p}}
\left( \begin{array}{c}
\chi_{s}^{(b)}(\bhp)\\\\
2 s \dfrac{|\bp|}{p^{0} + m_{p}}\, \chi_{s}^{(b)}(\bhp) \\
\end{array} \right),\\
& \chi_{s}^{(b)}(\bhp) = \dfrac{1 + 2s \,(\bsigma \cdot \bhp)}
{\sqrt{2(1-\bhp \cdot \bec)}}\;
\chi_{-s}^{(1)}\,,\\
& s \in \{+1/2,-1/2 \}\,.
\end{split}
\label{B_10}
\end{equation}
This gives
\begin{equation}
\begin{split}
& \chi_{1/2}^{(b)}(\bhp) = \dfrac{1}{\sqrt{2}}
\left( \begin{array}{c}
\dfrac{\hat{p}_{x} - i \hat{p}_{y}}{\sqrt{1 - \hat{p}_{z}}} \\\\
\sqrt{1 - \hat{p}_{z}} \\
\end{array} \right) =
\left( \begin{array}{c}
\cos\frac{\theta}{2}\; e^{-i \phi} \\\\
\sin\frac{\theta}{2} \\
\end{array} \right), \\
& \chi_{-1/2}^{(b)}(\bhp) = \dfrac{1}{\sqrt{2}}
\left( \begin{array}{c}
\sqrt{1 - \hat{p}_{z}} \\\\
-\dfrac{\hat{p}_{x} + i \hat{p}_{y}}{\sqrt{1 - \hat{p}_{z}}} \\
\end{array} \right) =
\left( \begin{array}{c}
\sin\frac{\theta}{2} \\\\
-\cos\frac{\theta}{2} \;e^{i \phi} \\
\end{array} \right).
\end{split}
\label{B_11}
\end{equation}
Comparing with (\ref{B_04}) we find for $0 < \theta < \pi$
\begin{equation}
\begin{split}
& \chi_{1/2}^{(b)}(\bhp) =   e^{-i \phi}\, \chi_{1/2}^{(a)}(\bhp)\,,\\
& \chi_{-1/2}^{(b)}(\bhp) = -e^{ i \phi}\, \chi_{-1/2}^{(a)}(\bhp)\,.
\end{split}
\label{B_12}
\end{equation}

With $u_{r}(p)$ from (\ref{B_06}) we find
\begin{equation}
\begin{split}
&\left(\bar{u}_{r}(p)\,u_{s}^{(h,b)}(p)\right) = 2 m_{p} \left( B^{(b)}_{rs}(\bhp) \right) \,,\\
&B^{(b)}(\bhp) = \left( B^{(b)}_{rs}(\bhp) \right) = \frac{1}{\sqrt{2}}
\left( \begin{array}{cc}
\dfrac{\hat{p}_{x}-i\hat{p}_{y}}{\sqrt{1 - \hat{p}_{z}}} 
& \;\, \sqrt{1 - \hat{p}_{z}} \\ 
\sqrt{1 - \hat{p}_{z}} 
& \;\, -\dfrac{\hat{p}_{x}+i \hat{p}_{y}}{\sqrt{1 - \hat{p}_{z}}} \\
\end{array} \right)\\
&\qquad \qquad \qquad \qquad \quad=
\left( \begin{array}{cc}
\cos\frac{\theta}{2}\; e^{-i \phi} & \;\, \sin\frac{\theta}{2}\\ 
\sin\frac{\theta}{2}              & \;\, -\cos\frac{\theta}{2}\; e^{i \phi} \\
\end{array} \right).
\end{split}
\label{B_13}
\end{equation}
Defining creation operators analogous to (\ref{B_08}) 
\begin{equation}
\begin{split}
a^{\dagger}_{h,b}(\bp,s) = a^{\dagger}_{r}(\bp) \,B^{(b)}_{rs}(\bhp)\,,
\end{split}
\label{B_14}
\end{equation}
we get
\begin{equation}
\begin{split}
&u_{s}^{(h,b)}(p)\, \bar{u}_{s}^{(h,b)}(p) = \slash{p} + m_{p}\,,\\
&a^{\dagger}_{h,b}(\bp,s)\, \bar{u}_{s}^{(h,b)}(p) = 
 a^{\dagger}_{r}(\bp)\, \bar{u}_{r}(p)\,.
\end{split}
\label{B_15}
\end{equation}

Now we go to antiprotons. For this we use the charge-conjugation matrix
\begin{equation}
\begin{split}
& S(C) = i \gamma^{2} \gamma^{0} = -i \gamma_{2} \gamma_{0} =
\left( \begin{array}{cc}
0 & \;\, -\varepsilon \\
-\varepsilon & \;\, 0 \\
\end{array} \right),\\
& \varepsilon = 
\left( \begin{array}{cc}
0  & \;\, 1 \\
-1 & \;\, 0 \\
\end{array} \right);
\end{split}
\label{B_16}
\end{equation}
see for instance chapter~4 of \cite{Nachtmann:1990ta}.
We have
\begin{equation}
\begin{split}
& S(C)=\bar{S}(C)=-S^{-1}(C)=-S(C)^{\dagger}=-S(C)^{T}\,,\\
& S^{-1}(C) \gamma^{\mu} S(C) = -\gamma^{\mu\,T}\,.
\end{split}
\label{B_17}
\end{equation}
We define the antiproton spinors as
\begin{equation}
\begin{split}
\bar{v}_{r}(p)=u_{r}^{T}(p)\, S(C) = 
-\sqrt{p^{0}+m_{p}} 
\left( -\chi_{r}^{(1)T} \varepsilon \frac{\bsigma \cdot \bp}{p^{0}+m_{p}}\;,\;
\chi_{r}^{(1)T} \varepsilon \right)
\,,
\end{split}
\label{B_18}
\end{equation}
\begin{equation}
\begin{split}
\bar{v}_{s}^{(h,a)}(p)=u_{s}^{T\,(h,a)}(p)\, S(C) = 
-\sqrt{p^{0}+m_{p}} 
\left( \chi_{s}^{(a)T}(\bhp) \,\varepsilon\, 2s\,\frac{|\bp|}{p^{0}+m_{p}}\;,\;
\chi_{s}^{(a)T}(\bhp) \,\varepsilon \right)
\,,
\label{B_19}
\end{split}
\end{equation}
\begin{equation}
\begin{split}
\bar{v}_{s}^{(h,b)}(p)=u_{s}^{T\,(h,b)}(p)\, S(C) = 
-\sqrt{p^{0}+m_{p}} 
\left( \chi_{s}^{(b)T}(\bhp) \,\varepsilon\, 2s\,\frac{|\bp|}{p^{0}+m_{p}}\;,\;
\chi_{s}^{(b)T}(\bhp) \,\varepsilon \right)
\,;
\label{B_20}
\end{split}
\end{equation}
see (\ref{B_06}), (\ref{B_02}), and (\ref{B_10}).

The creation operators for antiprotons are in the standard basis
\begin{equation}
\begin{split}
b^{\dagger}_{r}(\bp) = U(C) \, a^{\dagger}_{r}(\bp) \, U^{-1}(C)\,,
\end{split}
\label{B_21}
\end{equation}
where $U(C)$ is the charge-conjugation operator.
Analogously we define the creation operators for antiprotons of definite helicity
\begin{equation}
\begin{split}
b^{\dagger}_{h,a}(\bp,s) = U(C) \, a^{\dagger}_{h,a}(\bp,s) \, U^{-1}(C) =
b^{\dagger}_{r}(\bp) \, B_{rs}^{(a)}(\bhp)\,,
\label{B_22}
\end{split}
\end{equation}
\begin{equation}
\begin{split}
b^{\dagger}_{h,b}(\bp,s) = U(C) \, a^{\dagger}_{h,b}(\bp,s) \, U^{-1}(C) =
b^{\dagger}_{r}(\bp) \, B_{rs}^{(b)}(\bhp)\,,
\label{B_23}\\
\end{split}
\end{equation}
where we used (\ref{B_08}) and (\ref{B_14}).

With this we get
\begin{equation}
\begin{split}
&v_{s}^{(h,a)}(p) \, \bar{v}_{s}^{(h,a)}(p) = \slash{p} - m_{p}\,,\\
&v_{s}^{(h,b)}(p) \, \bar{v}_{s}^{(h,b)}(p) = \slash{p} - m_{p}\,,
\end{split}
\label{B_24}
\end{equation}
\begin{equation}
\begin{split}
&v_{s}^{(h,a)}(p) \, b^{\dagger}_{h,a}(\bp,s) = v_{r}(p) \, b^{\dagger}_{r}(\bp)\,,\\
&v_{s}^{(h,b)}(p) \, b^{\dagger}_{h,b}(\bp,s) = v_{r}(p) \, b^{\dagger}_{r}(\bp)\,.
\end{split}
\label{B_25}
\end{equation}

\begin{figure}[!ht]
\includegraphics[width=0.61\textwidth]{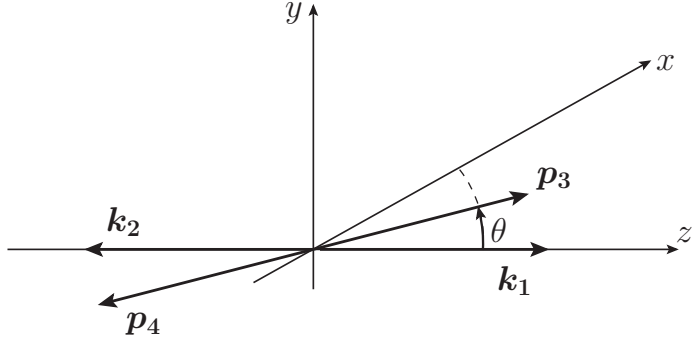}
\caption{\label{fig:cm_system}
\small
The reaction $\gamma \gamma \to p \bar{p}$ in the c.m. system.
}
\end{figure}

Now we come to the reaction (\ref{2to2_reaction}).
We consider (\ref{2to2_reaction}) in the c.m. system
with the $x$-$z$ plane giving the reaction plane; see Fig.~\ref{fig:cm_system}.
The usual kinematic variables are given by~(\ref{2to2_kinematic_app}).
Let $\bea$, $\beb$, $\bec$ be the Cartesian unit vectors
in the reference system of Fig.~\ref{fig:cm_system}. 
Then $k_{1}^{0} = k_{2}^{0} = p_{3}^{0} = p_{4}^{0} = \frac{1}{2}\sqrt{s}$
and the momenta of the particles are
\begin{eqnarray}
&& \bka = - \bkb = 
|\bka|\, \bec \,,\nonumber\\
&& \bpp = - \bppbar = 
|\bpp|\left(\sin\theta \,\bea + \cos\theta \,\bec\right) \,,\nonumber\\
&&|\bka| = \frac{1}{2}\sqrt{s}\,,\nonumber\\
&&|\bpp| = \frac{1}{2}\sqrt{s - 4 m_{p}^{2}} \,.
\label{B_29}
\end{eqnarray}
As polarization vectors for the incoming photons of definite helicity
we choose
\begin{equation}
\begin{split}
&\bepsilona^{(\pm)} = \mp \frac{1}{\sqrt{2}}\left(\bea \pm i\,\beb\right)\,,\\
&\bepsilonb^{(\pm)} = \mp \frac{1}{\sqrt{2}}\left(-\bea \pm i\,\beb\right)\,.
\end{split}
\label{B_31}
\end{equation}
The corresponding photon creation operators are
\begin{equation}
\begin{split}
&a^{\dagger}(\bkj,m) = \bepsilonj^{(m)} \,\ba^{\dagger}(\bkj)\,,\\
&j=1,2\,, \quad m = \pm 1\,.
\end{split}
\label{B_32}
\end{equation}

For the proton we choose the helicity basis $a$, for the antiproton
the basis $b$. From (\ref{B_08}) and (\ref{B_23})
we have for the corresponding creation operators
\begin{equation}
\begin{split}
a^{\dagger}_{h,a}(\bpp,s) = a^{\dagger}_{r}(\bpp) \,B^{(a)}_{rs}(\bhpp)\,,
\end{split}
\label{B_33}
\end{equation}
\begin{equation}
\begin{split}
b^{\dagger}_{h,b}(\bppbar,s) = b^{\dagger}_{r}(\bppbar) \,B^{(b)}_{rs}(\bhppbar)\,.
\end{split}
\label{B_34}
\end{equation}
Note that in calculating $B^{(a)}_{rs}(\bhpp)$ from (\ref{B_07})
we have to make the replacements
$\theta \to \theta$, $\phi \to 0$.
Calculating $B^{(b)}_{rs}(\bhppbar)$ from (\ref{B_13})
we have to make the replacements
$\theta \to \pi-\theta$, $\phi \to \pi$.

The symmetries of the reaction (\ref{2to2_reaction}) are the following.
The parity ($P$) transformation followed by a rotation by $\pi$
around the positive $y$-axis:
\begin{equation}
\begin{split}
U_{2}(\pi) U(P)\,.
\end{split}
\label{B_35}
\end{equation}
The charge-conjugation ($C$) transformation followed by a rotation
by $\pi$ around the positive $y$-axis:
\begin{equation}
\begin{split}
U_{2}(\pi) U(C)\,.
\end{split}
\label{B_36}
\end{equation}
From the transformation laws of the standard creation operators (see e.g. \cite{Nachtmann:1990ta})
and from the relations (see (\ref{B_07}), (\ref{B_13}))
\begin{equation}
\begin{split}
&\varepsilon^{T} \,B^{(a)}(\bhpp) \,\varepsilon = B^{(a)}(\bhpp)\,,\\
&\varepsilon^{T} \,B^{(b)}(\bhppbar) \,\varepsilon =-B^{(b)}(\bhppbar)\,,
\end{split}
\label{B_37}
\end{equation}
\begin{equation}
\begin{split}
&\varepsilon \,B^{(a)}(\bhpp)=-B^{(b)}(\bhppbar)\,\sigma_{3}\,,\\
&\varepsilon \,B^{(b)}(\bhppbar)= B^{(a)}(\bhpp)\,\sigma_{3}\,,
\end{split}
\label{B_38}
\end{equation}
we get the transformation laws for the helicity creation operators shown in Table~\ref{tab:transformation}.
\begin{table}[!ht]
\begin{tabular}{|c|c|c|}
\hline
$A^{\dagger}$
& $U_{2}(\pi) U(P) A^{\dagger} U^{-1}(P) U_{2}^{-1}(\pi)$ 
& $U_{2}(\pi) U(C) A^{\dagger} U^{-1}(C) U_{2}^{-1}(\pi)$ \\
\hline
$a^{\dagger}_{s}(\bpp)$& 
$-a^{\dagger}_{r}(\bpp) \,\varepsilon_{rs}$& 
$-b^{\dagger}_{r}(\bppbar) \,\varepsilon_{rs}$\\
$b^{\dagger}_{s}(\bppbar)$& 
$ b^{\dagger}_{r}(\bppbar) \,\varepsilon_{rs}$& 
$-a^{\dagger}_{r}(\bpp) \,\varepsilon_{rs}$\\
$a^{\dagger}_{h,a}(\bpp,s)$& 
$-a^{\dagger}_{h,a}(\bpp,r)\, \varepsilon_{rs}$& 
$ b^{\dagger}_{h,b}(\bppbar,r)\, (\sigma_{3})_{rs}$\\
$b^{\dagger}_{h,b}(\bppbar,s)$& 
$-b^{\dagger}_{h,b}(\bppbar,r)\, \varepsilon_{rs}$& 
$-a^{\dagger}_{h,a}(\bpp,r)\, (\sigma_{3})_{rs}$\\
$a^{\dagger}(\bka,m)$& 
$-a^{\dagger}(\bka,-m)$& 
$-a^{\dagger}(\bkb,m)$\\
$a^{\dagger}(\bkb,m)$& 
$-a^{\dagger}(\bkb,-m)$& 
$-a^{\dagger}(\bka,m)$\\
\hline
\end{tabular}
\caption{Transformation properties of creation operators for protons, 
antiprotons, and photons under the transformations (\ref{B_35}) and (\ref{B_36}).}
\label{tab:transformation}
\end{table}

We define now the helicity states for the reaction (\ref{2to2_reaction})
using (\ref{B_32}), (\ref{B_33}) and (\ref{B_34}) as
\begin{eqnarray}
&&| \gamma(\bka,m_{1}), \gamma(\bkb,m_{2})\rangle =
a^{\dagger}(\bka,m_{1}) a^{\dagger}(\bkb,m_{2}) |0\rangle \,,\nonumber\\
&& m_{1}, m_{2} \in \{+1,-1 \}\,,
\label{B_39}\nonumber\\
&&| p(\bpp,s_{3}), \bar{p}(\bppbar,s_{4})\rangle =
a^{\dagger}_{h,a}(\bpp,s_{3}) b^{\dagger}_{h,b}(\bppbar,s_{4}) |0\rangle \,,\nonumber\\
&& s_{3}, s_{4} \in \{+1/2,-1/2 \}\,.
\label{B_40}
\end{eqnarray}
The transformation laws of these states are shown in Table~\ref{tab:transformation_2}.
\begin{table}[!ht]
\begin{tabular}{|c|c|c|}
\hline
$|\;\rangle$
& $U_{2}(\pi) U(P) |\;\rangle$ 
& $U_{2}(\pi) U(C) |\;\rangle$ \\
\hline
$| p(\bpp,s_{3}), \bar{p}(\bppbar,s_{4})\rangle$& 
$| p(\bpp,r_{3}), \bar{p}(\bppbar,r_{4})\rangle\,
\varepsilon_{r_{3}s_{3}} \varepsilon_{r_{4}s_{4}}$& 
$| p(\bpp,r_{3}), \bar{p}(\bppbar,r_{4})\rangle\,
(\sigma_{3})_{r_{3}s_{4}} (\sigma_{3})_{r_{4}s_{3}}$ \\
$| \gamma(\bka,m_{1}), \gamma(\bkb,m_{2})\rangle$& 
$| \gamma(\bka,-m_{1}), \gamma(\bkb,-m_{2})\rangle$& 
$| \gamma(\bka,m_{2}), \gamma(\bkb,m_{1})\rangle$ \\
\hline
\end{tabular}
\caption{Transformation laws of the states
(\ref{B_39}) under the transformations (\ref{B_35}) and (\ref{B_36}).}
\label{tab:transformation_2}
\end{table}

Finally we come to the helicity amplitudes for the reaction (\ref{2to2_reaction})
\begin{eqnarray}
&&\langle 
p(\bpp,s_{3}), \bar{p}(\bppbar,s_{4})
|{\cal T}| 
\gamma(\bka,m_{1}), \gamma(\bkb,m_{2})\rangle \equiv
\langle 2s_{3},2s_{4}|{\cal T}| m_{1},m_{2} \rangle \,,\nonumber\\
&& 2s_{3},2s_{4},m_{1},m_{2} \in \{+1,-1 \}\,,
\label{B_41}
\end{eqnarray}
where we use the convenient shorthand notation of (\ref{hel_amp}).
There are 16 helicity amplitudes.
The symmetry $U_{2}(\pi) U(P)$ (\ref{B_35}) gives the relation,
using Table~\ref{tab:transformation_2},
\begin{eqnarray}
\langle 2s_{3},2s_{4}|{\cal T}| m_{1},m_{2} \rangle =
\langle 2r_{3},2r_{4}|{\cal T}| -m_{1},-m_{2} \rangle
\varepsilon_{r_{3}s_{3}} \varepsilon_{r_{4}s_{4}}\,.
\label{B_42}
\end{eqnarray}
From the symmetry of $U_{2}(\pi) U(C)$ (\ref{B_36}) we get
\begin{eqnarray}
\langle 2s_{3},2s_{4}|{\cal T}| m_{1},m_{2} \rangle =
\langle 2r_{3},2r_{4}|{\cal T}| m_{2},m_{1} \rangle
(\sigma_{3})_{r_{3}s_{4}} (\sigma_{3})_{r_{4}s_{3}}\,.
\label{B_43}
\end{eqnarray}

The relations (\ref{B_42}) and (\ref{B_43})
are written explicitly for the helicity amplitudes in Table~\ref{tab:helicity_amp}.
From this we find that there are only 6 independent helicity amplitudes
for (\ref{2to2_reaction}) which we choose as follows:
\begin{eqnarray}
&&\psi_{1}(s,t) = \langle ++ |{\cal T}| ++ \rangle \,,\nonumber \\
&&\psi_{2}(s,t) = \langle ++ |{\cal T}| -- \rangle \,,\nonumber \\
&&\psi_{3}(s,t) = \langle +- |{\cal T}| +- \rangle \,,\nonumber \\
&&\psi_{4}(s,t) = \langle +- |{\cal T}| -+ \rangle \,,\nonumber \\
&&\psi_{5}(s,t) = \langle ++ |{\cal T}| +- \rangle \,,\nonumber \\
&&\psi_{6}(s,t) = \langle +- |{\cal T}| ++ \rangle \,.
\label{B_44}
\end{eqnarray}
%

\begin{table}[!ht]
\begin{tabular}{|c|r|r|r|}
\hline
         & $U_{2}(\pi) U(P)$ & $U_{2}(\pi) U(C)$ &\\
\hline
$\langle ++|{\cal T}|++ \rangle$&$\langle --|{\cal T}|-- \rangle$&$\langle ++|{\cal T}|++ \rangle$&$\psi_{1}$ \\
$\langle +-|{\cal T}|++ \rangle$&$-\langle -+|{\cal T}|-- \rangle$&$-\langle -+|{\cal T}|++ \rangle$&$\psi_{6}$ \\
$\langle -+|{\cal T}|++ \rangle$&$-\langle +-|{\cal T}|-- \rangle$&$-\langle +-|{\cal T}|++ \rangle$&$-\psi_{6}$ \\
$\langle --|{\cal T}|++ \rangle$&$\langle ++|{\cal T}|-- \rangle$&$\langle --|{\cal T}|++ \rangle$&$\psi_{2}$ \\
$\langle ++|{\cal T}|+- \rangle$&$\langle --|{\cal T}|-+ \rangle$&$\langle ++|{\cal T}|-+ \rangle$&$\psi_{5}$ \\
$\langle +-|{\cal T}|+- \rangle$&$-\langle -+|{\cal T}|-+ \rangle$&$-\langle -+|{\cal T}|-+ \rangle$&$\psi_{3}$ \\
$\langle -+|{\cal T}|+- \rangle$&$-\langle +-|{\cal T}|-+ \rangle$&$-\langle +-|{\cal T}|-+ \rangle$&$-\psi_{4}$ \\
$\langle --|{\cal T}|+- \rangle$&$\langle ++|{\cal T}|-+ \rangle$&$\langle --|{\cal T}|-+ \rangle$&$\psi_{5}$ \\
$\langle ++|{\cal T}|-+ \rangle$&$\langle --|{\cal T}|+- \rangle$&$\langle ++|{\cal T}|+- \rangle$&$\psi_{5}$ \\
$\langle +-|{\cal T}|-+ \rangle$&$-\langle -+|{\cal T}|+- \rangle$&$-\langle -+|{\cal T}|+- \rangle$&$\psi_{4}$ \\
$\langle -+|{\cal T}|-+ \rangle$&$-\langle +-|{\cal T}|+- \rangle$&$-\langle +-|{\cal T}|+- \rangle$&$-\psi_{3}$ \\
$\langle --|{\cal T}|-+ \rangle$&$\langle ++|{\cal T}|+- \rangle$&$\langle --|{\cal T}|+- \rangle$&$\psi_{5}$ \\
$\langle ++|{\cal T}|-- \rangle$&$\langle --|{\cal T}|++ \rangle$&$\langle ++|{\cal T}|-- \rangle$&$\psi_{2}$ \\
$\langle +-|{\cal T}|-- \rangle$&$-\langle -+|{\cal T}|++ \rangle$&$-\langle -+|{\cal T}|-- \rangle$&$\psi_{6}$ \\
$\langle -+|{\cal T}|-- \rangle$&$-\langle +-|{\cal T}|++ \rangle$&$-\langle +-|{\cal T}|-- \rangle$&$-\psi_{6}$ \\
$\langle --|{\cal T}|-- \rangle$&$\langle ++|{\cal T}|++ \rangle$&$\langle --|{\cal T}|-- \rangle$&$\psi_{1}$ \\
\hline
\end{tabular}
\caption{Helicity amplitudes for $\gamma \gamma \to p \bar{p}$ (\ref{2to2_reaction})
and their symmetry relations.}
\label{tab:helicity_amp}
\end{table}

With this we have obtained a complete overview of the general constraints
of the helicity amplitudes of $\gamma \gamma \to p \bar{p}$
following from rotational, parity, and charge-conjugation invariance
of strong and electromagnetic interactions.

Finally we note that the same analysis applies to any reaction
\begin{eqnarray}
\gamma + \gamma \to B + \bar{B} \,,
\label{2to2_general}
\end{eqnarray}
where $B$ stands for a spin $1/2$ baryon.
We only have to replace in all our formulas $m_{p}$ by $m_{B}$.
Interesting examples may be $B = \Lambda, \Sigma^{+}$, $\Lambda^{+}_{c}$.
\footnote{The $\Lambda$ baryon has a magnetic moment
$\mu_{\Lambda} = -0.613 \pm 0.004 \, \mu_{N}$ \cite{Olive:2016xmw}.
Thus, the reaction $\gamma \gamma \to \Lambda \bar{\Lambda}$
can proceed through the analogue of the diagrams
of Fig.~\ref{fig:diagram_2to2}~(a) and (b).}
The polarization of these baryons can be
obtained from their decay distributions.
\section{The $lS$ coupling scheme and helicity amplitudes for the reaction $\gamma \gamma \to f_{2} \to p \bar{p}$}
\label{section:Appendix2}

In this Appendix we discuss the relation of $lS$ couplings
to the helicity amplitudes for the reaction $\gamma \gamma \to f_{2} \to p \bar{p}$.
Here $l$ stands for the orbital angular momentum and
$S$ for the total spin of the $p \bar{p}$ system.

Let us see in how many ways one can construct a $p \bar{p}$ state
with $J^{PC} = 2^{++}$.
The partial-wave analysis (which is perfectly relativistic)
says that we can combine the spins
of $p$ and $\bar{p}$ to give the total spin $S = 0,1$.
Now we must combine this with the orbital angular momentum $l$ to 
the total angular momentum $J=2$.
This gives the four possibilities listed in Table~\ref{tab:lS_coupllings}.
In general we have the parity of $p \bar{p}$ state
$P = (-1)^{l+1}$ ($p$ and $\bar{p}$ have opposite intrinsic parity) 
and charge-conjugation $C = (-1)^{l+S}$.
There are, thus, two possible $(l,S)$ couplings for $f_{2}(2^{++}) \to p \bar{p}$:
$(1,1)$ and $(3,1)$.
\begin{table}[!ht]
\begin{tabular}{|c|c|c|}
\hline
$l$ & $S$ & $J^{PC}$ \\
\hline
2& 0& $2^{-+}$ \\
1& 1& $2^{++}$ \\
2& 1& $2^{--}$ \\
3& 1& $2^{++}$ \\
\hline
\end{tabular}
\caption{The $l$ and $S$ values leading to $p \bar{p}$ states with $J=2$.}
\label{tab:lS_coupllings}
\end{table}

We shall now analyze the $lS$ content of the $f_{2} p \bar{p}$ couplings
(\ref{A01}) and (\ref{A02}).

Let $u_{r}(p)$, $v_{r}(p)$ be the usual Dirac spinors with
spin in $\pm z$ direction for $r = \pm 1/2$; 
see (\ref{B_06}) and (\ref{B_18}).
For these we find in the c.m. system of reaction (\ref{2to2_reaction})
the matrix elements of the vertex functions
$\Gamma^{(f_{2}p\bar{p})(j)}_{\kappa \lambda}$ ($j = 1,2$)
(see (\ref{A03}), (\ref{A04}))
with $P^{(2)\, \kappa \lambda, \kappa' \lambda'}$
the spin~2 projector (the term in square brackets in (\ref{prop_f2})) as follows.
For $j=1$ we get
\begin{equation}
\begin{split}
& P^{(2)\, \kappa \lambda, \kappa' \lambda'}(p_{3}+p_{4})
\bar{u}_{r_{3}}(p_{3})
\Gamma^{(f_{2}p\bar{p})(j)}_{\kappa' \lambda'}(p_{3},p_{4})\,
v_{r_{4}}(p_{4}) = 0 \\
& {\rm unless} \;\kappa = k, \,\lambda = l,\quad k, l \in \{ 1,2,3 \}\,,
\end{split}
\label{C_01}
\end{equation}
\begin{equation}
\begin{split}
& P^{(2)\, kl, \kappa' \lambda'}(p_{3}+p_{4})
\bar{u}_{r_{3}}(p_{3})\, \Gamma^{(f_{2}p\bar{p})(1)}_{\kappa' \lambda'}(p_{3},p_{4})\,
v_{r_{4}}(p_{4}) \\
&=-\frac{4g^{(1)}_{f_{2}pp}}{M_{0}} F^{(f_{2} p \bar{p})(1)}[(p_{3}+p_{4})^{2}] \\
&\quad \;\times \chi_{r_{3}}^{\dagger}
\Big\lbrace 
-p_{3}^{\,0} 
\Big[ 
\frac{1}{2}p_{3}^{\,k} \sigma^{l} + \frac{1}{2}p_{3}^{\,l}\sigma^{k} 
- \frac{1}{3}\delta^{kl}(\bpp \cdot \bsigma)
\Big]\\
&\quad \; +\Big[ 
\frac{1}{2}p_{3}^{\,k} p_{3}^{\,l}+ \frac{1}{2}p_{3}^{\,l}p_{3}^{\,k} 
- \frac{1}{3}\delta^{kl}|\bpp|^{2}
\Big]
\frac{1}{p_{3}^{\,0}+m_{p}} (\bpp \cdot \bsigma)
\Big\rbrace 
\varepsilon \chi_{r_{4}}^{*}
\,.
\end{split}
\label{C_02}
\end{equation}
Here and in the following we set $\chi_{r} \equiv \chi_{r}^{(1)}$;
see (\ref{B_06}) and (\ref{B_18}). For $j=2$ we get
\begin{equation}
\begin{split}
& P^{(2)\, \kappa \lambda, \kappa' \lambda'}(p_{3}+p_{4})
\bar{u}_{r_{3}}(p_{3})
\Gamma^{(f_{2}p\bar{p})(2)}_{\kappa' \lambda'}(p_{3},p_{4})\,
v_{r_{4}}(p_{4}) = 0 \\
& {\rm for} \;\kappa = 0, \,\lambda \,{\rm arbitrary} \,{\rm and}\;
              \kappa \,{\rm arbitrary}, \lambda = 0 \,;
\end{split}
\label{C_03}
\end{equation}
\begin{equation}
\begin{split}
& P^{(2)\, kl, \kappa' \lambda'}(p_{3}+p_{4})
\bar{u}_{r_{3}}(p_{3})\, \Gamma^{(f_{2}p\bar{p})(2)}_{\kappa' \lambda'}(p_{3},p_{4})\,
v_{r_{4}}(p_{4}) \\
&=-\frac{8g^{(2)}_{f_{2}pp}}{M_{0}^{2}} F^{(f_{2} p \bar{p})(2)}[(p_{3}+p_{4})^{2}] \,
\Big[ p_{3}^{\,k} p_{3}^{\,l} - \frac{1}{3}\delta^{kl}|\bpp|^{2} \Big]
\bpp \cdot \chi_{r_{3}}^{\dagger} \bsigma \varepsilon \chi_{r_{4}}^{*}
\,.
\end{split}
\label{C_04}
\end{equation}

The $l$-$S$ amplitudes are as follows.
For $l=1$, $S=1$ we have
\begin{equation}
\begin{split}
{\cal A}^{kl}_{(1,1)}
=
\chi_{r_{3}}^{\dagger}
\Big[ 
\frac{1}{2}p_{3}^{\,k} \sigma^{l} + \frac{1}{2}p_{3}^{\,l}\sigma^{k} 
- \frac{1}{3}\delta^{kl}(\bpp \cdot \bsigma)
\Big]
\varepsilon \chi_{r_{4}}^{*}
\,.
\end{split}
\label{C_05}
\end{equation}

The traceless symmetric ($l=3$) tensor is
\begin{equation}
\begin{split}
T^{klm}_{3}
= p_{3}^{\,k} p_{3}^{\,l} p_{3}^{\,m}
-\frac{1}{5} |\bpp|^{2} 
\Big(
\delta^{kl} p_{3}^{\,m} + 
\delta^{km} p_{3}^{\,l} + 
\delta^{lm} p_{3}^{\,k}
\Big)
\,.
\end{split}
\label{C_06}
\end{equation}
This gives, for instance, with $\theta$ as defined in Fig.~\ref{fig:cm_system}
and $P_{3}$ the Legendre polynomial
\begin{equation}
\begin{split}
T^{klm}_{3} e_{z}^{\,k} e_{z}^{\,l} e_{z}^{\,m} 
= |\bpp|^{2} \,\frac{2}{5}
\Big(
\frac{5}{2} \cos^{3}\theta - 
\frac{3}{2} \cos\theta
\Big)
=|\bpp|^{2} \,\frac{2}{5} \,P_{3}(\cos\theta)
\,.
\end{split}
\label{C_07}
\end{equation}

The $l=3$, $S=1$ the amplitude is
\begin{equation}
\begin{split}
&T^{klm}_{3} 
\chi_{r_{3}}^{\dagger} \sigma^{m} \varepsilon \chi_{r_{4}}^{*}
= {\cal A}^{kl}_{(3,1)}\\
&= \Big[ 
p_{3}^{\,k} p_{3}^{\,l}
- \frac{1}{3}\delta^{kl}|\bpp|^{2}
\Big]
\chi_{r_{3}}^{\dagger} (\bpp \cdot \bsigma) \varepsilon \chi_{r_{4}}^{*}\\
& \quad \; -\frac{2}{5} |\bpp|^{2} 
\chi_{r_{3}}^{\dagger} 
\Big[ 
\frac{1}{2}p_{3}^{\,k} \sigma^{l} + \frac{1}{2}p_{3}^{\,l}\sigma^{k} 
- \frac{1}{3}\delta^{kl}(\bpp \cdot \bsigma)
\Big] 
\varepsilon \chi_{r_{4}}^{*}
\,.
\end{split}
\label{C_08}
\end{equation}

From (\ref{C_02}), (\ref{C_04}), (\ref{C_05}) and (\ref{C_08})
we get the $l$-$S$ decomposition of our couplings $j=1$ and 2 as follows:
\begin{equation}
\begin{split}
& P^{(2)\, kl, \kappa' \lambda'}(p_{3}+p_{4})\,
\bar{u}_{r_{3}}(p_{3})\, \Gamma^{(f_{2}p\bar{p})(1)}_{\kappa' \lambda'}(p_{3},p_{4})\,
v_{r_{4}}(p_{4}) \\
&=-\frac{4g^{(1)}_{f_{2}pp}}{M_{0}} F^{(f_{2} p \bar{p})(1)}[(p_{3}+p_{4})^{2}] \\
&
\quad \; \times \Big\lbrace 
-\Big( \frac{3}{5}p_{3}^{\,0} + \frac{2}{5} m_{p} \Big) \,{\cal A}^{kl}_{(1,1)}
+ \frac{1}{p_{3}^{\,0} + m_{p}} \,{\cal A}^{kl}_{(3,1)}
\Big\rbrace
\,,
\end{split}
\label{C_09}
\end{equation}
\begin{equation}
\begin{split}
& P^{(2)\, kl, \kappa' \lambda'}(p_{3}+p_{4})\,
\bar{u}_{r_{3}}(p_{3})\, \Gamma^{(f_{2}p\bar{p})(2)}_{\kappa' \lambda'}(p_{3},p_{4})\,
v_{r_{4}}(p_{4}) \\
&=-\frac{8g^{(2)}_{f_{2}pp}}{M_{0}^{2}} F^{(f_{2} p \bar{p})(2)}[(p_{3}+p_{4})^{2}] \\
&
\quad \; \times \Big\lbrace 
\frac{2}{5}\Big( (p_{3}^{\,0})^{2} - m_{p}^{2} \Big) \,{\cal A}^{kl}_{(1,1)}
+ {\cal A}^{kl}_{(3,1)}
\Big\rbrace
\,.
\end{split}
\label{C_10}
\end{equation}
Note that -- for $\chi_{r_{3}}$ and $\chi_{r_{4}}$
not depending on $\theta$ --
${\cal A}^{kl}_{(1,1)}$ clearly has only $l=1$
and ${\cal A}^{kl}_{(3,1)}$ clearly has only $l=3$;
see (\ref{C_05}) and (\ref{C_08}), respectively.

But now we can go to the helicity amplitudes.
All we have to do is to replace the two-component spinors
as follows
\begin{equation}
\begin{split}
& \chi_{r_{3}} \to \chi_{s_{3}}^{(a)}(\bhpp)
\;{\rm from \;(\ref{B_04})\;
with \;the \;replacements}\; \theta \to \theta, \,\phi \to 0\,,\\
&\chi_{r_{4}} \to \chi_{s_{4}}^{(b)}(-\bhpp)
\;{\rm from \;(\ref{B_11})\;
with \;the \;replacements}\; \theta \to \pi-\theta,\, \phi \to \pi
\,.
\end{split}
\label{C_11}
\end{equation}
Note that these spinors depend on $\theta$.

We get
\begin{equation}
\begin{split}
&
\left( \chi_{s_{3}}^{(a)\dagger} \varepsilon \;\chi_{s_{4}}^{(b)*} \right)
= \left( \begin{array}{cc}
1 &  0 \\
0 & -1 \\
\end{array} \right)
= \left( \sigma^{3}_{s_{3}s_{4}} \right),
\\
&
\left( \chi_{s_{3}}^{(a)\dagger} \sigma^{1}\varepsilon \;\chi_{s_{4}}^{(b)*} \right)
= \left( \begin{array}{cc}
\sin\theta & -\cos\theta \\
\cos\theta &  \sin\theta \\
\end{array} \right),
\\
&
\left( \chi_{s_{3}}^{(a)\dagger} \sigma^{2}\varepsilon \;\chi_{s_{4}}^{(b)*} \right)
= \left( \begin{array}{cc}
0 & i \\
i & 0 \\
\end{array} \right),
\\
&
\left( \chi_{s_{3}}^{(a)\dagger} \sigma^{3}\varepsilon \;\chi_{s_{4}}^{(b)*} \right)
= \left( \begin{array}{cc}
 \cos\theta & \sin\theta \\
-\sin\theta & \cos\theta \\
\end{array} \right),
\\
&
\left( \chi_{s_{3}}^{(a)\dagger} \bhpp \cdot \bsigma \varepsilon \;\chi_{s_{4}}^{(b)*} \right)
= \delta_{s_{3}s_{4}}
\,,
\\
&
\; \bhpp = \bpp/|\bpp|
\,.
\end{split}
\label{C_12}
\end{equation}
Inserting these expressions in (\ref{C_02}) and (\ref{C_04})
we see that the $\bpp$ dependence, that is,
the $\theta$ dependence of the amplitudes
will in general be changed.
Take, for instance, (\ref{C_04})
which is a combination of $l =3$ plus $l=1$;
see (\ref{C_05}) and (\ref{C_08}).
With the replacements (\ref{C_11}) we get from (\ref{C_12})
\begin{equation}
\begin{split}
& 
\Big[ p_{3}^{\,k} p_{3}^{\,l} - \frac{1}{3}\delta^{kl}|\bpp|^{2} \Big]
\chi_{s_{3}}^{(a)\dagger} \bpp \cdot \bsigma \varepsilon \chi_{s_{4}}^{(b)*}
=
\Big[ p_{3}^{\,k} p_{3}^{\,l} - \frac{1}{3}\delta^{kl}|\bpp|^{2} \Big]
|\bpp|\,\delta_{s_{3}s_{4}}
\,.
\end{split}
\label{C_13}
\end{equation}
From $l=3$ plus $l=1$ we go, effectively, to $l=2$.

The replacements (\ref{C_11}) lead from (\ref{C_02}) and (\ref{C_04}),
using the expression for the diagram for $\gamma \gamma \to f_{2} \to p \bar{p}$
[see Fig.~\ref{fig:diagram_2to2}~(c)],
to the helicity amplitudes (\ref{hel0_ppbar}) and (\ref{hel2_ppbar}).

\section{Phase conventions}
\label{section:Appendix3}
For the hand-bag contribution, Sec.~\ref{sec:hand_bag},
we must take into account different phase conventions
used in \cite{Diehl:2002yh} relative to ours,
as explained in Appendix~\ref{section:Appendix1}.
In \cite{Diehl:2002yh} the orientation of the particle
momenta corresponds to a rotation by $\frac{\pi}{2} - \theta$
relative to the momenta in Fig.~\ref{fig:cm_system}.
Considering this we find that their spinors for
proton and antiproton correspond to our
$u_{s_{3}}^{(h,a)}(p_{3})$ and $-2s_{4}\,v_{s_{4}}^{(h,b)}(p_{4})$, respectively.
The phase conventions for the photons are not stated explicitly in \cite{Diehl:2002yh}.
From a comparison 
\footnote{We thank M.~Diehl for correspondence on this point.}
of the calculations (22) and (23) of \cite{Diehl:2002yh}
with the corresponding ones with our conventions we conclude that
the $| \gamma(k_{1},\pm), \gamma(k_{2},\mp) \rangle$ states of \cite{Diehl:2002yh}
have an extra minus sign compared to ours.
Taking everything together we obtain (\ref{hb_amplitudes}) for the amplitudes.

\bibliography{refs}

\end{document}